\documentclass[a4paper,11pt]{article}
\pdfoutput=1

\usepackage{jheppub}
\usepackage[T1]{fontenc} 
\usepackage{bm}
\usepackage{mathrsfs}
\usepackage{bbold}
\usepackage{amsbsy}
\usepackage{braket}
\usepackage{amssymb}
\usepackage{diagbox}
\usepackage{multirow}
\usepackage{array}
\usepackage{graphicx}
\usepackage[center]{subfigure}
\usepackage{subfigure}
\usepackage{amsmath}
\usepackage{amsfonts}
\usepackage{xcolor}
\usepackage{makecell}

\newcommand{\be}{\begin{equation}}
\newcommand{\ee}{\end{equation}}
\newcommand{\bea}{\begin{eqnarray}}
\newcommand{\eea}{\end{eqnarray}}

\title{\boldmath On the Replica Problem in Supersymmetric SYK Models}

\author[a,1]{Xian-Hui Ge,\note{Corresponding author.}}
\author[a]{Chenhao Zhang}

\affiliation[a]{Department of Physics, College of Sciences, Shanghai University,\\99 Shangda Road, 200444 Shanghai, China}
\emailAdd{gexh@shu.edu.cn}
\emailAdd{zhangchenhao@shu.edu.cn}

\abstract{We investigate the replica problem for  Sachdev-Ye-Kitaev (SYK) models. First, we consider $n-$replicas  of the non-supersymmetric SYK model, finding that this $n$-replica model is solvable only under specific conditions. We then introduce the $\mathcal{N}=1$ supersymmetry and utilize the superconformal symmetry to develop a ``multi-ordered trick" that covers the replica structure. By incorporating ordered off-diagonal couplings, we study the resulting thermal phase structure under higher-order interactions. The Lorentzian time dynamics is analyzed, and we plot the time evolution of the effective action. Furthermore, we investigate emergent superconformal symmetry in the low-energy limit of the replicated theory. In the superconformal limit, we propose an ordered super-Schwarzian action and derive reparameterization relations for the ordered coordinates. Corresponding constraints are derived for holographic matching to $\mathcal{N}=1$ super-Jackiw-Teitelboim(SJT) gravity. We numerically calculate the modular thermodynamics (modular entropy, the $n$-dependent relative entropy, and the entanglement capacity) using fully diagonal methods. Our result provide a framework for studying $n$-replica wormholes with supersymmetric SYK model.}

\begin{document}
\maketitle
\flushbottom

\section{Introduction} \label{intro}
\quad The Sachdev-Ye-Kitaev (SYK) model~\cite{1,2,3,4}, a solvable model of $q$-body interacting Majorana fermions with random Gaussian couplings, exhibits remarkable holographic properties. Its emergent infrared conformal symmetry realizes the nearly $\mathrm{AdS}_2$/nearly $\mathrm{CFT}_1$  ($\mathrm{NAdS}_2/\mathrm{NCFT}_1$) duality~\cite{5,6}, with connections to black hole physics~\cite{7,8}. The supersymmetric extensions of the SYK model with $\mathcal{N} \geq 1$ supersymmetry~\cite{9,10,11} incorporate dynamical bosonic degrees of freedom while preserving the essential features of solvability and emergent conformal symmetry. Notably, $T\bar{T}$ deformations in these models exhibit emergent supersymmetric conformal symmetry, which admits a holographic interpretation in terms of $\mathcal{N} = 1,2$ supersymmetric $\mathrm{NAdS}_2$ spacetimes endowed with Grassmann-valued geometric structures~\cite{12,13,14,BRSZ,TT}. Recent works~\cite{15,16,17,18,180,1801} have further illuminated the rich infrared dynamics and correlation structures arising in these theories.

Recently, there has been a large discussion on the generalized island formula for entanglement entropy derived from the black hole information paradox. Ref.\cite{1908.10996} builds on the island formula derived from the black hole information paradox~\cite{Hawking:1976}. Mathematically, this is expressed through the replica wormhole method in gravity path integral formulation~\cite{Penington:2019,Almheiri:2019}. For dynamical gravity, the implementation of replica trick derives from the entropy formula~\cite{Penington:2020,Faulkner:2013}, revealing replica wormhole saddle points in the geometric path integral~\cite{Page:1993,Page:2013}. The island formula fundamentally encodes the von Neumann entropy of Hawking radiation (or thermal dynamics). The recent progresses of n-dependence relative entropy proposed in \cite{R1,R2,301,R3} highlight the importance of the replica parameter `$n$'. However, to extract more complete state information, we must also employ additional measures like Rényi entropy or extended modular entropy.

In this paper, we study off-diagonal couplings between distinct supersymmetric SYK (SSYK) systems that preserve both supersymmetry and exact solvability~\cite{OFF}. For the case of $\mathcal{N} = 1$, such couplings can be interpreted as supersymmetric traversable wormhole interactions, and are shown to induce nontrivial patterns of entanglement while preserving holographic duality. Our analysis demonstrates how these couplings alter the spectral statistics and entanglement structure, revealing new insights into the interplay between supersymmetry, quantum chaos, and emergent spacetime geometry. Moreover, we extend the SYK wormhole construction to arbitrary $n$ replicas. By introducing first-order interaction terms in the supersymmetric SYK model, we preserve exact solvability. The SSYK model with wormhole effects (e.g., \cite{20,21,22,23,24,25}) enables further analysis of this extension. Our model is primarily based on the $\mathcal{N}=1$ coupled SYK model~\cite{OFF}, and we explore its low-energy effective action~\cite{28,29}, thermal phase structure, and time evolution. While prior work established a minimal $\mathcal{N}\!=\!1$ Jackiw-Teitelboim(JT) gravity supersymmetrization with robust holographic properties~\cite{30,31}, we introduce additional constraints to this framework.

This work generalizes the two-site SYK wormhole construction (eternal traversable wormholes~\cite{1804.00491}, thermal bath evaporation~\cite{1909.10637}, and double cone configurations~\cite{Penington:2019,1708.00871,1806.06840}) to $n$-site interactions, resolving replica dynamics through the SYK framework. We demonstrate that the $\mathcal{N}\!=\!1$ supersymmetry suffices to encapsulate all $n$-replica interactions simultaneously, whereas SYK chains \cite{2410.23397} would require analytically intractable higher-order off-shell terms.

The first goal of this paper is to present a discussion on solvable SYK replica wormholes by using the ``multi-ordered trick``. We focus on the natural fractal symmetry in the $\mathcal{N}\!=\!1$ SSYK model: first-order interactions automatically generate all $n$-replica terms. We focus on how the $n$-replica interactions as replica wormhole influence the thermal phase structures, statistical properties. We also consider the real time dynamics and obtain the effective action with time-evolution. Furthermore, we evaluate the modular entropy, n-dependent relative entropy, and the entanglement capacity.

The second objective is to investigate JT gravity as a gravity dual to $n$-copies of SYK. While prior studies have examined quantum extremal surfaces and information recovery in JT gravity~\cite{2406.16339,2209.11774}, we focus specifically on $\mathcal{N}=1$ supersymmetric JT gravity, which naturally incorporates replica structures. We begin with the emergent superconformal symmetry, extracting three key constraints of the $\mathcal{N}=1$ SSYK model that persist under both global symmetries and in the low-energy limit. These constraints are imposed on $\mathcal{N}=1$ super-JT gravity, which is intrinsically embedded in supersymmetric $\text{AdS}_2$ spacetime.

This paper is organized as follows. In Section~\ref{S2}, we expand the two copies of SYK models to $n$-replica via the permutation of the indices, and we obtain a special solution in the thermal limit. In Section~\ref{S3}, we focus on the $\mathcal{N}\!=\!1$ supersymmetric SYK model, which naturally incorporates the $n$-replica problem while maintaining solvability. We propose a modified $\mathcal{N}\!=\!1$ supersymmetric SYK model that simultaneously handles both copies and interaction orders. In Section~\ref{S4}, we analyze the fractal properties of the $\mathcal{N}\!=\!1$ supersymmetric SYK model and supersymmetry breaking, explaining how this model resolves the replica problem. In Section~\ref{S5}, we examine multi-boundary superconformal symmetry in the low-energy limit and constrain super-JT gravity to establish holographic correspondence. Section~\ref{S6} develops the fully diagonal method and finite-element solutions for replicated thermodynamics, including modular entropy, relative entropy, and capacity of entanglement. Section~\ref{S7} presents conclusions and discussion.

\section{Special SYK with many replicas}\label{S2}

\quad In this section, we provide a method to extend the two-coupled system (as shown in \cite{Penington:2019}) to arbitrary $n$-replicated systems. Using permutation theory, we can reduce these problems to a two-coupled situation. However, this process only works under special conditions. 

\subsection{Expand the n-replicated interaction}
\quad To study SYK systems with replica wormhole solutions, we begin with considering one SYK system
\begin{equation}
H_{a}^{\alpha}=\sum_{1 \leq i_{a1}< \cdot \cdot < i_{aq} \leq N_{a}} {J_{i_{a1}...i_{aq}}} \psi_{i_{a1}}^{\alpha}\cdot \cdot \cdot \psi_{i_{aq}}^{\alpha},\label{SYK}
\end{equation}
For a system of $n$ SYK replicas, we can define a more general fermion operator with three identical indices
\begin{equation}
\begin{split}
\psi _{i_{an}}^{\alpha}:&i_{an}\in \left\{ 1,\cdot \cdot \cdot ,N_a \right\} ={\rm flavor},
\\
a\in \left\{ 1,2,\cdot \cdot \cdot \right\} &={\rm interacting~ physical~ system},
\\
\alpha& \in \left\{ 1,\cdot \cdot \cdot ,n \right\} ={\rm replicas}. 
\end{split}
\end{equation}

We initiate calculations with two replicas ($n=2$), incorporating the replicated interaction term $V$ between systems. The intrareplica interactions take the form
\begin{equation}
V^{\alpha \alpha '}=\sum_{\begin{array}{c}
	1\leqslant i_{11}<...<i_{1\bar{q}}\leqslant N\\
	1\leqslant i_{21}<...<i_{2\bar{q}}\leqslant N\\
\end{array}}{\bar{J}_{i_1...i_{\bar{q}},...,i_1...i_{\bar{q}}}}\psi _{i_{11}}^{\alpha}...\psi _{i_{1\bar{q}}}^{\alpha}\psi _{i_{21}}^{\alpha '}...\psi _{i_{2\bar{q}}}^{\alpha '},\label{rand}
\end{equation}
where the parameters $J$(random parameter in \eqref{SYK}) and $\bar{J}$(coupling parameter in \eqref{rand}) are averaged over their random distributions
\begin{equation}
\begin{split}
\left< J_{i_{a1}...i_{aq}}J_{i_{a1}^\prime...i_{aq}^\prime} \right> &=J^2\delta _{i_{a1},i'_{a1}}...\delta _{i_{aq},i'_{aq}}\delta _{a,a'}\frac{\left( q-1 \right) !}{(N_a)^q},
\\
\left< \bar{J}_{i_{11}...i_{1\bar{q}},i_{21}...i_{2\bar{q}}}\bar{J}_{i_{11}^\prime...i_{1\bar{q}}^\prime,i_{21}^\prime...i_{2\bar{q}^\prime}} \right> &=\bar{J}^2\delta _{i_{11},i_{11}^\prime}...\delta _{i_{1\bar{q}},i_{1\bar{q}}^\prime}\delta _{i_{21},i_{21}^\prime}...\delta _{i_{2\bar{q}},i_{2\bar{q}}^\prime}\frac{\left( \bar{q}! \right) ^2}{q(N_1N_2)^{\bar{q}}}.
\end{split}
\end{equation}
The interaction between two replicas decomposes into four distinct components:
\begin{equation}
V=\int_{C_1}{d\tau\left( V^{11}+V^{22} \right)}+\int_{C_2}{d\tau\left( V^{12}+V^{21} \right)}=\int_C{d\tau\sum_{\alpha \gamma}{V^{\alpha \gamma}}g^{\alpha \gamma}\left( \tau \right)},\label{rand1}
\end{equation}
which encode all possible replicated interactions. We introduce a $\tau$-dependent density function $g^{\alpha \gamma}$ to parameterize the integration along the contours $C_1$ and $C_2$ (as illustrated in Figures~\ref{fig:1} and \ref{fig:2}). We find that these interactions exhibit a $2q$-ordered structure. The tensor $g^{\alpha \gamma}$, with all possible index combinations, characterizes interactions between distinct replica sectors occurring at specific Euclidean times. The corresponding saddle-point equations admit two classes of solutions: the \emph{diagonal} solutions ($V^{11}$ and $V^{22}$), and the \emph{wormhole} solutions ($V^{12}$ and $V^{21}$).

The total Hamiltonian comprises two identical SYK Hamiltonians with replica indices and an interaction term
\begin{equation}
H_{total}=H_{L}+H_{R}+V,
\end{equation}
where the action can be expressed as
\begin{equation}
I=\int_C{d\tau}\sum_{\alpha =1...n}{\sum_{a=1,2}{\sum_{i_{an}}{\left( \psi _{i_{an}}^{\alpha}\partial _{\tau}\psi _{i_{an}}^{\alpha}+H_{a}^{\alpha} \right)}+\int_C{d\tau \,\,V}}}.
\end{equation}
The second term represents the replicated interaction. Its disorder average yields the effective action in terms of Green's functions $V\rightarrow \mathcal{V} $
\begin{equation}
\mathcal{V} =\frac{1}{2}\frac{\bar{J}^2}{q}\int_C{d\tau _1d\tau _2}\sum_{\alpha \alpha \gamma \gamma}{G_{L}^{\alpha \alpha '}}\left( \tau _1,\tau _2 \right) ^{\bar{q}}g^{\alpha \gamma}\left( \tau _1 \right) g^{\alpha '\gamma '}\left( \tau _2 \right) G_{R}^{\gamma \gamma '}\left( \tau _1,\tau _2 \right) ^{\bar{q}}.\label{interac}
\end{equation}
Here we use $\mathcal{V}$ to represent the disordered-averaged interactions \eqref{rand1} via two-point function. The mathematic form can be naturally interpreted as the multiples of the separated copies, and the $n$-th replica problem will be considered subsequently. These interactions between two replica systems and could be extented, we will give a further discussion in the following subsections. 

We will use the parameter $\mathcal{V}$ to denote the averaged interaction strength, and take the three-replica interaction terms as an illustrative example for further discussion in the next subsection.

\subsection{SYK with 3 replicas}

\quad The different replicas interact through multiple coupling channels. The disorder-averaged solution is most efficiently represented using Green's functions. We employ a technique that decomposes interactions with two-point correlators, enabling subsequent reconstruction from these fundamental components.

\begin{figure}[!t]
\begin{minipage}{0.38\linewidth}
\centerline{\includegraphics[width=11cm]{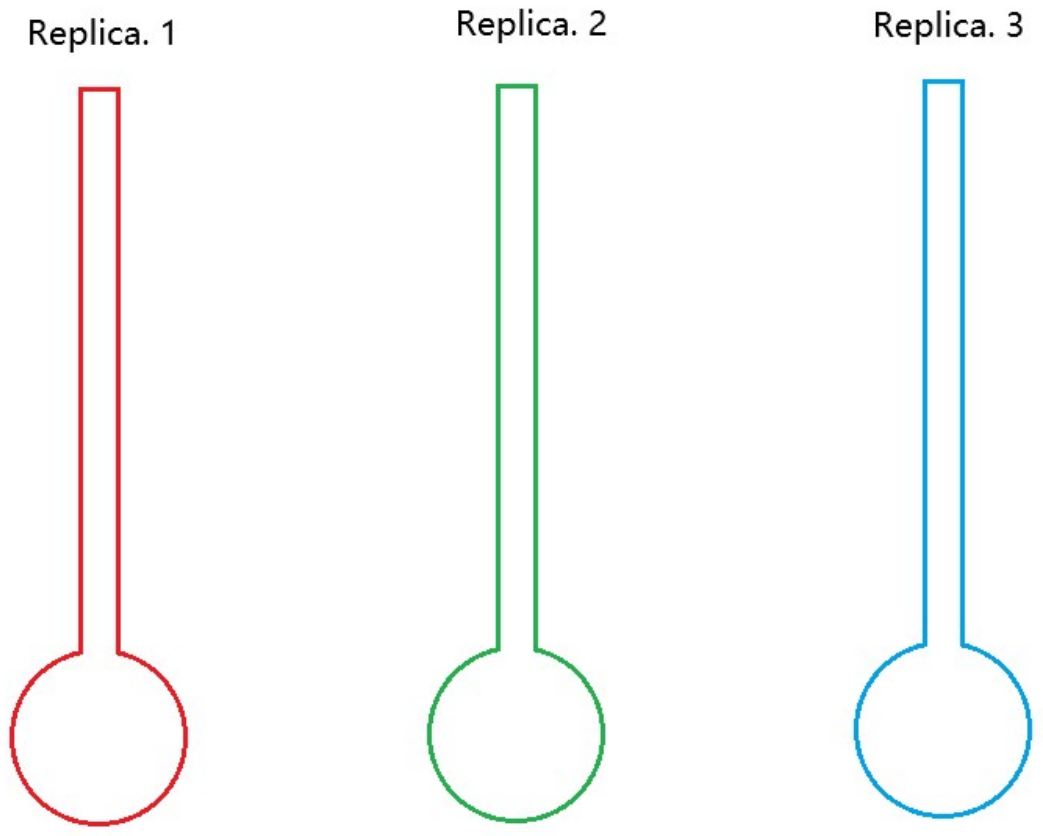}}
\centerline{(a)}
\end{minipage}
\hfill
\begin{minipage}{0.38\linewidth}
\centerline{\includegraphics[width=11cm]{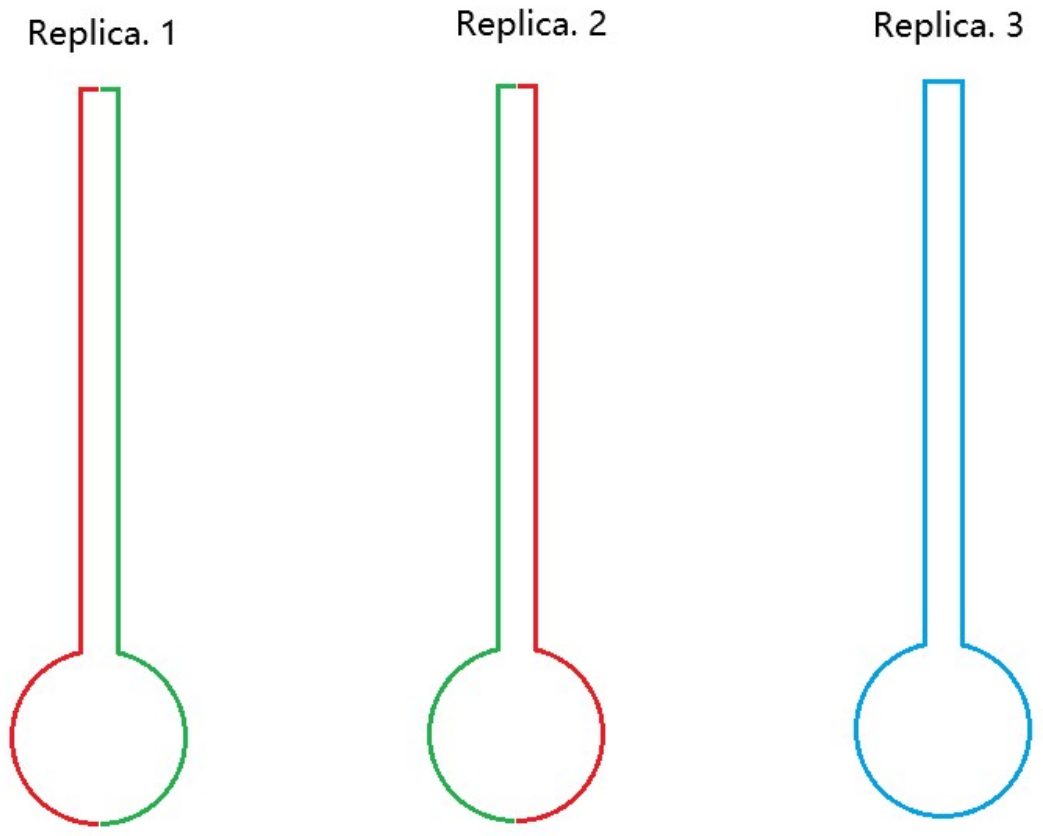}}
\centerline{(b)}
\end{minipage}
\vfill
\begin{minipage}{0.38\linewidth}
\centerline{\includegraphics[width=11cm]{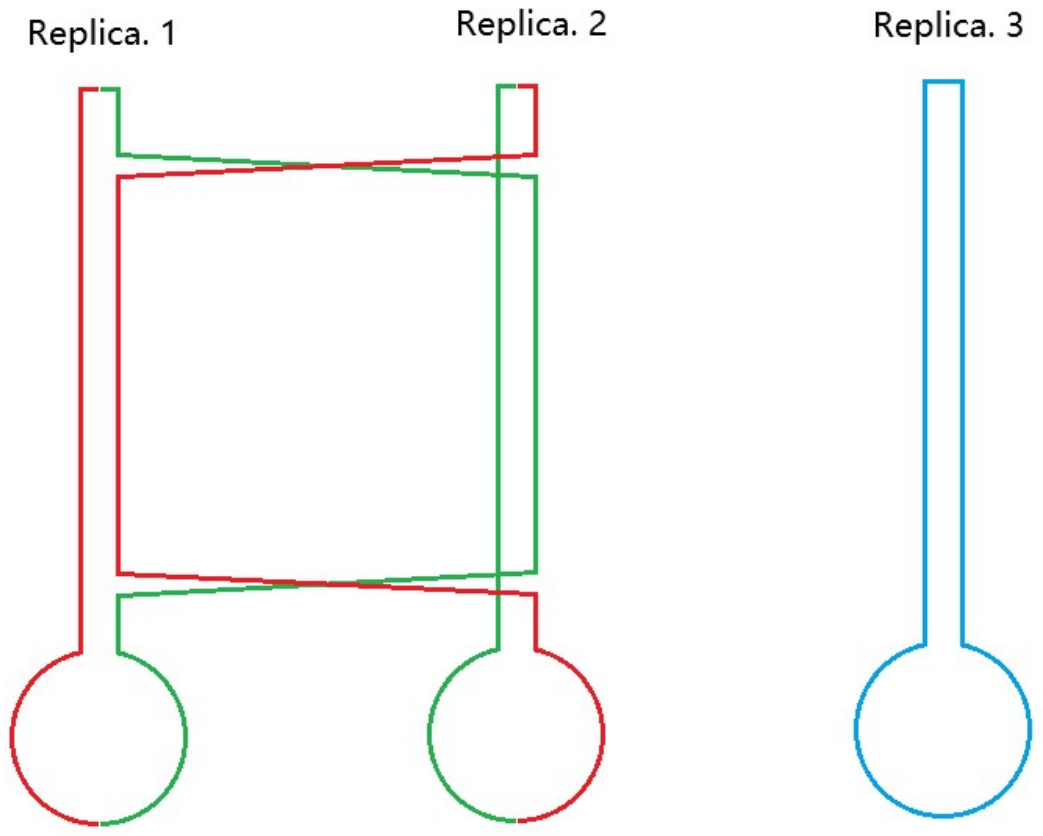}}
\centerline{(c)}
\end{minipage}
\hfill
\begin{minipage}{0.38\linewidth}
\centerline{\includegraphics[width=11cm]{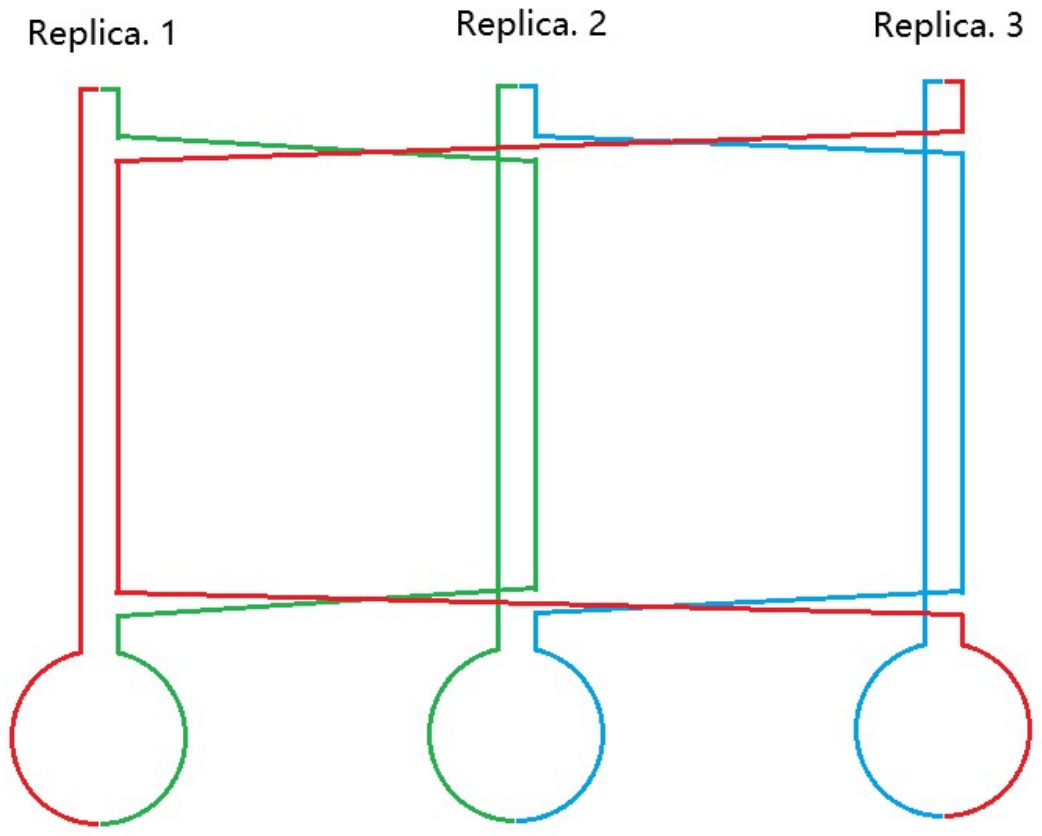}}
\centerline{(d)}
\end{minipage}
\caption{\label{fig:1}  The path integral contours for the action with 3-replicas: (a) Contour without interaction between replica. 1 and replica. 2. (b) Contour with diagonal interactions $V^{11}$ and $V^{22}$. (c) Contour with non-diagonal interactions $V^{12}$ and $V^{21}$. (d) 3-replica interaction contour $V^{123}$. }
\end{figure}

For example, we consider the three-replica interaction terms 
\begin{equation}
V^{\alpha _1\alpha _2\alpha _3}=\sum_{\begin{array}{c}
	1\leqslant i_{11}<...<i_{1\bar{q}}\leqslant N\\
	1\leqslant i_{21}<...<i_{2\bar{q}}\leqslant N\\
	1\leqslant i_{31}<...<i_{3\bar{q}}\leqslant N\\
\end{array}}{\bar{J}_{i_{11}...i_{1\bar{q}},...,i_{n1}...i_{n\bar{q}}}}\psi _{i_{11}}^{\alpha _1}...\psi _{i_{1\bar{q}}}^{\alpha _1}...\psi _{i_{31}}^{\alpha _3}...\psi _{i_{3\bar{q}}}^{\alpha _3},
\end{equation}
where the three-replica interaction terms incorporates random couplings
\begin{equation}
\left< \bar{J}_{i_{11}...i_{1\bar{q}}...i_{31}...i_{3\bar{q}}}\bar{J}_{i'_{11}...i'_{1\bar{q}}...i'_{31}...i'_{3\bar{q}}} \right> =J^2\delta _{i_{11}i'_{11}}...\delta _{i_{1\bar{q}}i'_{1\bar{q}}}...\delta _{i_{31}i'_{31}}...\delta _{i_{3\bar{q}}i'_{3\bar{q}}}\frac{\left( \bar{q}! \right) ^3}{\bar{N}}.
\end{equation}
We provide a complete enumeration of all possible terms of the 3-replica interaction contributing to the interaction potential \( V \) within this framework 
\begin{equation}
\begin{split}
V&=\int_{C_1}{\left( V^{11}+V^{22}+V^{33} \right)}+\int_{C_2}{\left( V^{12}+V^{21}+V^{13}+V^{31}+V^{23}+V^{32} \right)}
\\
&+\int_{C_2}{\left( V^{123}+V^{213}+V^{132}+V^{312}+V^{231}+V^{321} \right)}=\int_C{\sum_{\alpha \beta \gamma}{V^{\alpha \beta \gamma}}g^{\alpha \beta \gamma}\left( \tau \right)}.\label{V123}
\end{split}
\end{equation}
We introduce the function \( g^{\alpha \beta \gamma}(\tau) \) to encode interactions among replica sectors labeled by \( \alpha, \beta, \gamma \), with explicit Euclidean-time dependence through \( \tau \). The ordering of the indices specifies the sequence of interaction processes in replica space. The associated integration contour for this interaction is illustrated in Figure~\ref{fig:1}. Other replicated interactions can be represented in an analogous manner, including the two-replica couplings \(V^{23}\) and \(V^{32}\), \(V^{13}\) and \(V^{31}\), the diagonal terms \(V^{22}\), \(V^{11}\), and \(V^{33}\), as well as higher-order processes such as \(V^{132}\), \(V^{321}\), and \(V^{213}\). Here, the indices of \(V\) correspond to elements of the permutation group acting on the replica labels.

We can express the 3-replica interaction term via the disorder-averaged result of \eqref{V123}, that is 
\begin{equation}
\begin{split}
\mathcal{V} =\frac{1}{2}\frac{\bar{J}^2}{q}\int_C{d\tau _1d\tau _2}&\sum_{\alpha \alpha \beta \beta \gamma \gamma}{\left( G_{L}^{\alpha \alpha '}\left( \tau _1,\tau _2 \right) ^{\bar{q}/2}g^{\alpha \beta}\left( \tau _1 \right) g^{\alpha '\beta '}\left( \tau _2 \right) G_{R}^{\beta \beta '}\left( \tau _1,\tau _2 \right) ^{\bar{q}/2} \right)}
\\
&\left( G_{L}^{\beta \beta '}\left( \tau _1,\tau _2 \right) ^{\bar{q}/2}g^{\beta \gamma}\left( \tau _1 \right) g^{\beta '\gamma '}\left( \tau _2 \right) G_{R}^{\gamma \gamma '}\left( \tau _1,\tau _2 \right) ^{\bar{q}/2} \right) 
\\
&\left( G_{L}^{\gamma \gamma '}\left( \tau _1,\tau _2 \right) ^{\bar{q}/2}g^{\gamma \alpha}\left( \tau _1 \right) g^{\gamma '\alpha '}\left( \tau _2 \right) G_{R}^{\alpha \alpha '}\left( \tau _1,\tau _2 \right) ^{\bar{q}/2} \right) .
\end{split}
\end{equation}

Note that each interaction operator appears twice, corresponding to the path integral with interactions at $\tau_1$ and $\tau_2$. This requires modifying the Green's function indices by $\bar{q}/2$. This mathematical form explains why the 3-replica systems coexists with 2-replica parts. Here we use the $g^{\alpha \beta}\left( \tau_1 \right)$ and $g^{\alpha' \beta'}\left( \tau_2 \right)$ to denote the two-point function $G^{\alpha \alpha '}$ interact with $G^{\beta \beta '}$ at a certain Lorentzian time $\tau_1$ and $\tau_2$. The function $g^{\alpha \beta \gamma}\left( \tau \right)$ will return to the $g^{\alpha \beta}\left( \tau \right)$ when the number of indices reduces to 2.

The coefficient $g^{\alpha \alpha}$ in the so-called full diagonal solution in \cite{Penington:2019} is identical to the time-independent density matrix and does not contribute to the replica interaction term $\mathcal{V}$. In this condition, the multiple interaction terms $\mathcal{V}$ do not contribute to time dependence.  We can use the equation to describe both 2-replica and 3-replica interaction terms. For example, the condition $\alpha=\gamma$ in a 3-replica system describes possible 2-replica contributions.

\begin{figure}[!t]
\begin{minipage}{0.38\linewidth}
\centerline{\includegraphics[width=8cm]{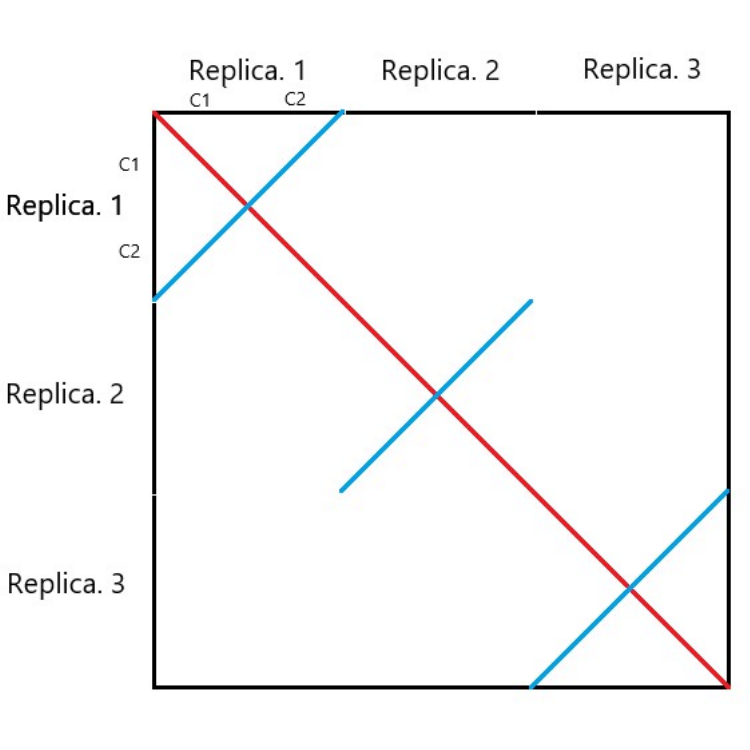}}
\centerline{(a)}
\end{minipage}
\hfill
\begin{minipage}{0.38\linewidth}
\centerline{\includegraphics[width=8cm]{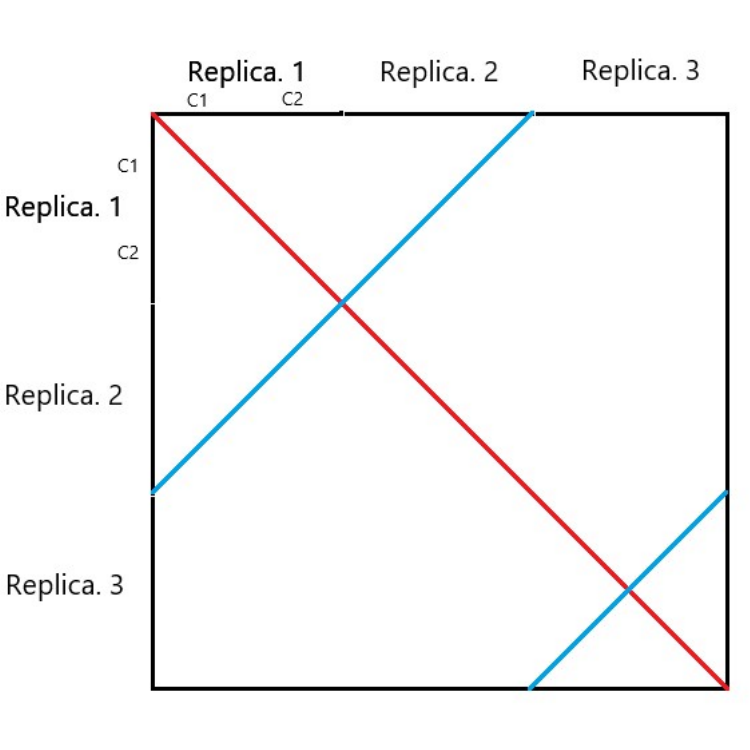}}
\centerline{(b)}
\end{minipage}
\vfill
\begin{minipage}{0.38\linewidth}
\centerline{\includegraphics[width=8cm]{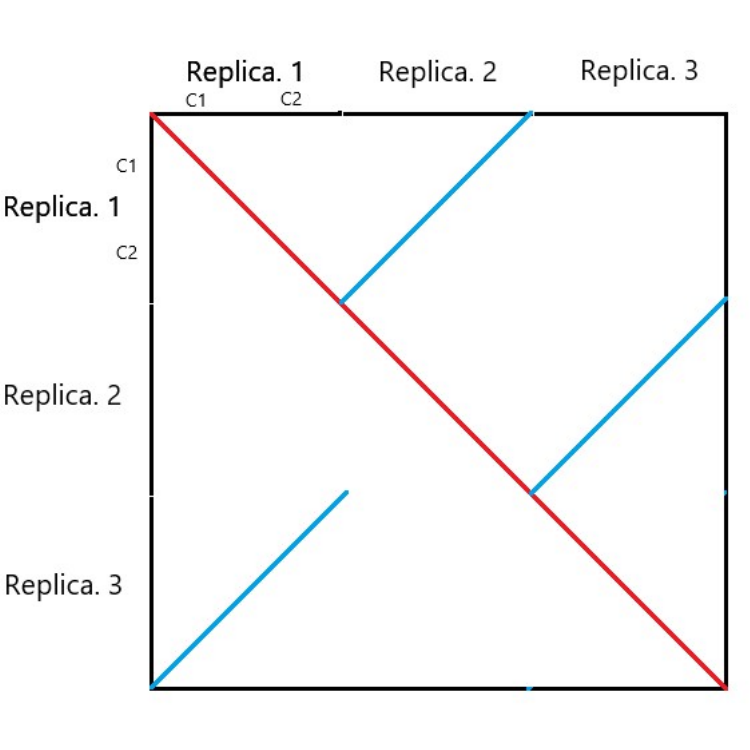}}
\centerline{(c)}
\end{minipage}
\hfill
\begin{minipage}{0.38\linewidth}
\centerline{\includegraphics[width=8cm]{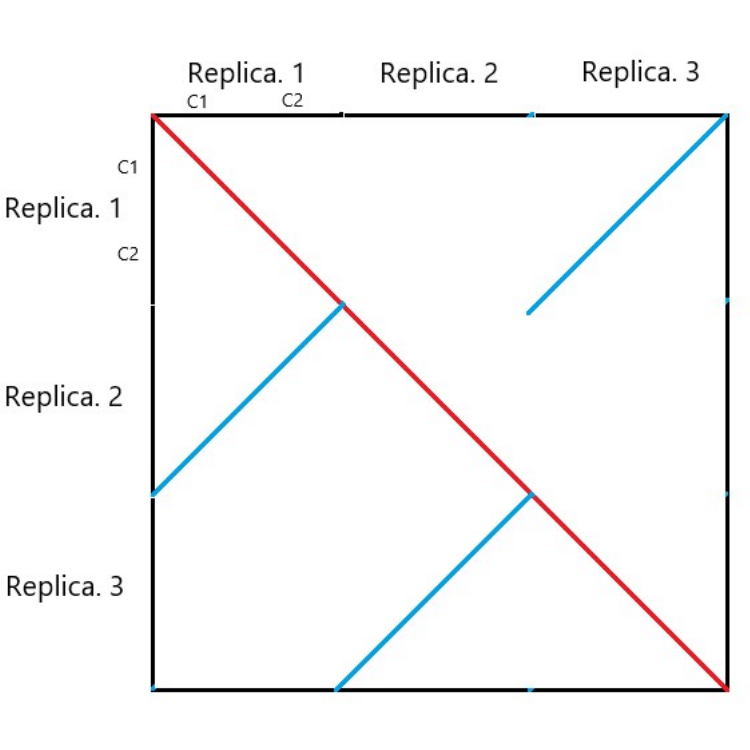}}
\centerline{(d)}
\end{minipage}
\caption{\label{fig:2} The Hamiltonian for 3-replica SYK as a function of Lorentzian time contours: The red line represents the SYK contribution, while the blue line represents the interaction contribution. (a) Diagonal interaction within replicas. (b) Non-diagonal interaction between replica. 1 and replica. 2. (c) 3-replica non-diagonal interaction $V^{123}$. (d) 3-replica non-diagonal interaction $V^{132}$. }
\end{figure}

We illustrate the contribution of replicated interactions to the total Hamiltonian in Figure~\ref{fig:2}, explicitly incorporating time dependence. This depiction aligns with the Green's function heat map presented in previous work~\cite{Penington:2019}. By adjusting the density matrix \( g \), one can generate different diagrammatic forms of the interactions.

\subsection{SYK with n replicas}
\quad We can extend the theory to larger replica cases. For example, the 4-replica interaction terms can be expressed as
\begin{equation}
V=\int_{C_2}{d\tau \left( V^{\alpha \beta} \right)}+\int_{C_2}{d\tau \left( V^{\alpha \beta \gamma} \right)}+\int_{C_2}{d\tau \left( V^{\alpha \beta \gamma \kappa} \right)}=\int_C{d\tau \sum_{\alpha \beta \gamma \kappa}{V^{\alpha \beta \gamma \kappa}}g^{\alpha \beta \gamma \kappa}\left( \tau \right)}.
\end{equation}
This interaction involves both 2,3,4-replica interaction terms, where $\alpha \beta \gamma \kappa=1,2,3,4$. Here we use the notation $g^{\alpha \beta \gamma \kappa}\left( \tau \right)$ to represent the interaction between different replicas with indices $\alpha \beta \gamma\kappa$.

The thermal interaction is given by corresponding disorder-averaged Green's function
\begin{equation}
\begin{split}
\mathcal{V} =\frac{1}{2}\frac{\bar{J}^2}{q}\int_C{d\tau _1d\tau _2}&\sum_{\alpha \alpha \beta \beta \gamma \gamma \kappa \kappa}{\left( G_{L}^{\alpha \alpha '}\left( \tau _1,\tau _2 \right) ^{\bar{q}/2}g^{\alpha \beta}\left( \tau _1 \right) g^{\alpha '\beta '}\left( \tau _2 \right) G_{R}^{\beta \beta '}\left( \tau _1,\tau _2 \right) ^{\bar{q}/2} \right)}
\\
&\left( G_{L}^{\beta \beta '}\left( \tau _1,\tau _2 \right) ^{\bar{q}/2}g^{\beta \gamma}\left( \tau _1 \right) g^{\beta '\gamma '}\left( \tau _2 \right) G_{R}^{\gamma \gamma '}\left( \tau _1,\tau _2 \right) ^{\bar{q}/2} \right) 
\\
&\left( G_{L}^{\gamma \gamma '}\left( \tau _1,\tau _2 \right) ^{\bar{q}/2}g^{\gamma \kappa}\left( \tau _1 \right) g^{\gamma '\kappa '}\left( \tau _2 \right) G_{R}^{\kappa \kappa '}\left( \tau _1,\tau _2 \right) ^{\bar{q}/2} \right) 
\\
&\left( G_{L}^{\kappa \kappa '}\left( \tau _1,\tau _2 \right) ^{\bar{q}/2}g^{\kappa \alpha}\left( \tau _1 \right) g^{\kappa '\alpha '}\left( \tau _2 \right) G_{R}^{\alpha \alpha '}\left( \tau _1,\tau _2 \right) ^{\bar{q}/2} \right) ,
\end{split}
\end{equation}

This result generalizes to $n$-replica systems with arbitrary indices
\begin{equation}
\begin{split}
V&=\int_{C_2}{d\tau \left( V^{\alpha_1\alpha_2} \right)}+\int_{C_2}{d\tau \left( V^{\alpha_1\alpha_2\alpha_3} \right)}+...+\int_{C_2}{d\tau \left( V^{\alpha_1\alpha_2...\alpha_n} \right)}
\\
&=\int_C{d\tau \sum_{\alpha_1\alpha_2...\alpha_n}{V^{\alpha_1\alpha_2...\alpha_n}}g^{\alpha_1\alpha_2...\alpha_n}\left( \tau \right)}.\label{interac2}
\end{split}
\end{equation}
and the disorder-averaged results
\begin{equation}
\begin{split}
&\mathcal{V} =\frac{1}{2}\frac{\bar{J}^2}{q}\int_C{d\tau _1d\tau _2}\sum_{\alpha_1\alpha_2...\alpha_n}{\left( G_{L}^{\alpha_1\alpha_1'}\left( \tau _1,\tau _2 \right) ^{\bar{q}/2}g^{\alpha_1\alpha_2}\left( \tau _1 \right) g^{\alpha_1'\alpha_2'}\left( \tau _2 \right) G_{R}^{\alpha_2\alpha_2'}\left( \tau _1,\tau _2 \right) ^{\bar{q}/2} \right)}
\\
&\left( G_{L}^{\alpha_2\alpha_2'}\left( \tau _1,\tau _2 \right) ^{\bar{q}/2}g^{\alpha_2\alpha_3}\left( \tau _1 \right) g^{\alpha_2'\alpha_3'}\left( \tau _2 \right) G_{R}^{\alpha_3\alpha_3'}\left( \tau _1,\tau _2 \right) ^{\bar{q}/2} \right) 
\\
&...\left( G_{L}^{\alpha_{\left( n-1 \right)} \alpha'_{\left( n-1 \right)}}\left( \tau _1,\tau _2 \right) ^{\bar{q}/2}g^{\alpha_{\left( n-1 \right)} \alpha_n}\left( \tau _1 \right) g^{\alpha_{\left( n-1 \right)} '\alpha_n'}\left( \tau _2 \right) G_{R}^{\alpha_n\alpha_n'}\left( \tau _1,\tau _2 \right) ^{\bar{q}/2} \right) .\label{varv}
\end{split}
\end{equation}

The interactions can be simply expanded into arbitrary $n$-replica systems
\begin{equation}
V^{\alpha _1...\alpha _a}=\sum_{\begin{array}{c}
	1\leqslant i_{11}<...<i_{1\bar{q}}\leqslant N\\
	1\leqslant i_{21}<...<i_{2\bar{q}}\leqslant N\\
	...\\
	1\leqslant i_{a1}<...<i_{a\bar{q}}\leqslant N\\
\end{array}}{\bar{J}_{i_{11}...i_{1\bar{q}},...,i_{a1}...i_{a\bar{q}}}^{\alpha _1...\alpha _a}}\psi _{i_{11}}^{\alpha _1}...\psi _{i_{1\bar{q}}}^{\alpha _1}...\psi _{i_{a1}}^{\alpha _a}...\psi _{i_{a\bar{q}}}^{\alpha _a},
\end{equation}
where we have considered a possible $q^n$ ordered interaction within $\alpha_1$ to $\alpha_a$ (the $a$-th copy among the $n$ replicas) .

The corresponding replica-coupling parameters are also randomly distributed and satisfy the statistical consistency conditions \eqref{rand}. Furthermore, the disorder-averaged Green's function as path integrals for $n$ replicas can be expressed with ansatz
\begin{equation}
\left< Tr\left( \rho ^n\left( t \right) \right) \right> =\frac{1}{Z\left( \beta \right)}\int{D\varSigma _{aa'}^{\alpha \alpha'}DG_{aa'}^{\alpha \alpha'}}e^{-I}.
\end{equation}
The thermal effective action can be derived as
\begin{equation}
\begin{split}
I&=-\log Pf\left( \partial _{\tau}\delta _{aa'}^{\alpha \alpha'}-\varSigma _{aa'}^{\alpha \alpha'} \right) 
\\
&+\frac{1}{2}\int_C{d\tau _1d\tau _2}\left( \varSigma _{aa'}^{\alpha \alpha'}\left( \tau _1,\tau _2 \right) G_{aa'}^{\alpha \alpha'}\left( \tau _1,\tau _2 \right) -\frac{J^2}{q}G_{aa'}^{\alpha \alpha ,q}\left( \tau _1,\tau _2 \right) \right) -\mathcal{V} .\label{1action}
\end{split}
\end{equation}
Thus, we obtain the saddle-point equations of motion in the thermal limit
\begin{equation}
G_{aa'}^{\alpha \alpha'}=\left( \partial _{\tau}\delta _{aa'}^{\alpha \alpha'}-\varSigma _{aa'}^{\alpha \alpha'} \right) ^{-1},
\
\varSigma _{aa'}^{\alpha \alpha'}\left( \tau _1,\tau _2 \right) =J^2G_{aa'}^{\alpha \alpha ,q-1}\left( \tau _1,\tau _2 \right) +\partial _G\mathcal{V} .
\end{equation}

Since the system exists a $Z_(n)$ symmetry and replica indices become indistinguishable, it is easy to check that the numerical result of interaction term $\mathcal{V}$ would be preserved under this special condition \eqref{cond}. Consequently, the contributions from each order sum to numerically equal results under this condition, allowing the derivation of the action for arbitrary copies or orders of replicated theory,
\begin{equation}
I_n\left( \bar{q}\left( n \right) ,\bar{J}\left( n \right) \right) =I_2\left( \bar{q}\left( 2 \right) ,\bar{J}\left( 2 \right) \right) ,\label{CON2}
\end{equation}
where the fermionic parameters $\bar{N}$ in the averaged constants have been absorbed into the coupling constants $J$ and $\bar{J}$, while the $I_n$ represent the n-replica action in \eqref{1action}. 

It is straightforward to show that any $n$-replica result can be converted to a 2-replica equivalent through a transformation trick. This effect can be fully absorbed by multiplying the result by the symmetry factor $n!$.

For example, the indices of $n$-replica interaction relationships should form a symmetric permutation group. Under replica-diagonal conditions
\begin{equation}
G^{\alpha \alpha '}\left( \tau _1,\tau _2 \right) =\delta ^{\alpha \alpha '}G\left( \tau _1-\tau _2 \right) .
\end{equation}
When the interaction parameters can be shifted as
\begin{equation}
\bar{J}^2\left( n \right) =\frac{2\bar{J}^2\left( 2 \right)}{n!}
, \
\frac{\bar{q}\left( n \right)}{n}=\frac{\bar{q}\left( 2 \right)}{2},\label{cond}
\end{equation}
where we use $\bar{J}\left( n \right)$ and $\bar{q}\left( n \right)$ to denote the $n$-th ordered coupling parameter and coupling order. The replica-diagonal solution's action takes the form
\begin{equation}
I\supset \int_C{d\tau _1d\tau _2}G\left( \tau _1-\tau _2 \right) ^{2\bar{q}}=\int_C{d\tau _1d\tau _2}G\left( \tau _1-\tau _2 \right) ^{\bar{q}}G\left( \tau _1-\tau _2 \right) ^{\bar{q}},\label{ACT}
\end{equation}
where the density function $g\left(\tau\right)$ is trivial. The result would preserve as in 2-replica situation under the condition \eqref{CON2}. Physically, since the mathematical form of interaction \eqref{interac} preserves under the expansion \eqref{interac2}, the $n$ replicas SYK models are indistinguishable.

However, the method in this section is difficult to numerically obtain, due to the order of $q/2$ in \eqref{varv}. We can only obtain the result via the numerical result of 2-replica \eqref{CON2}. In this case, it would be necessary to develop a new method to obtain the numerical result exactly. Since the $\mathcal{N}=1$ supersymmetric SYK model have natural properties to cover the $n$-replica structure, we will attempt to resolving the replica problem with supersymmetric SYK models this in $\mathcal{N}=1$ in the following sections.

\section{Exploring the exactly n-replica SYK solution with supersymmetry}\label{S3}
\quad In this section, we first briefly review $\mathcal{N}=1$ supersymmetric SYK model and the off-diagonal coupled model. On this basis, we introduce the fractal structures as the indices of the bosons and the corresponding supersymmetric Green's function, which cover the replica signature of the $n$-replicas system. We derive the effective action in the thermal limit, and further obtain Green's function via saddle point equation, and further analyze the thermal phase structures with free energy. We can also obtain the dynamical properties by investigating the total action with dependence of Lorentzaion time.

\subsection{The multi-ordered trick}
\quad The supersymmetric SYK model (SSYK)\cite{9} processes enable exploration of random matrix theories with simultaneous replicated solutions. It is defined from a supercharge with random couplings
\begin{equation}
Q=i^{\frac{q-1}{2}}\sum_{i_1i_2\dotsi_q}{C_{i_1i_2...i_q}\psi ^{i_1}\psi ^{i_2}}\dots\psi ^{i_q},\label{SC0}
\end{equation}
$\psi^i$ represents Majorana fermions on sites $1\dots N$. In the following, we consider the situation $q=3$ as an example. The corresponding coupling $C_{ijk}$ is an $N \times N \times N$ Gaussian random antisymmetric tensor, with its scale determined by the energy-dimensioned constant $J$, where the random couplings are disorder-averaged through
\begin{equation}
\overline{C_{ijk}}=0,~~~~~\
\overline{C_{ijk}^{2}}=\frac{2\mathcal{J}}{N^2}.
\end{equation}
The $N=1$ Hamiltonian is proportional to the square of the supercharge.
\begin{equation}
\mathcal{H} =Q^2=E_0+\sum_{1\leqslant i<j<k\leqslant N}{\mathcal{J}_{ijkl}\psi ^i}\psi ^j\psi ^k\psi ^l,
\end{equation}
while the averaged constant and the random matrix $\mathcal{J}_{ijkl}$ are given by
\begin{equation}
E_0=\sum_{1\leqslant i<j<k\leqslant N}{C_{ijk}^{2}},
\
\mathcal{J}_{ijkl}=-\frac{1}{8}\sum_a{C_{aij}C_{kla}}.
\end{equation}
Here, $E_0$ is a disorder-averaged constant, whereas $\mathcal{J}_{ijkl}$ constitutes random matrices via $C_{ijk}$.

For simplicity, we introduce an auxiliary bosonic field $b_i$.
\begin{equation}
\left\{ Q,\psi ^i \right\} =Q\psi ^i=i\sum_{1\leqslant j<k\leqslant N}{C_{ijk}}\psi ^j\psi ^k\equiv b_i,
\
Qb^i=H\psi ^i=\partial _{\tau}\psi ^i.
\end{equation}
The Lagrangian contains kinetic terms for both fermionic and auxiliary bosonic field operators, along with random interactions
\begin{equation}
\mathcal{L} =\sum_i{\left[ \frac{1}{2}\psi ^i\partial _{\tau}\psi ^i-\frac{1}{2}b^ib^i+i\sum_{1\leqslant j<k\leqslant N}{C_{ijk}}b^i\psi ^j\psi ^k \right]}.
\end{equation}

For simplicity, it defines a superfield $\varPsi$ with Grassmann variable $\theta$ in superspace\cite{9}. This formulation simultaneously constrains both bosonic and fermionic components when describing the supersymmetric theory, without loss of generality. The superfield is defined as
\begin{equation}
\varPsi \left( \tau ,\theta \right) =\psi \left( \tau \right) +\theta b\left( \tau \right),
\end{equation}
and introduce the super-covariant derivative
\begin{equation}
D_{\theta}\equiv \partial _{\theta}+\theta \partial _{\tau},
\
D_{\theta}^{2}=\partial _{\tau}.
\end{equation}
We define the bilinear superfield correlation function, which contains the components of fermionic-fermionic($\psi \psi$), fermionic-bosonic($\psi b$), bosonic-fermionic($b \psi$), and bosonic-bosonic($bb$).
\begin{align}
\mathcal{G} \left( \tau ,\theta ;\tau ',\theta ' \right) &=\left< \varPsi \left( \tau ,\theta \right) \varPsi \left( \tau ',\theta ' \right) \right> =\left< \left( \psi \left( \tau \right) +\theta b\left( \tau \right) \right) \left( \psi \left( \tau ' \right) +\theta 'b\left( \tau ' \right) \right) \right> \nonumber
\\
&\equiv G_{\psi \psi}\left( \tau ,\tau ' \right) +\sqrt{2}\theta G_{b\psi}\left( \tau ,\tau ' \right) -\sqrt{2}\theta 'G_{\psi b}\left( \tau ,\tau ' \right) +2\theta \theta 'G_{bb}\left( \tau ,\tau ' \right) .
\end{align}
We then return to the original SSYK model and reformulate its effective action in superspace as
\begin{equation}
S_{EFF}=\int{d\theta d\tau \left( -\frac{1}{2}\varPsi ^iD_{\theta}\varPsi ^i \right)}+\frac{\mathcal{J}}{3N^2}\int{d\theta _1d\tau _1d\theta _2d\tau _2\left( \varPsi ^i\varPsi ^i \right) ^3}.\label{SEFF}
\end{equation}
This action can be diagonalized into the standard $\mathcal{N}=0$ SYK form using superfields and covariant derivatives. On this base, we can define the interaction in superspace, and make the total action to describe a holographic wormhole.

We propose an interaction term analogous to that in Maldacena-Qi(MQ) theory~\cite{1804.00491}, formulated in superspace for consistency. When extended to $\mathcal{N}=1$ SYK, the coupled theory combines two $\mathcal{N}=1$ SYK models with an interaction term\cite{OFF}
\begin{equation}
H_{int}=i\mu \int{d\theta}\left( \varPsi _L\varPsi _R-\varPsi _R\varPsi _L \right) =i\mu \left( \psi _Lb_R-b_L\psi _R-\psi _Rb_L+b_R\psi _L \right). \label{COU}
\end{equation}
In this setup, the Grassmann integral (or the derivative \( \partial_\theta \)) is introduced to implement a supersymmetric delta function constraint in the context of a single-sided superposition. We further utilize variations of Green's function $G$ and self-energy $\Sigma$ in the effective action to derive the equations of motion including the replica coupling term.

\begin{equation}
\begin{split}
D_{\theta}\mathcal{G} _{AB}\left( \tau ,\tau ' \right) -\sum_C{\left( -i\mu \epsilon _{AC}\partial _{\theta}\mathcal{G} _{AB}\left( \tau ,\tau ' \right) -\int{d\tau ''d\theta ''}\varSigma _{AC}\left( \tau ,\tau '' \right) \mathcal{G} _{BC}\left( \tau ,\tau '' \right) \right)}&
\\
=\delta _{AB}\left( \theta -\theta ' \right) \delta \left( \tau -\tau ' \right) .&
\end{split}
\end{equation}

We can consider multiple copies of the supersymmetric SYK model, indexed by parameter $n$, and rewrite both the supercharges in \eqref{SC0} and the corresponding Hamiltonian.
\begin{equation}
Q_n=i^{\frac{q-1}{2}}\sum_{i_1i_2...i_q}{C_{i_1i_2...i_q}^{(n)}\psi _{n}^{i_1}\psi _{n}^{i_2}}...\psi _{n}^{i_q},
\
H^{(n)}=\left( Q_n \right) ^2.\label{repQ}
\end{equation}
We also label the fermions $\psi_{n}^{i}$ and analogously define the random coupling $C_{i_1i_2...i_q}^{(n)}$. We focus on the situation $q=3$. For a single system, the extended supersymmetric bosonic fields are
\begin{equation}
\left\{ Q_n,\psi _{n}^{i} \right\} =i\sum_{1\leqslant j<k\leqslant N}{C_{ijk}}\psi _{n}^{j}\psi _{n}^{k}\equiv b_{n}^{i}.\label{SC1}
\end{equation}
The formalism for single-replica ordered systems is given by 
\begin{equation}
\mathcal{L} _n=\sum_n{\sum_i{\left[ \frac{1}{2}\psi _{n}^{i}\partial _{\tau}\psi _{n}^{i}-\frac{1}{2}b_{n}^{i}b_{n}^{i}+i\sum_{1\leqslant j<k\leqslant N}{C_{ijk}^{(n)}}b_{n}^{i}\psi _{n}^{j}\psi _{n}^{k} \right]}},
\end{equation}
which can be constructed by adding the index $n$.

Then, we highlight a central component of this section: a complete replicated theory must remain compatible with the underlying permutation symmetry among the copies. To realize this, we propose an extension of the $\mathcal{N}=1$ supersymmetric SYK model by incorporating what we refer to as a ``fractal symmetry,'' which organizes fermionic degrees of freedom across replicas in a hierarchical and ordered manner. This symmetry can be naturally embedded into an extended ``fractal-like symmetry'' structure endowed with an intrinsic ordering. In this section, we will introduce the definition of the model, a further discussion of ``fractal symmetry'' is given in the next section. We can label the fermions in SYK systems(or supercharges) with the index $n$ to represent the copies
\begin{equation}
\psi ^i\rightarrow \psi _{n}^{i}.
\end{equation}
In order to represent the replicated theory, we can define the fermions that has extended indices
\begin{equation}
\psi _{n}^{i}\rightarrow \psi _{n}^{i}+\psi _{m}^{i}\rightarrow \psi _{n}^{i}+\psi _{m}^{i}+...\psi _{l}^{i}=\psi_{nm..l}^{i}.\label{fer}
\end{equation}
The supercharges with indices are consist of the fermions with indices \eqref{repQ}
\begin{equation}
Q\rightarrow Q_n,
\end{equation}
and expanded the indices as \eqref{fer}
\begin{equation}
Q_n\rightarrow Q_{nm}\rightarrow Q_{nm...l} .
\end{equation}

For example, we proceed to expand the supercharge sector to include broader 2-replica components. 
\begin{equation}
Q_{nm}=i^{\frac{q-1}{2}}\sum_{nm}{\sum_{i_1i_2...i_q}{C_{i_1i_2...i_q}^{(nm)}\psi _{nm}^{i_1}\psi _{nm}^{i_2}}...\psi _{nm}^{i_q}.}
\end{equation}
In order to describe interactions between replicas, we extend the fermionic sector \( n \) to include fields labeled by replica indices \( n \) and \( m \), capturing the full structure of the two-replica system.  The fermionic operators are given as the sum over both indices in \eqref{fer}. 
\begin{equation}
\psi _{nm}^{i}=\psi _{n}^{i}+\psi _{m}^{i}.
\end{equation}
Thus, we define additional bosonic operators through the anticommutation relations between the extended supercharges and the original fermions 
\begin{equation}
\left\{ Q_{nm},\psi _{n}^{i} \right\} =\left\{ Q_{nm},\psi _{m}^{i} \right\}\equiv b_{nm}^{i}.\label{SC2}
\end{equation}
A key structural feature is that the fermionic replica index $n$ must align with one of the bosonic replica indices $nm$ to yield a supersymmetric contribution. Configurations violating this condition are canceled by disorder averaging, making them physically insignificant in the replica path integral formulation. In the following discussions, we will label the physical quantities with indices $nm$ as ``second ordered'' quantities, and so forth.

We therefore construct an alternative SSYK formulation by summing over indices $n$ and $m$, where we have define the total sector with the index $nm$.
\begin{equation}
\mathcal{L} _{nm}=\sum_{nm}{\sum_i{\left[ \frac{1}{2}\psi _{n}^{i}\partial _{\tau}\psi _{n}^{i}-\frac{1}{2}b_{nm}^{i}b_{nm}^{i}+i\sum_{1\leqslant j<k\leqslant N}{C_{ijk}^{(nm)}}b_{nm}^{i}\psi _{nm}^{j}\psi _{nm}^{k} \right]}}.
\end{equation}
In this construction, we simply enlarge the fermionic sector without introducing any essential modification to the models. The fermions retain the same characteristics as those in the original SSYK model, and their dynamics remain governed by the same underlying principles.

\begin{figure}[!t]
\begin{minipage}{0.43\linewidth}
\centerline{\includegraphics[width=9cm]{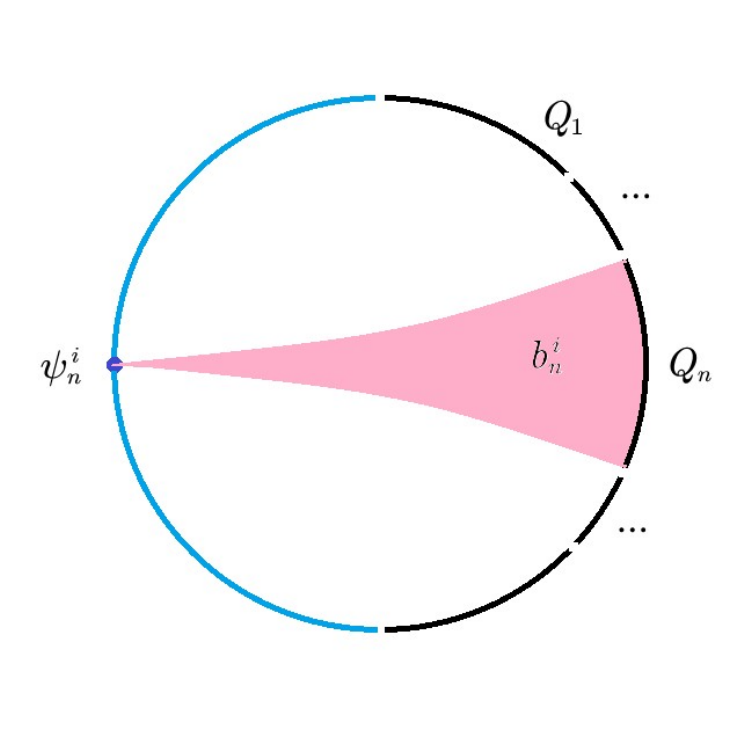}}
\centerline{(a)}
\end{minipage}
\hfill
\begin{minipage}{0.43\linewidth}
\centerline{\includegraphics[width=9cm]{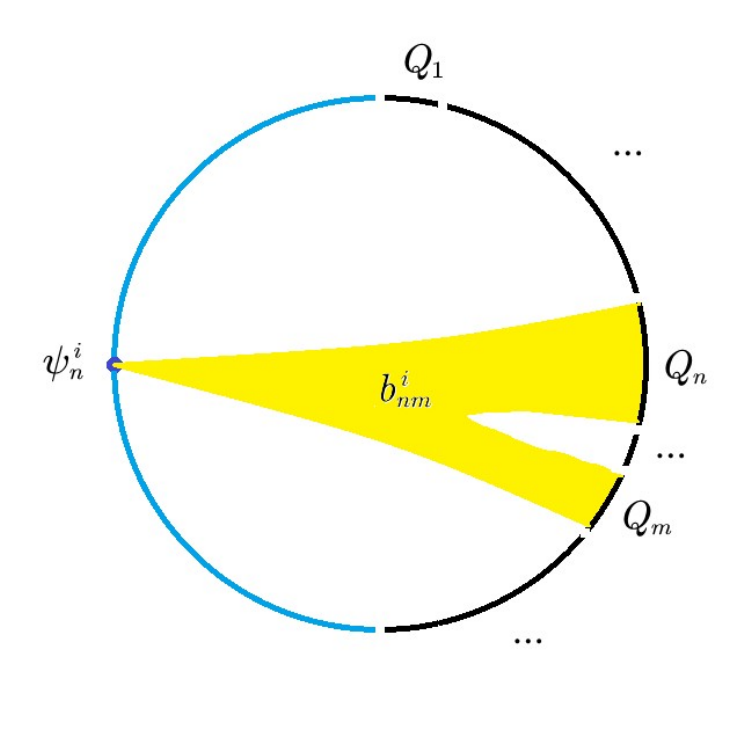}}
\centerline{(b)}
\end{minipage}

\caption{\label{fig:3} The replicated supercharge structure and the auxiliary bosons are represented as a chord diagrams (a) The first-order definition(as shown in ~\eqref{SC1}). (b) The second-order definition(as shown in ~\eqref{SC2}). }
\end{figure}
As illustrated in Figure~\ref{fig:3}(a), the structure of our supersymmetric theory is depicted using a chord diagram. On the left-hand side (LHS), the blue line represents multiple instances of random-matrix systems that share a common coupling tensor $C_{ijk}$, which defines the structure of the supercharges. On the right-hand side (RHS), the black line corresponds to the conjugated SYK systems equipped with internal interactions. These interactions are encoded by chords that stretch from the LHS to the RHS, indicating the entanglement and coupling between the supercharge sector and the conjugate SYK dynamics.

The pink region is defined as the segment that begins with a fermion labeled by index $n$ on the left-hand side (LHS) and terminates at a supercharge on the right-hand side (RHS), where the supercharge also carries the index $n$ to preserve consistency. This region encodes two distinct interactions: one between the single fermion $\psi_{n}^{i}$ on the LHS and the conjugate supercharge system on the RHS, and the other corresponding to the conventional SYK-type coupling $C_{ijk}$ that operates entirely within the RHS. We interpret this composite interaction region—highlighted in pink—as the effective propagation of an auxiliary bosonic field $b_{n}^{i}$.

As shown in Figure~\ref{fig:3}(b), we extend the fermionic sector to include both $\psi_{n}^{i}$ and $\psi_{m}^{i}$, and accordingly generalize the definition of the supercharges from $Q_{n}$ to encompass both $Q_{n}$ and $Q_{m}$. This extension gives rise to new composite supercharges denoted by $Q_{nm}$, which represent random couplings involving the full set of fermions. Furthermore, as indicated in the yellow region, we define the associated conjugated bosonic fields $b_{nm}^{i}$ through the anticommutation relations of the extended supercharges.

Since the supercharges form fully random matrices, the bosons naturally contain the full information of the SYK system. A first-order supersymmetric interaction is sufficient to generate the replica wormholes when we introduce additional left and right indices in the coupled SSYK model\cite{OFF}, which can be expressed as
\begin{equation}
H_{total}=H_L+H_R+H_{int}.
\end{equation}
The single-side total Hamiltonian $H_{A}$ is given by
\begin{equation}
   H_{A}=H_n+H_{nm}+...+H_{nm...l}=Q_{n}^{2}+Q_{nm}^{2}+...+Q_{nm...l}^{2},
\end{equation}
where the parameter $H_{A}$ refer to $H_{L}$ and $H_{R}$. 
One can extend the construction to arbitrary replica orders by enlarging the index range in both the fermionic fields and the supercharge definitions. The generalized form of the supercharge takes the form
\begin{equation}
Q_{nm\ldots l} = i^{\frac{q-1}{2}} \sum_{nm\ldots l} \sum_{i_1 i_2 \ldots i_q} C_{i_1 i_2 \ldots i_q}^{(nm\ldots l)} \psi_{nm\ldots l}^{i_1} \psi_{nm\ldots l}^{i_2} \cdots \psi_{nm\ldots l}^{i_q},
\end{equation}
where the indices $n, m, \ldots, l$ label different replica sectors. The corresponding bosonic fields are defined through the anticommutation of these extended supercharges 
\begin{equation}
\left\{ Q_{nm...l},\psi _{n}^{i} \right\} \equiv b_{nm...l}^{i}.
\end{equation}
We can also express the Language with arbitrary orders of indices $nm\cdots l$.
\begin{equation}
\mathcal{L} _{nm...l}=\sum_{nm...l}{\sum_i{\left[ \frac{1}{2}\psi _{n}^{i}\partial _{\tau}\psi _{n}^{i}-\frac{1}{2}b_{nm...l}^{i}b_{nm...l}^{i}+i\sum_{1\leqslant j<k\leqslant N}{C_{ijk}^{(nm...l)}}b_{nm...l}^{i}\psi _{nm...l}^{j}\psi _{nm...l}^{k} \right]}}.
\end{equation}
For simplicity, we use a special representation of the indexed fermions with random coupling constants.
\begin{equation}
\psi _{nm...l}^{i}=\psi _{n}^{i}+\psi _{m}^{i}+...+\psi _{l}^{i}.
\end{equation}
The complete Language is obtained by summing over all copies and orders of the single SSYK model.
\begin{equation}
\mathcal{L} =\mathcal{L} _n+\mathcal{L} _{nm}+...+\mathcal{L} _{nm...l}.
\end{equation}
The original coupled SSYK model is now extended to arbitrary replicas, where the interaction Hamiltonian
\begin{align}
H_{int}&=i\mu ^{(n)}\int{d\theta}\varPsi _{L,n}^{i}\varPsi _{R,n}^{i}+i\mu ^{(nm)}\int{d\theta}\varPsi _{L,nm}^{i}\varPsi _{R,nm}^{i}+...+i\mu ^{(nm...l)}\int{d\theta}\varPsi _{L,nm...l}^{i}\varPsi _{R,nm...l}^{i}\nonumber
\\
&=i\mu ^{(n)}\left( \psi _{L,n}b_{R,n}-b_{L,n}\psi _{R,n}-\psi _{R,n}b_{L,n}+b_{R,n}\psi _{L,n} \right) \label{156151}
\\
&+i\mu ^{(nm)}\left( \psi _{L,n}b_{R,nm}-b_{L,nm}\psi _{R,n}-\psi _{R,n}b_{L,nm}+b_{R,nm}\psi _{L,n} \right) \nonumber
\\
&+...+i\mu ^{(nm...l)}\left( \psi _{L,n}b_{R,nm...l}-b_{L,nm}\psi _{R,n}-\psi _{R,n}b_{L,nm...l}+b_{R,nm...l}\psi _{L,n} \right), \nonumber 
\end{align}
where the first-order term with $\mu^{(n)}$, second-order term $\mu^{(nm)}$, and arbitrary-order term $\mu^{(nm\cdots l)}$ are independently distributed in \eqref{COU}. The direct relationship between first-order bosons $b_n$, $b_m$ and the second-order boson $b_{mn}$ is forbidden in both the single-sided model and the interaction terms.

In the context of replicated SSYK models, replica wormhole configurations emerge as correlated saddle points in the path integral, characterized by nontrivial off-diagonal Green’s functions between different replica sectors. However, the stability of such wormhole solutions is sensitive to the structure of the interaction terms, particularly the couplings that connect distinct replicas. 

Inter-replica interactions generically introduce fluctuations that couple different replica trajectories in Euclidean time, which can lift the wormhole saddle or destabilize it into disconnected geometries. To suppress these destabilizing effects and ensure that the wormhole configuration remains a dominant saddle in the low-energy limit, we impose constraints on the interaction tensors of the higher-order SSYK model. Specifically, we retain only intra-replica couplings and systematically eliminate all multi-replica interaction terms:
\begin{equation}
C_{i_1 i_2 \dots i_q}^{(n)} \neq 0, \quad
C_{i_1 i_2 \dots i_q}^{(nm)} \rightarrow 0, \quad
C_{i_1 i_2 \dots i_q}^{(nm \dots l)} \rightarrow 0 ,
\end{equation}
where \( C_{i_1 i_2 \dots i_q}^{(n)} \) denotes the $q$-fermion coupling tensor confined to the $n$-th replica, while \( C^{(nm)} \) and \( C^{(nm \dots l)} \) represent two-replica and multi-replica interaction tensors, respectively. This truncation preserves the wormhole saddle by decoupling extraneous replica-mixing channels that would otherwise destabilize the semiclassical geometry.

In this case, the total model return to the copies of original SSYK model when the interaction parameters $\mu^{(n)}$ vanish. The simplified effective theory can be obtained
\begin{equation}
\begin{split}
\mathcal{L} &=\mathcal{L} _n+\mathcal{L} _{nm}+...+\mathcal{L} _{nm...l}
\\
&=\sum_{nm...l}{\sum_i{\left[ \frac{1}{2}\psi _{n}^{i}\partial _{\tau}\psi _{n}^{i}-\frac{1}{2}b_{n}^{i}b_{n}^{i}-\frac{1}{2}b_{nm}^{i}b_{nm}^{i}-\frac{1}{2}b_{nm...l}^{i}b_{nm...l}^{i}+i\sum_{1\leqslant j<k\leqslant N}{C_{ijk}^{(n)}}b_{n}^{i}\psi _{n}^{j}\psi _{n}^{k} \right]}}.
\end{split}
\end{equation}
For this effective action in superspace, we find that the total action from \eqref{SEFF}
\begin{align}
S_{EFF}&=\int{d\theta d\tau \left( -\frac{1}{2}\varPsi _{n}^{i}D_{\theta}\varPsi _{n}^{i} \right)}+\frac{\mathcal{J}^{(n)}}{3N^2}\int{d\theta _1d\tau _1d\theta _2d\tau _2\left( \varPsi _{n}^{i}\varPsi _{n}^{i} \right) ^3}\nonumber
\\
&+\int{d\theta d\tau \left( -\frac{1}{2}\varPsi _{nm}^{i}D_{\theta}\varPsi _{nm}^{i} \right)}+\frac{\mathcal{J}^{(nm)}}{3N^2}\int{d\theta _1d\tau _1d\theta _2d\tau _2\left( \varPsi _{nm}^{i}\varPsi _{nm}^{i} \right) ^3}
\\
&+...+\int{d\theta d\tau \left( -\frac{1}{2}\varPsi _{nm...l}^{i}D_{\theta}\varPsi _{nm...l}^{i} \right)}+\frac{\mathcal{J}^{(nm...l)}}{3N^2}\int{d\theta _1d\tau _1d\theta _2d\tau _2\left( \varPsi _{nm...l}^{i}\varPsi _{nm...l}^{i} \right) ^3}.\nonumber
\end{align}
The corresponding disorder-averaged random matrix must satisfy the imposed constraints.
\begin{equation}
\overline{\left( C_{ijk}^{(n)} \right) ^2}=\frac{2\mathcal{J}^{(n)}}{N^2},
\
\overline{\left( C_{ijk}^{(nm)} \right) ^2}=\frac{2\mathcal{J}^{(nm)}}{N^2}\rightarrow 0,
\
\overline{\left( C_{ijk}^{(nm...l)} \right) ^2}=\frac{2\mathcal{J}^{(nm...l)}}{N^2}\rightarrow 0,
\end{equation}
where we define the superfield composed of both single-index fermions and multi-index bosons
\begin{equation}
\varPsi _{n}^{i}=\psi _{n}^{i}+\theta b_{n}^{i},
\
\varPsi _{nm}^{i}=\psi _{n}^{i}+\theta b_{nm}^{i},
\
\varPsi _{nm...l}^{i}=\psi _{n}^{i}+\theta b_{nm...l}^{i}.
\end{equation}

In order to describe the disorder-averaged result, we can define the two-point supersymmetric Green's function within the superfield formalism. For example, the first order two-point Green's function
\begin{equation}
\begin{split}
\mathcal{G} ^{\left( n \right)}\left( \tau ,\theta ;\tau ',\theta ' \right) &=\left< \varPsi _n\left( \tau ,\theta \right) \varPsi _n\left( \tau ',\theta ' \right) \right> =\left< \left( \psi_n \left( \tau \right) +\theta b_n\left( \tau \right) \right) \left( \psi_n \left( \tau ' \right) +\theta 'b_n\left( \tau ' \right) \right) \right> 
\\
&\equiv G_{\psi \psi}\left( \tau ,\tau ' \right) +\sqrt{2}\theta G_{b\psi}^{\left( n \right)}\left( \tau ,\tau ' \right) -\sqrt{2}\theta 'G_{n,\psi b}^{\left( n \right)}\left( \tau ,\tau ' \right) +2\theta \theta 'G_{bb}^{\left( n \right)}\left( \tau ,\tau ' \right) .\nonumber
\end{split}
\end{equation}
The second-order definition of Green's function takes the form
\begin{equation}
\begin{split}
\mathcal{G} ^{\left( nm \right)}\left( \tau ,\theta ;\tau ',\theta ' \right) &=\left< \varPsi _{nm}\left( \tau ,\theta \right) \varPsi _{nm}\left( \tau ',\theta ' \right) \right> =\left< \left( \psi_n \left( \tau \right) +\theta b_{nm}\left( \tau \right) \right) \left( \psi_n \left( \tau ' \right) +\theta 'b_{nm}\left( \tau ' \right) \right) \right> 
\\
&\equiv G_{\psi \psi}\left( \tau ,\tau ' \right) +\sqrt{2}\theta G_{b\psi}^{\left( nm \right)}\left( \tau ,\tau ' \right) -\sqrt{2}\theta 'G_{\psi b}^{\left( nm \right)}\left( \tau ,\tau ' \right) +2\theta \theta 'G_{bb}^{\left( nm \right)}\left( \tau ,\tau ' \right) .\nonumber
\end{split}
\end{equation}
Note that arbitrary order superfield Green's functions systematically capture the dynamical content of the theory
\begin{equation}
\begin{split}
\mathcal{G} _{nm...l}\left( \tau ,\theta ;\tau ',\theta ' \right) &=\left< \varPsi _{nm...l}\left( \tau ,\theta \right) \varPsi _{nm...l}\left( \tau ',\theta ' \right) \right> =\left< \left( \psi_n \left( \tau \right) +\theta b_{nm...l}\left( \tau \right) \right) \left( \psi_n \left( \tau ' \right) +\theta 'b_{nm...l}\left( \tau ' \right) \right) \right> 
\\
&\equiv G_{\psi \psi}\left( \tau ,\tau ' \right) +\sqrt{2}\theta G_{b\psi}^{\left( nm...l \right)}\left( \tau ,\tau ' \right) -\sqrt{2}\theta 'G_{\psi b}^{\left( nm...l \right)}\left( \tau ,\tau ' \right) +2\theta \theta 'G_{bb}^{\left( nm...l \right)}\left( \tau ,\tau ' \right) .\nonumber
\end{split}
\end{equation}
Having accounted for interactions, the effective action can be expanded in terms of Green's functions up to finite order
\small
\begin{equation}
\begin{split}
&\frac{S_{eff}}{N}=-\log Pf\left( D_{ha} \right) +\log\det \left( \frac{D_{in}}{\left( iw_n \right) ^2} \right) +\sum_{A,B=L,R}{\frac{1}{2}}\int{d\tau d\tau '}\left[ \varSigma _{\psi \psi ,AB}^{\left( n \right)}\left( \tau ,\tau ' \right)G_{\psi \psi ,AB}^{\left( n \right)}\left( \tau ,\tau ' \right) \right. 
\\
& +\varSigma _{\psi b,AB}^{\left( n \right)}\left( \tau ,\tau ' \right) G_{\psi b,AB}^{\left( n \right)}\left( \tau ,\tau ' \right) +\varSigma _{b\psi ,AB}^{\left( n \right)}\left( \tau ,\tau ' \right) G_{b\psi ,AB}^{\left( n \right)}\left( \tau ,\tau ' \right) +\varSigma _{bb,AB}^{\left( n \right)}\left( \tau ,\tau ' \right) G_{bb,AB}^{\left( n \right)}\left( \tau ,\tau ' \right)
\\
& -\left( \delta _{AB}+\left( 1-\delta _{AB} \right) \left( -1 \right) ^{\left( q-1 \right) /2} \right) \mathcal{J}^{(n)}\left( G_{\psi \psi ,AB}^{\left( n \right) q-1}\left( \tau ,\tau ' \right) G_{bb,AB}^{\left( n \right)}\left( \tau ,\tau ' \right) -\frac{\left( q-1 \right)}{2}G_{\psi \psi ,AB}^{\left( n \right) q-2}\left( \tau ,\tau ' \right) \right. 
\\
&\left. \left.  G_{\psi b,AB}^{\left( n \right) ;A}\left( \tau ,\tau ' \right) G_{b\psi ,AB}^{\left( n \right) ;S}\left( \tau ,\tau ' \right) -\frac{\left( q-1 \right)}{2}G_{\psi \psi ,AB}^{\left( n \right) q-2}\left( \tau ,\tau ' \right) G_{\psi b,AB}^{\left( n \right) ;S}\left( \tau ,\tau ' \right) G_{b\psi ,AB}^{\left( n \right) ;A}\left( \tau ,\tau ' \right) \right) \right] 
\\
&+...+\sum_{A,B=L,R}{\frac{1}{2}}\int{d\tau d\tau '}\left[ \varSigma _{\psi \psi ,AB}^{\left( nm...l \right)}\left( \tau ,\tau ' \right) G_{\psi \psi ,AB}^{\left( nm...l \right)}\left( \tau ,\tau ' \right) \right. +\varSigma _{\psi b,AB}^{\left( nm...l \right)}\left( \tau ,\tau ' \right) G_{\psi b,AB}^{\left( nm...l \right)}\left( \tau ,\tau ' \right)
\\
& +\varSigma _{b\psi ,AB}^{\left( nm...l \right)}\left( \tau ,\tau ' \right) G_{b\psi ,AB}^{\left( nm...l \right)}\left( \tau ,\tau ' \right) +\varSigma _{bb,AB}^{\left( nm...l \right)}\left( \tau ,\tau ' \right) G_{bb,AB}^{\left( nm...l \right)}\left( \tau ,\tau ' \right) -\left( \delta _{AB}+\left( 1-\delta _{AB} \right) \left( -1 \right) ^{\left( q-1 \right) /2} \right) 
\\
&\mathcal{J}^{(nm...l)}\left( G_{\psi \psi ,AB}^{\left( nm...l \right) q-1}\left( \tau ,\tau ' \right) G_{bb,AB}^{\left( nm...l \right)}\left( \tau ,\tau ' \right) -\frac{\left( q-1 \right)}{2}G_{\psi \psi ,AB}^{\left( nm...l \right) q-2}\left( \tau ,\tau ' \right) G_{\psi b,AB}^{\left( nm...l \right) ;A}\left( \tau ,\tau ' \right) \right. 
\\
&\left. \left. G_{b\psi ,AB}^{\left( nm...l \right) ;S}\left( \tau ,\tau ' \right) -\frac{\left( q-1 \right)}{2}G_{\psi \psi ,AB}^{\left( nm...l \right) q-2}\left( \tau ,\tau ' \right) G_{\psi b,AB}^{\left( nm...l \right) ;S}\left( \tau ,\tau ' \right) G_{b\psi ,AB}^{\left( nm...l \right) ;A}\left( \tau ,\tau ' \right) \right) \right] .
\end{split}
\end{equation}
The matrix is given by
{\footnotesize
\begin{equation}
\begin{split}
&D_{ha}=\\
&\left( \begin{matrix}
	-iw_n-\varSigma _{LL,\psi \psi}&		-0&		-0&		-i\mu ^{(n)}-\varSigma _{LR,\psi b}^{n,A}&		-0&		-i\mu '^{(nm)}-\varSigma _{LR,\psi b}^{nm,A}\\
	-0&		-iw_n-\varSigma _{RR,\psi \psi}&		i\mu ^{(n)}-\varSigma _{RL,\psi b}^{n,A}&		-0&		i\mu '^{(nm)}-\varSigma _{RL,\psi b}^{A}&		-0\\
	-0&		i\mu ^{(n)}-\varSigma _{LR,b\psi}^{n,A}&		-1-0&		-0&		-0&		-0\\
	-i\mu ^{(n)}-\varSigma _{RL,b\psi}^{n,A}&		-0&		-0&		-1-0&		-0&		-0\\
	-0&		i\mu '^{(nm)}-\varSigma _{LR,b\psi}^{nm,A}&		-0&		-0&		-1-0&		-0\\
	-i\mu '^{(nm)}-\varSigma _{RL,b\psi}^{nm,A}&		-0&		-0&		-0&		-0&		-1-0\\
\end{matrix} \right) ,
\end{split}
\end{equation}
\begin{equation}
\begin{split}
&D_{in}=\\
&\left( \begin{matrix}
	-iw_n-0&		-0&		-0&		i\mu ^{(n)}-\varSigma _{LR,\psi b}^{n,S}&		-0&		i\mu '^{(nm)}-\varSigma _{LR,\psi b}^{nm,S}\\
	-0&		-iw_n-0&		-i\mu ^{(n)}-\varSigma _{RL,\psi b}^{n,S}&		-0&		-i\mu '^{(nm)}-\varSigma _{RL,\psi b}^{nm,S}&		-0\\
	-0&		-i\mu ^{(n)}-\varSigma _{LR,b\psi}^{n,S}&		-1-\varSigma _{LL,bb}^{(n)}&		-0&		-0&		-0\\
	i\mu ^{(n)}-\varSigma _{RL,b\psi}^{n,S}&		-0&		-0&		-1-\varSigma _{RR,bb}^{(n)}&		-0&		-0\\
	-0&		-i\mu '^{(nm)}-\varSigma _{LR,b\psi}^{nm,S}&		-0&		-0&		-1-\varSigma _{LL,bb}^{(nm)}&		-0\\
	i\mu '^{(nm)}-\varSigma _{RL,b\psi}^{nm,S}&		-0&		-0&		-0&		-0&		-1-\varSigma _{RR,bb}^{(nm)}\\
\end{matrix} \right) .
\end{split}
\end{equation}
}

\begin{figure}[!t]
\begin{minipage}{0.44\linewidth}
\centerline{\includegraphics[width=9cm]{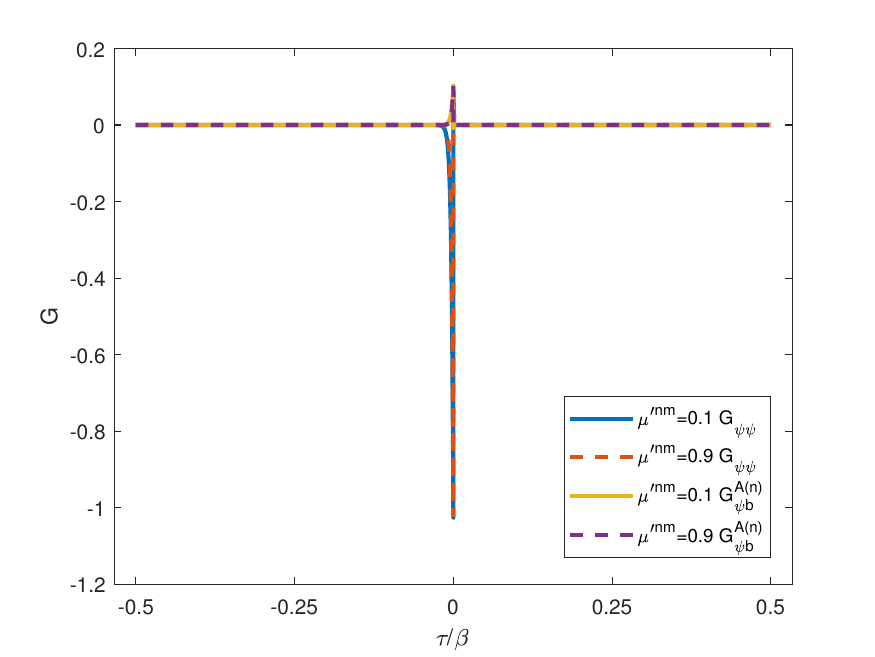}}
\centerline{(a)}
\end{minipage}
\hfill
\begin{minipage}{0.44\linewidth}
\centerline{\includegraphics[width=9cm]{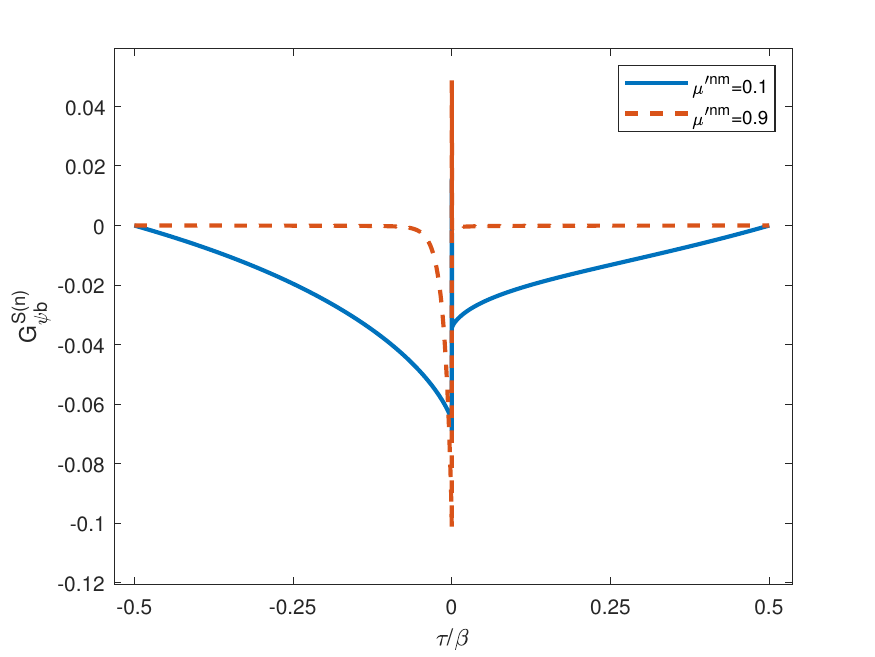}}
\centerline{(b)}
\end{minipage}
\vfill
\begin{minipage}{0.44\linewidth}
\centerline{\includegraphics[width=9cm]{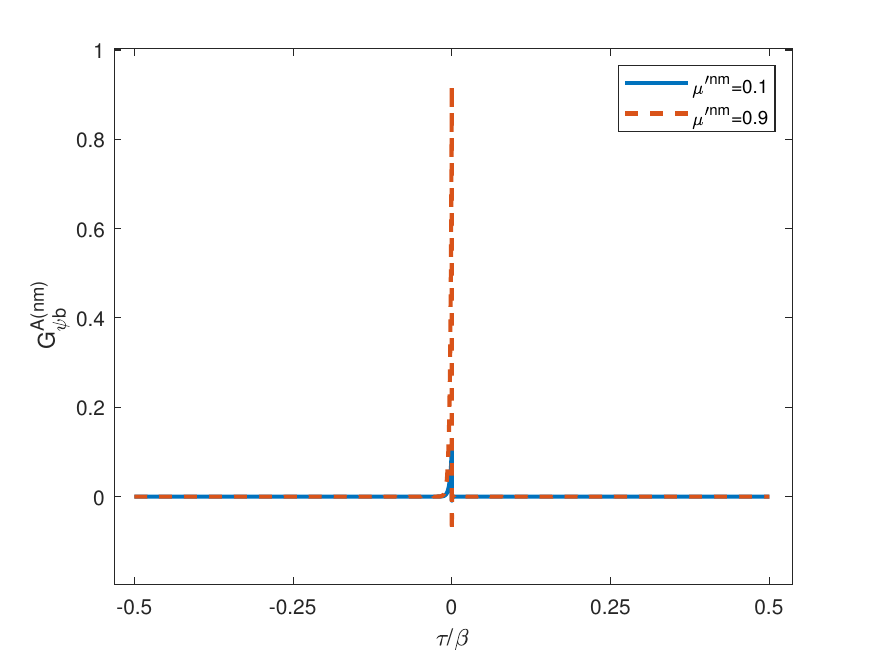}}
\centerline{(c)}
\end{minipage}
\hfill
\begin{minipage}{0.44\linewidth}
\centerline{\includegraphics[width=9cm]{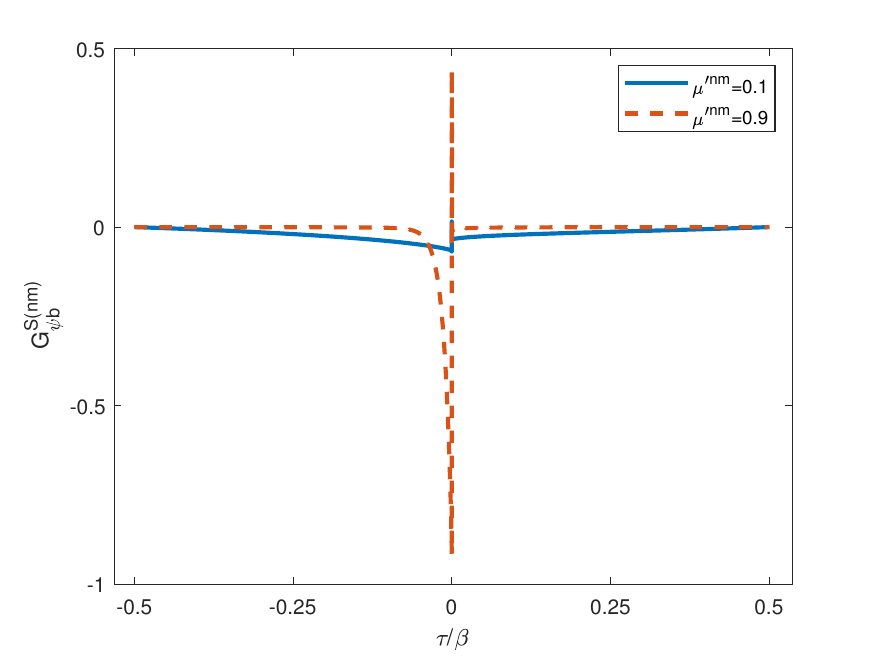}}
\centerline{(d)}
\end{minipage}
\caption{\label{fig:4} Supersymmetric Green's functions in components with fixed parameters $\mathcal{J}=3$, $\mu^{(n)}=0.1$, $T=0.01$, $q=5$, and additional parameters $\mu'^{(nm)}=0.1,0.9$ (a) Single-sided fermionic Green's function $G_{\psi\psi,AA}$ and first-order multi-sided anti-symmetric fermion-boson correlation function $G_{\psi b,LR}^{A(n)}$, (b) First-order multi-sided symmetric fermion-boson correlation function $G_{\psi b,LR}^{S(n)}$ (c) Additional multi-sided anti-symmetric fermion-boson correlation function $G_{\psi b,LR}^{A(nm)}$ (d) Additional multi-sided symmetric fermion-boson correlation function $G_{\psi b,LR}^{S(nm)}$}
\end{figure}

The numerical results in Figure~\ref{fig:4} show the supersymmetric Green's functions in components with fixed parameters $\mathcal{J}=3$, $\mu^{(n)}=0.1$, $T=0.01$ and additional parameters $\mu'^{(nm)}=0.1,0.9$. In Figure~\ref{fig:4}(a), comparison with previous work reveals that the additional interaction parameters only slightly affect $G_{\psi\psi,AA}$ and $G_{\psi b,LR}^{A(n)}$. Figure~\ref{fig:4}(b) demonstrates that the first-order multi-sided symmetric fermion-boson correlation function $G_{\psi b,LR}^{S(n)}$ is sensitive to the additional parameter $\mu'^{(nm)}$. The antisymmetric and symmetric fermion boson components of $G_{\psi b,LR}^{(nm)}$ are shown in Figures~\ref{fig:4}(c) and~\ref{fig:4}(d). Upon including the coupling term, these off-diagonal components are similar to the first-order solutions discussed in our previous work. The Green's functions plotted here can be analogized to quantization with an additional time dimension. A detailed discussion of their physical significance, including the phase diagram, will be provided in a subsequent section. The full numerical procedure is outlined in Appendix~\ref{ap1}.

Within the two-body effective action, interaction terms such as $G_{\psi b,AB}$ enter the Schwinger-Dyson equations and, while nonvanishing, still describe nearly free particles. In the large-$N$ limit, higher-order correlation functions, such as $\mathcal{G}_{nm}$, $\mathcal{G}_{nm\cdots l}$, and their components, become asymptotically equivalent and indistinguishable. This permits a redefinition of the couplings via the transformation. In our following discussion, we involve the reparameterization
\begin{equation}
C_{n}^{2}\mathcal{J}^{(nm)}+\cdots+C_{n}^{n}\mathcal{J}^{(nm\cdots l)} \rightarrow \mathcal{J}'^{(nm)}, \quad 
C_{n}^{2}\mu^{(nm)}+\cdots+C_{n}^{n}\mu^{(nm\cdots l)} \rightarrow \mu'^{(nm)},\label{rep}
\end{equation}
where the summation runs over all possible higher-replica indices.

One can verify that imposing the conditions $G_{\psi\psi,LR}=0$, $G_{bb,LR}=0$, and $G_{\psi b,AA}=0$ leads to a stable solution. A noteworthy feature is that the self-energy of the off-diagonal component $\Sigma_{\psi b,LR}$ tends to vanish, i.e., $\Sigma_{\psi b,LR} \rightarrow 0$, indicating that the corresponding propagator $G_{\psi b,LR}$ describes degrees of freedom that are effectively free or weakly interacting. In our calculation, several Green's function components identically vanish, which helps isolate the non-zero terms that can subsequently be absorbed into the second-order contributions\eqref{rep}.

In this formulation, reparameterized couplings $\mathcal{J}'^{(nm)}$ and $\mu'^{(nm)}$ effectively capture the contributions of the entire tower of higher-order interactions. Importantly, this reparameterization scheme is valid only in the thermal limit, where the system approaches equilibrium and replica symmetry simplifies the effective dynamics.
\begin{figure}[!t]
\begin{minipage}{0.48\linewidth}
\centerline{\includegraphics[width=8cm]{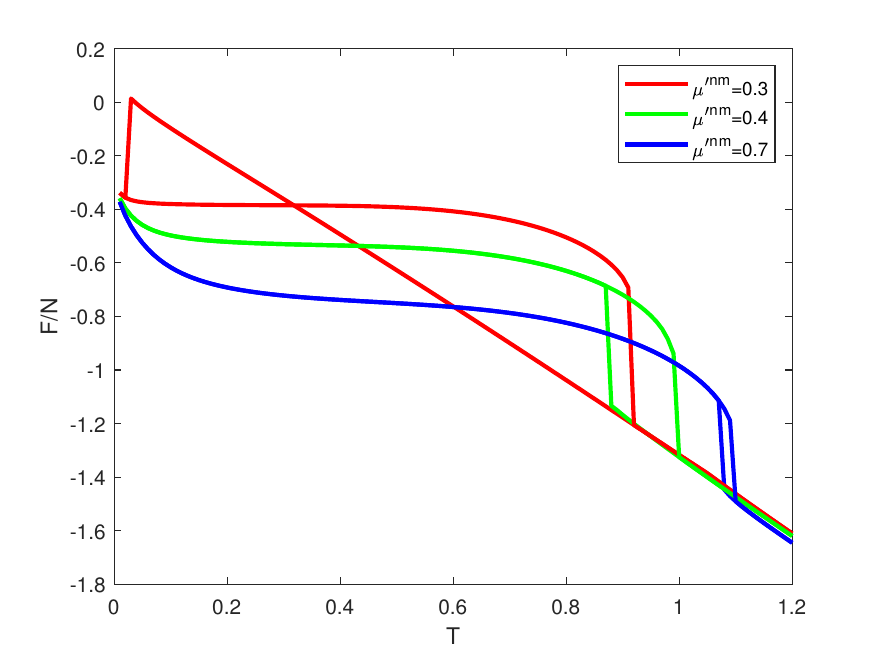}}
\centerline{(a)}
\end{minipage}
\hfill
\begin{minipage}{0.48\linewidth}
\centerline{\includegraphics[width=8cm]{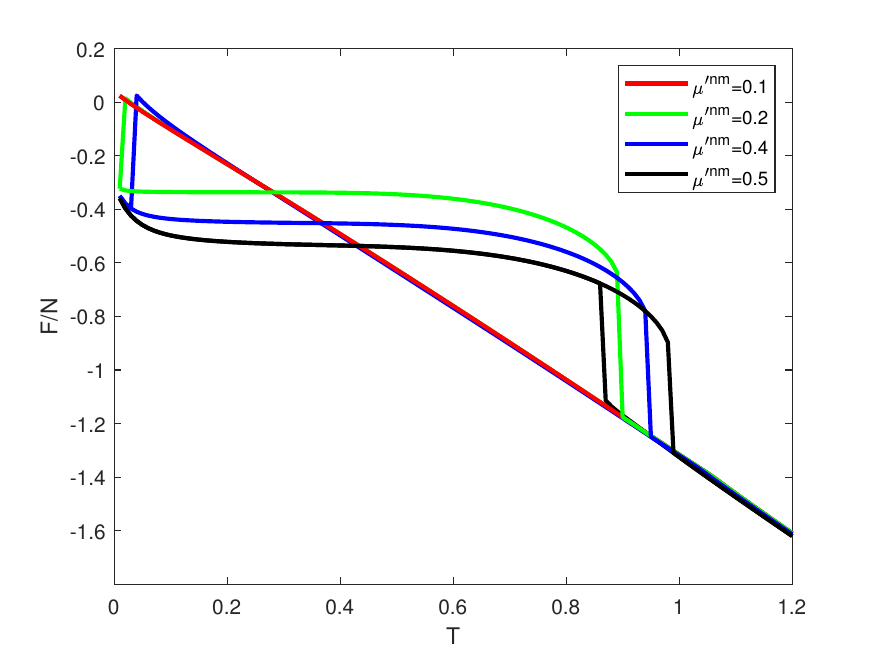}}
\centerline{(b)}
\end{minipage}
\begin{minipage}{0.48\linewidth}
\centerline{\includegraphics[width=8cm]{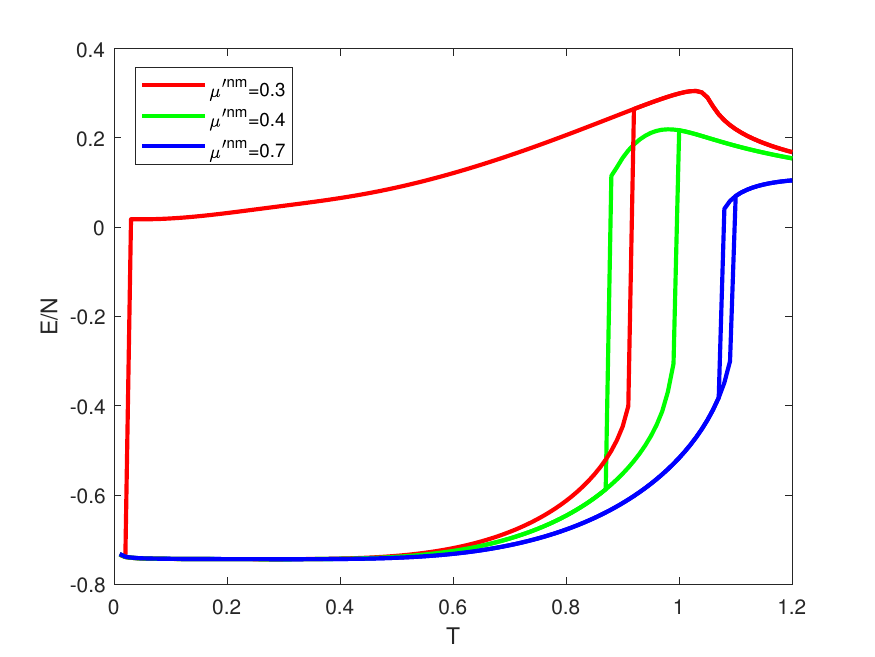}}
\centerline{(c)}
\end{minipage}
\hfill
\begin{minipage}{0.48\linewidth}
\centerline{\includegraphics[width=8cm]{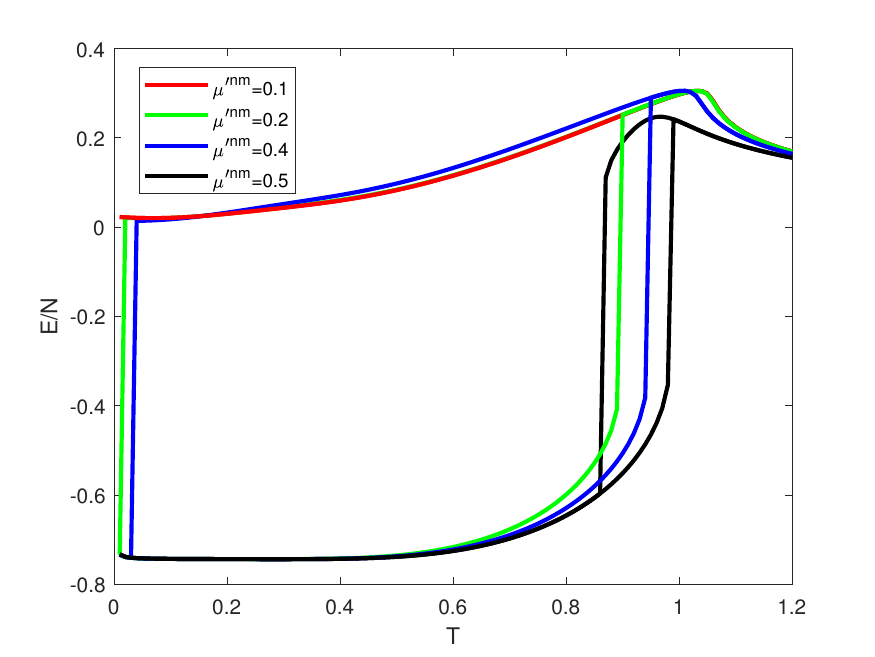}}
\centerline{(d)}
\end{minipage}

\caption{\label{fig:5} Free energy and energy of $\mathcal{N}=1$ supersymmetric SYK model with fixed $\mathcal{J}=3$, $q=3$, extended to second-order interactions $\mu^{(n)}$ and $\mu'^{(nm)}$ (a)Free energy with $\mu^{(n)}=0$ and $\mu'^{(nm)}=0.3,0.4,0.7$, (b) Free energy with $\mu^{(n)}=0.3$ and $\mu'^{(nm)}=0.1,0.2,0.4,0.5$, (c) Energy with $\mu^{(n)}=0$ and $\mu'^{(nm)}=0.3,0.4,0.7$, (d) Energy with $\mu^{(n)}=0.3$ and $\mu'^{(nm)}=0.1,0.2,0.4,0.5$.}
\end{figure}

The free energy and energy as functions of temperature are plotted in Figure~\ref{fig:5}. We first cool the system from its highest temperature and then heat it back to its original temperature, considering $q=3$ couplings. The phase transition observed in ~\cite{OFF} shows free energy behavior analogous to the Hawking-Page transition in standard MQ models. We interpret the low-temperature phase as a wormhole configuration that evolves into two gapless systems with increasing temperature.

However, the introduction of supersymmetric interaction with $\mu'^{(nm)}$ modifies this picture: the contributions to the free energy gradually transform the 'wormhole phase' into a state from the eternal wormhole solutions in existing one, which conclude the both contributions of two-side wormhole with $\mu^{(n)}$ and the contribution of multi-side wormhole with $\mu'^{(nm)}$. Furthermore, this effect also influences the phase transition (as shown in Figure~\ref{fig:5}(a)). Figure~\ref{fig:5}(b) displays that the energy as a function of inverse temperature $\beta$. A similar energy behavior is observed in Figures~\ref{fig:5}(c) and~\ref{fig:5}(d), showing comparable ensemble results. Outside the transition region, the curves remain smooth. As Figure~\ref{fig:5} demonstrates, the phase transition emerges in the range $0.1<\mu'^{(nm)}<0.2$ with increasing coupling strength. This system exhibits another distinct transition structure when $0.4<\mu'^{(nm)}<0.5$ (physically representing the same phase transition, though without equal partition functions and analogous to band theory). For the original phase transition at fixed $\mu^{(n)}=0.3$, the additional coupling $\mu'^{(nm)}$ plays a role similar to introducing $\mu=0.3$ (as demonstrated in previous work). In the previous work, the second-order transition ends at a critical point ($T_c=1.15$, $\mu_c=0.7$). In our calculation in this work, if we have the first-order parameter $\mu^{(n)}_c=0.3$, when the second-order coupling arrives $\mu'^{(nm)}_c=0.84$, the critical point appears at $T_c=1.15$. 

\subsection{Lorentzian correlation functions and Numerical approach}
\quad In this subsection, we briefly review the supersymmetric Lorentzian correlation functions~\cite{OFF}. For both $\mathcal{N}=1$ and $\mathcal{N}=2$ supersymmetric SYK models, the Lorentz-Wightman correlation functions can be derived with superfield~\cite{35,36,37}. We write the Green functions as follows
\begin{equation}
\begin{split}
\mathcal{G} _{AB}^{>}\left( t_1,t_2 \right) =-i\mathcal{G} _{AB}\left( it_{1}^{-},it_{2}^{+} \right) &=-i\underset{\epsilon \rightarrow -0}{\lim}\mathcal{G} _{AB}\left( it_1+\epsilon ,it_2-\epsilon \right) ,
\\
\mathcal{G} _{AB}^{<}\left( t_1,t_2 \right) =-i\mathcal{G} _{AB}\left( it_{1}^{+},it_{2}^{-} \right) &=-i\underset{\epsilon \rightarrow +0}{\lim}\mathcal{G} _{AB}\left( it_1-\epsilon ,it_2+\epsilon \right) ,
\\
\mathcal{G} _{AB}^{R}\left( t_1,t_2 \right) =\vartheta \left( t_1-t_2 \right) &\left( \mathcal{G} _{AB}^{>}\left( t_1,t_2 \right) -\mathcal{G} _{AB}^{<}\left( t_1,t_2 \right) \right) ,
\\
\mathcal{G} _{AB}^{A}\left( t_1,t_2 \right) =\vartheta \left( t_2-t_1 \right) &\left( \mathcal{G} _{AB}^{>}\left( t_1,t_2 \right) -\mathcal{G} _{AB}^{<}\left( t_1,t_2 \right) \right) .
\end{split}
\end{equation}
The Schwinger-Dyson equations in Lorentzian time are defined as
\begin{equation}
iD_{t_1}G_{AB}^{>}\left( t_1,t_2 \right) =\mu\epsilon _{AC}\partial _{\theta}G_{CB}^{>}\left( t_1,t_2 \right) +\int{dtd\theta \left( \varSigma _{AC}^{R}\left( t_1,t \right) \mathcal{G} _{CB}^{>}\left( t,t_2 \right) +\varSigma _{AC}^{>}\left( t_1,t \right) \mathcal{G} _{CB}^{A}\left( t,t_2 \right) \right)}\,,
\end{equation}
\begin{equation}
  \begin{split}
iD_{t_1}G_{AB}^{R}\left( t_1,t_2 \right) -\mu \epsilon_{AC}\partial _{\theta}G_{CB}^{R}\left( t_1,t_2 \right) -\int{dtd\theta \left( \varSigma _{AC}^{R}\left( t_1,t \right) \mathcal{G} _{CB}^{R}\left( t,t_2 \right) +\varSigma _{AC}^{A}\left( t_1,t \right) \mathcal{G} _{CB}^{A}\left( t,t_2 \right) \right)}
\\
=\delta _{AB}\left( \bar{\theta}_1-\bar{\theta}_2 \right) \delta \left( t_1-t_2 \right) \,,\nonumber
  \end{split}  
\end{equation}
The corresponding Lorentzian equations of motion are given by
\begin{equation}
\begin{split}
&\varSigma _{AB}^{>}\left( t_1,t_2 \right) =\mathcal{J}\mathcal{G} _{AB}^{>2}\left( t_1,t_2 \right) \mathcal{G} _{BA}^{>}\left( t_2,t_1 \right),
\\
&\varSigma _{AB}^{R}\left( t_1,t_2 \right) =\vartheta \left( t_1-t_2 \right) \left( \varSigma _{AB}^{>}\left( t_1,t_2 \right) -\varSigma _{AB}^{<}\left( t_2,t_1 \right) \right).
\end{split}
\end{equation}
Furthermore, we label the processes with replica indices. First, we extend the parameter domains $n$ to accommodate multiple copies. The Green's functions are correspondingly labeled as
\begin{equation}
\begin{split}
\mathcal{G} _{AB}^{>}\left( t_1,t_2 \right) \rightarrow \mathcal{G} _{AB}^{\left( n \right) >}\left( t_1,t_2 \right) =-i\mathcal{G} _{AB}^{\left( n \right)}\left( it_{1}^{-},it_{2}^{+} \right) ,
\\
\mathcal{G} _{AB}^{<}\left( t_1,t_2 \right) \rightarrow \mathcal{G} _{AB}^{\left( n \right) <}\left( t_1,t_2 \right) =-i\mathcal{G} _{AB}^{\left( n \right)}\left( it_{1}^{+},it_{2}^{-} \right) .
\end{split}
\end{equation}
We can also express the following Green's function with orders
\begin{equation}
\begin{split}
\mathcal{G} _{AB}^{\left( nm...l \right) >}\left( t_1,t_2 \right) =-i\mathcal{G} _{AB}^{\left( nm...l \right)}\left( it_{1}^{-},it_{2}^{+} \right) ,
\\
\mathcal{G} _{AB}^{\left( nm...l \right) <}\left( t_1,t_2 \right) =-i\mathcal{G} _{AB}^{\left( nm...l \right)}\left( it_{1}^{+},it_{2}^{-} \right) .
\end{split}
\end{equation}

\begin{figure}[!t]
\begin{minipage}{0.38\linewidth}
\centerline{\includegraphics[width=9cm]{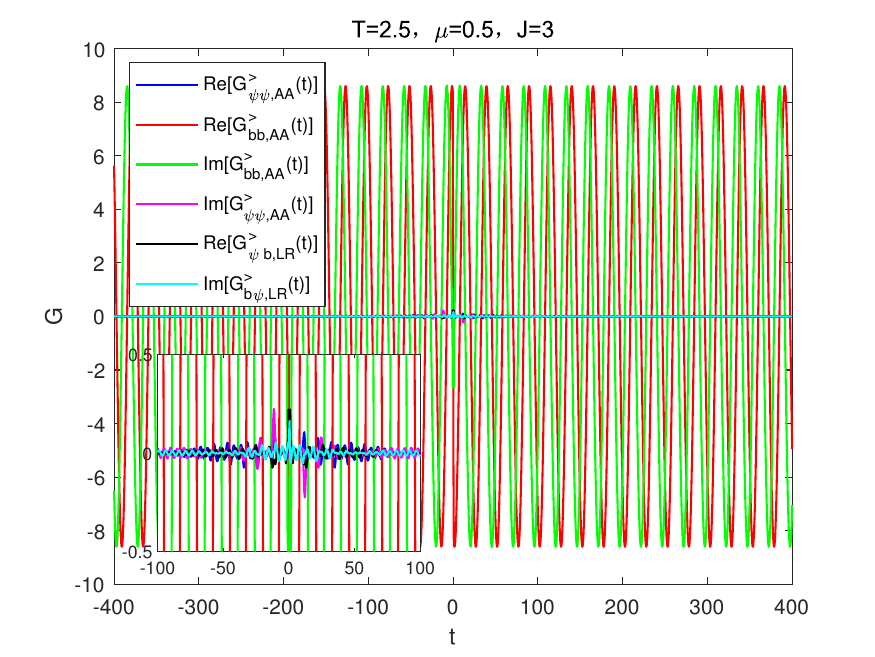}}
\centerline{(a)}
\end{minipage}
\hfill
\begin{minipage}{0.38\linewidth}
\centerline{\includegraphics[width=9cm]{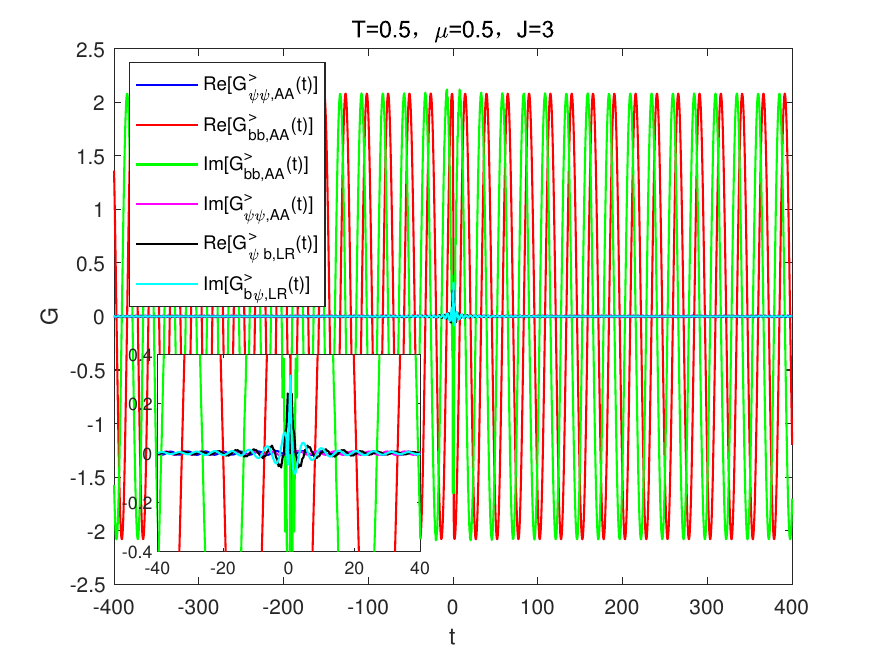}}
\centerline{(b)}
\end{minipage}

\caption{\label{fig:6} (a)Lorentzian Green's function with $\mu^{(n)}=0.5,T=2.5$(black hole phase).(b)Lorentzian Green's function with $\mu^{(n)}=0.5,T=0.5$(wormhole phase).}
\end{figure}

We can also apply the reparameterization technique to reduce higher-order terms to second order term, as discussed in the previous subsection. Since we have demonstrated that $\mu'^{(nm)}$ and $\mu^{(n)}$ exhibit similar behavior in the thermal limit, we set $\mu'^{(nm)}=0$ to simplify the Lorentzian Green's function calculation, as shown in Appendix~\ref{ap1}. We suppose that the interaction occurs at a specific time. Our numerical results for the Lorentzian Green's function components naturally exclude initial-time excitations, as the numerical treatment near $t=0$ would cause divergences in the bosonic retarded correlator.

This fermionic solution package gradually decays to zero, enabling the application of numerical long-time cutoff techniques. As Figure~\ref{fig:6} demonstrates, we obtain two distinct solution sets corresponding to black hole and wormhole configurations, consistent with our previous analysis\cite{OFF}. Specifically, Figure~\ref{fig:6}(a) displays the black hole phase solution, while Figure~\ref{fig:6}(b) shows the wormhole phase solution. Both solutions exhibit temporal oscillations but with different wavepacket decay rates. We have also provided the detailed processes in Appendix~\ref{ap1}.

To resolve this ambiguity, we observe a clear distinction in the action definitions. Building on this observation, we express the total action with explicit Lorentzian time $t$ dependence, which admits a supersymmetric expansion from \eqref{ACT}:
\begin{equation}
\begin{split}
I\left( t \right) &=I\left( 0 \right) +\int{dtd\theta}\left( 2H_0+H_{int} \right) 
\\
&=I\left( 0 \right) +2\int{dt}\left( \left( G_{\psi \psi}^{>}\left( t \right) \right) ^{q-1}G_{bb}^{>}\left( t \right) +\left( G_{\psi \psi}^{>}\left( t \right) \right) ^{q-2}G_{b\psi}^{>}\left( t \right) G_{\psi b}^{>}\left( t \right) \right) 
\\
&+i\mu \int{dt}\left( G_{\psi b}^{>}\left( t \right) -G_{b\psi}^{>}\left( t \right) \right) .
\end{split}
\end{equation}
We can also obtain the replicated action:
\begin{equation}
\begin{split}
I\left( t \right) &=I\left( 0 \right) +2\mathcal{J} \int{dt}\left( \left( G_{\psi \psi}^{\left( n \right) >}\left( t \right) \right) ^{q-1}G_{bb}^{\left( n \right) >}\left( t \right) +\left( G_{\psi \psi}^{\left( n \right) >}\left( t \right) \right) ^{q-2}G_{b\psi}^{\left( n \right) >}\left( t \right) G_{\psi b}^{\left( n \right) >}\left( t \right) \right) 
\\
&+i\mu ^{\left( n \right)}\int{dt}\left( G_{\psi b}^{\left( n \right) >}\left( t \right) -G_{b\psi}^{\left( n \right) >}\left( t \right) \right) +...2\mathcal{J} \int{dt}\left( \left( G_{\psi \psi}^{\left( nm \right) >}\left( t \right) \right) ^{q-1}G_{bb}^{\left( nm \right) >}\left( t \right) \right. 
\\
&\left. +\left( G_{\psi \psi}^{\left( nm \right) >}\left( t \right) \right) ^{q-2}G_{b\psi}^{\left( nm \right) >}\left( t \right) G_{\psi b}^{\left( nm \right) >}\left( t \right) \right) +i\mu \prime^{\left( nm \right)}\int{dt}\left( G_{\psi b}^{\left( nm \right) >}\left( t \right) -G_{b\psi}^{\left( nm \right) >}\left( t \right) \right) .
\end{split}
\end{equation}

\begin{figure}[!t]
\begin{minipage}{0.41\linewidth}
\centerline{\includegraphics[width=9cm]{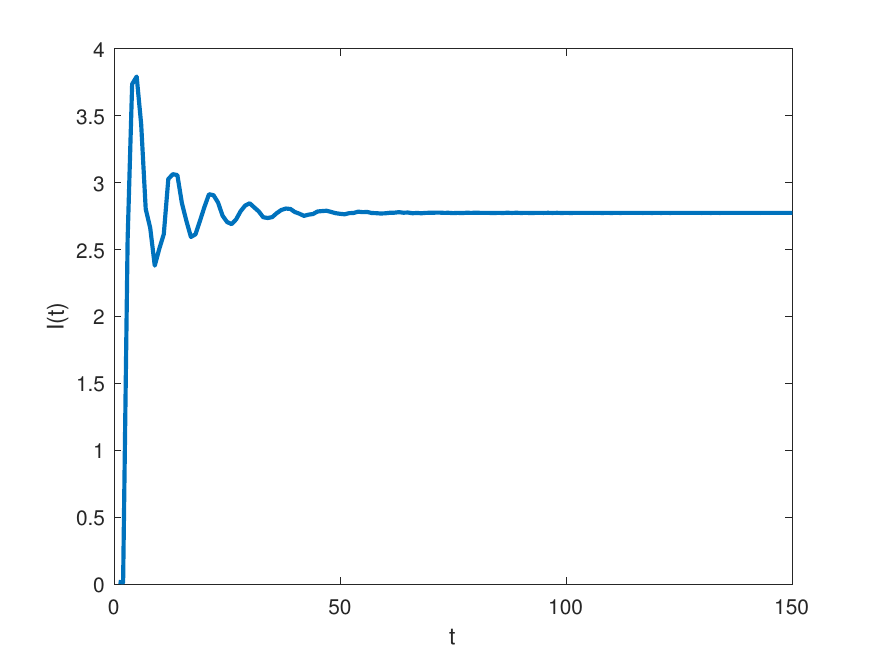}}
\centerline{(a)}
\end{minipage}
\hfill
\begin{minipage}{0.44\linewidth}
\centerline{\includegraphics[width=9cm]{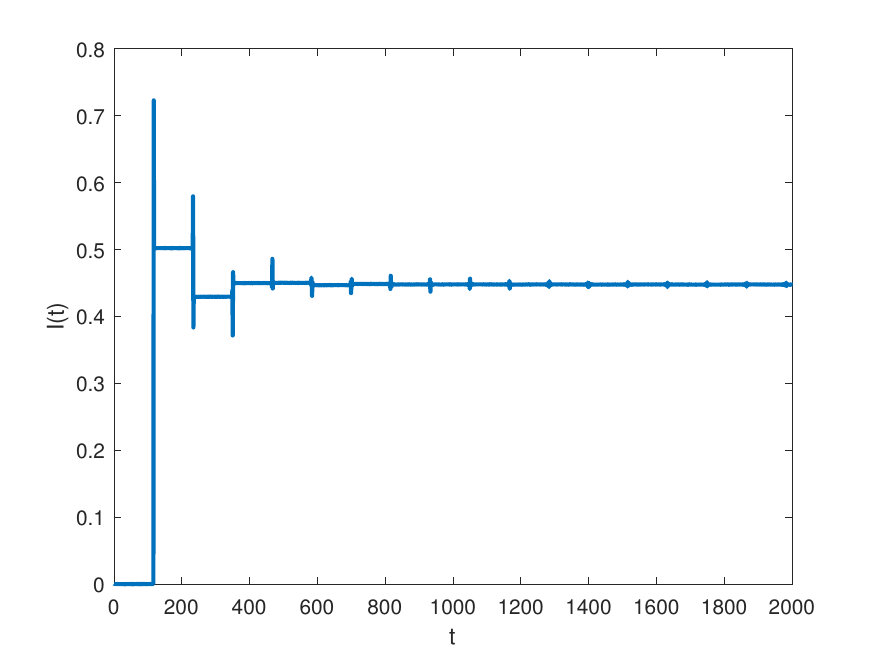}}
\centerline{(b)}
\end{minipage}

\caption{\label{fig:7} (a) Total action as a function of Lorentzian time with $\mu^{(n)}=0.5$, $T=2.5$ (black hole phase).(b)Total action as a function of Lorentzian time with $\mu^{(n)}=0.5$, $T=0.5$ (wormhole phase)}
\end{figure}

Our numerical results for the action are shown in Figure~\ref{fig:7}. The bosonic sector's sine solutions induce early-time oscillations in the action (arising from numerical Fourier artifacts that should average to a constant). Figure~\ref{fig:7}(a) demonstrates that the total action converges to a constant at late times. Notably, the black hole solution reaches its asymptotic value earlier than the wormhole configuration. In Figure~\ref{fig:7}(b), the wormhole solution exhibits multiple equilibrium states due to wavepacket interference effects. These action values share the same magnitude order as the entropy in statistical mechanics, displaying entanglement behavior.

\section{Fractal extension of supersymmetric SYK model}\label{S4}

\quad In this section, we briefly review the fractal structures inherent in the $\mathcal{N}=1$ supersymmetric SYK model and examine their physical relevance to our analysis. In contrast to the approach adopted in the previous section, the framework developed here uncovers additional features of direct physical interest.

A fractal is a geometric or mathematical structure characterized by self-similarity and scale invariance, exhibiting nontrivial detail at arbitrarily small scales~\cite{fr1,fr2}. Its defining property is that local substructures resemble the global form, with fine features persisting under successive magnifications. No matter how many times the structure is magnified, new patterns emerge, and the geometry does not become smooth. Such scale-invariant behavior naturally connects to the hierarchical organization of fermionic degrees of freedom in our supersymmetric SYK construction, where the ordered replica sectors display a self-similar arrangement across different energy scales.

For the $\mathcal{N}=1$ SYK model, signatures of self-similarity and fine-scale structure can be identified in the organization of fermionic degrees of freedom, even though the fermionic sites themselves form a discrete set. Our construction proceeds as follows: we first formulate the $\mathcal{N}=1$ supersymmetric SYK model, then extend it via the fractal mapping~\eqref{fer} to incorporate the replicated structure, and finally introduce the ordered coupling~\eqref{156151} to render these fractal extensions dynamically nontrivial. 

The fractal mapping~\eqref{fer} acts on the index space of fermionic variables, recursively generating self-similar patterns across different replica sectors while preserving the underlying supersymmetry. This mapping effectively embeds a hierarchical organization into the replicated system, allowing local substructures to inherit the symmetry properties of the whole. The ordered coupling~\eqref{156151}, in turn, links specific fermionic regions across different hierarchical levels, introducing interaction channels that break the trivial product structure of independent replicas. As a result, the replicated fractal configuration acquires new collective modes and nontrivial correlations, which are essential for realizing the wormhole-like connectivity discussed in later sections.

\subsection{Fractal symmetry in $\mathcal{N}=1$ SYK model}
\quad The $\mathcal{N}=1$ SYK model satisfies the following fundamental relations. We perform the expansion of the fermionic boundary matrices as
\begin{equation}
\psi_{nm...l}^{i} = \psi_{n}^{i} + \psi_{m}^{i} +... + \psi_{l}^{i}. 
\end{equation}
Note that the supercharge remains invariant under this transformation, implying
\begin{equation}
\begin{split}
Q_{nm...l}=i^{\frac{q-1}{2}}\sum_{i_1i_2...i_q}{C_{i_1i_2...i_q}^{(nm...l)}\psi _{nm...l}^{i_1}\psi _{nm...l}^{i_2}}...\psi _{nm...l}^{i_q}
\\
\rightarrow Q=i^{\frac{q-1}{2}}\sum_{i_1i_2...i_q}{\sum_{nm...l}{C_{i_1i_2...i_q}^{}}\psi _{}^{i_1}\psi _{}^{i_2}}...\psi _{}^{i_q},
\end{split}
\end{equation}
where the fermionic label space has been extended through replica indices,
and the corresponding Hamiltonians remain invariant
\begin{equation}
H=Q_{nm...l}^{2}\rightarrow \sum_{nm...l}{Q^2},
\end{equation}
where the supercharges and Hamiltonian have been reformulated by extracting the fermionic index sum $nm\cdots l$. This construction absorbs the additional effects solely through fermionic sector extensions, maintaining the original SYK fermionic number $N$. We can define arbitrary number of processes and obtain another $\mathcal{N}=1$ SYK system with different fermionic sector. Here we have defined this property as the fractal symmetry of $\mathcal{N}=1$ SYK model. In this case, we can involve the self-similar structure infinitely in the model.

While these transformations appear trivial in both supersymmetric and non-supersymmetric SYK frameworks, the inclusion of supersymmetric interaction terms enables simultaneous treatment of models with varying index structures. As previously established, the $\mathcal{N}=1$ SYK model with interactions inherently exhibits fractal symmetry (unlike its non-supersymmetric counterpart, which cannot incorporate solvable multi-boundary correlations). In the previous section, it is also important to involve the supersymmetric interactions between the left and right sides with ordered.

\subsection{Additional crossterms via the supersymmetry breaking}

In this subsection, we analyze the essential connections between differently ordered surfaces. We mainly focus on the terms that correspond to supersymmetry breaking.

Following model definition, we can start with the two kinds of identical fermions and reconstruct the fractal structure again. Returning to the previous section, any specified special transition must satisfy the relation
\begin{equation}
\psi =\psi _1+\psi _2.
\end{equation}
We classify the conjugate bosonic fields $b_{12}$ according to indices 1 and 2
\begin{equation}
b_{12}=\left\{ Q_{12},\psi _1+\psi _2 \right\} =b_1+b_2+{\rm crossterm}.
\end{equation}
Returning to the interaction term, we obtain
\begin{equation}
H_{int}=H_{int,1}+H_{int,2}+{\rm crossterm}.
\end{equation}
The cross term here corresponds to the super-symmetry breaking parts (for example $\left\{Q_{1},\psi _2  \right\} $).

In the previous section, we have proved that when considering a single $\mathcal{N}=1$ SYK model, fractal generations must exist, with interactions distinct from the initial model. The interactions in the original model divide into three parts: those acting on replica. 1, those acting on replica. 2, and the cross term. This demonstrates a natural derivation of multi-connected diagrams in coupled supersymmetric SYK models. We have provided an interim conclusion, and we will give a further application on SJT gravity in the next section.

When the order exceeds $n=2$, supersymmetry breaks due to additional particle states (the fermionic sectors are reducible). The mismatch between identical fermions and distinct bosons actually violates supersymmetry. These extra bosons necessarily break supersymmetry, as introducing new regulators in the superfield theory leads to loss of physical information (e.g., replica indices). Furthermore, correlators between different boson types are physically forbidden. 
In gravitational theories, no complete mapping exists between different-order dilaton fields at the boundaries. We will further give a detailed discussion in the next sections.

Mathematically, the higher-order terms
\begin{equation}
G_{bb}^{\left( n,nm \right)}\left( \tau ,\tau ' \right) =\left< b_{n}^{i}\left( \tau \right) b_{nm}^{i}\left( \tau ' \right) \right> ,
\end{equation}
which are strictly forbidden in our theoretical framework. The inclusion of differently-ordered replicas necessarily breaks global supersymmetry, as we cannot maintain supersymmetric invariance. Given the random nature of SYK couplings, the sole controllable parameter is the interaction strength.

Additionally, at higher orders where the coupling $\mathcal{J} \to 0$ is suppressed, these terms in the thermal limit become physically indistinguishable. Our analysis demonstrates that no parameterization can fully describe both first- and second-order terms simultaneously. Nevertheless, terms beyond second order may remain renormalizable as near-free particle. We can give another expression of the supersymmetry breaking: We can not find a supersymmetric field globally to include bosons with different orders at the same time.

\subsection{Relationship between replicated thermodynamics and the Fractal symmetry}
\quad In this subsection, we provide a detailed discussion of the relationship between symmetric replicas (Section~II) and the fractal extension in supersymmetry (Section~III). We will also give an argument about the fractal symmetry in the $\mathcal{N}=1$ supersymmetric SYK model and replicated thermodynamics.

The model introduced in Section~II is orthogonally defined, meaning we directly introduce the parallel model and incorporate its random interactions. In the low-energy limit, the Section~II model comprises multiple independent SYK models, with the Hamiltonian summing over labeled systems. In Section~III, we introduce replicas through fractal symmetry of N=1 supersymmetry, expanding the fermionic definition domains with specific surface relationships. The model in Section~III only expands the fermion sector in the supercharges before extending to bosons. Compared to the model in the previous section, both maintain the same Hilbert space dimension order. We can therefore use identical fermionic irreducible representations for both spinor models. Here we define an alternative $\mathcal{N}=1$ SYK model incorporating both the total fermions and corresponding bosons through expanded supercharges.

Here we have also provided some important details about the supersymmetric model. After incorporating interactions with non-trivial orders, these theories become fundamentally distinct. The first physical effect manifests in the bosonic sector: we can define additional bosonic particles through supersymmetry breaking (where the summed fermions remain reducible while the newly-defined bosons are irreducible). These bosons lack direct correspondence to the Section~II models. Second, the interactions now contain supersymmetric terms connecting individual fermions and bosons, differing fundamentally from either the random couplings in Section~II or chain structures, as shown in Appendix~\ref{ap4}. When examining off-shell diagrams, the supersymmetric interactions may perturbatively reduce to purely fermionic interactions. We first make the model in Section~II supersymmetric (the $\mathcal{N}=1$ supersymmetry does not change the Hilbert space but makes the bosons dynamical). We also modify the interactions to be supersymmetric.

Compared with the supersymmetric model in this section, if we only introduce first-order MQ interactions\cite{1804.00491} between $n$-replicated non-supersymmetric SYK models, defining the details of replica wormholes becomes difficult. Although we can incorporate similar fractal relationships and first-order interactions in non-supersymmetric SYK models, we no longer find interaction terms with global properties like $b\psi$. As shown in the subsections before, we have sum that we can use the supersymmetric model to cover the many body structures with replicas.

\section{Multi-Boundary Supersymmetric JT Gravity and the Super-Schwarzian Flow}\label{S5}

\quad In this section, we investigate the low-energy limit of coupled supersymmetric SYK models and explore the associated constraints on $\mathcal{N}=1$ supersymmetric Jackiw–Teitelboim (SJT) gravity. These constraints are required for the bulk theory to remain holographically dual to the boundary SYK model. We begin by extracting the global structure of the replicated supersymmetric SYK system and then analyze its emergent superconformal symmetry in the infrared regime. These features provide guiding principles for constructing the corresponding multi-boundary SJT gravity with replica structure.

The effective bulk theory must obey three key principles emerging from the boundary replica construction: 
(i) a \emph{principle of parallelism}, which requires that all replicas share identical infrared dynamics, 
(ii) a \emph{principle of perpendicularity}, where coordinate transformations between ordered replicas involve additional flow terms that encode inter-replica interactions, and 
(iii) a \emph{principle of finite displacement}, which imposes constraints on the allowed separation between replicated boundary coordinates. These principles collectively inform the construction of the effective replicated supergravity action, as we will discuss in detail.

\subsection{Infrared Replicas and Emergent Superconformal Symmetry}

\quad The emergence of a holographic dual description is closely tied to the appearance of superconformal symmetry in the low-energy limit~\cite{5,6}. In particular, the $\mathcal{N}=1$ supersymmetric SYK model develops an approximate reparametrization symmetry described by the superconformal group $OSp(1|2)$. Previous studies~\cite{12} have shown that this group includes an emergent fermionic reparametrization that becomes dynamical in the infrared.

To describe this symmetry, we adopt generalized superspace coordinates and implement reparametrizations via superdiffeomorphisms~\cite{OFF}. As demonstrated in~\cite{9}, the interaction terms responsible for wormhole-like replica couplings naturally flow to the superconformal regime in the deep infrared. In our analysis, we derive explicit relations between Green's functions of different replica orders, demonstrating how they obey consistent transformation properties under $OSp(1|2)$ symmetry. These relations serve as boundary conditions for the dual bulk SJT theory 
\begin{equation}
G_{bb}\left( \tau ,\tau ' \right) =\left< b\left( \tau \right) b\left( \tau ' \right) \right> =-\partial _{\tau}\left< \psi \left( \tau \right) \psi \left( \tau ' \right) \right> =-\partial _{\tau}G_{\psi \psi}\left( \tau ,\tau ' \right),   
\end{equation}
with self-energy term
\begin{equation}
\varSigma _{\psi \psi}\left( \tau ,\tau ' \right) =-\partial _{\tau}\varSigma _{bb}\left( \tau ,\tau ' \right) .
\end{equation}
The conformal ansatz remains preserved as the higher-order terms play analogous roles in the symmetry structure.

In the holographic picture of the non-supersymmetric ($\mathcal{N}=0$) case, the emergent conformal symmetry in the low-energy limit becomes manifest. Previous work has demonstrated that the $\mathcal{N}=1$ supersymmetric algebra naturally incorporates an additional super-reparametrization symmetry~\cite{9}. We can express these reparametrization transformations in terms of generalized coordinates 
\begin{equation}
\tau \rightarrow \tau '\left( \tau ,\theta \right) ,
\
\theta \rightarrow \theta'\left( \tau ,\theta \right) .
\end{equation}
Under more general reparametrizations, correlation functions with certain dimensions should remain invariant under rescaled coordinate transformations.

In the low-energy limit, the corresponding Berezinian simplifies to a super-Jacobian derivative factor. This simplification gives rise to a nearly superconformal symmetry with a conformal dimension of $1/q$ 
\begin{equation}
Ber\left( \tau ',\theta ';\tau ,\theta \right) =D_{\theta}\theta '.
\end{equation}
In the IR limit, the correlation functions display superconformal symmetry, which encompasses invariance under covariant derivative transformations 
\begin{equation}
\mathcal{G} \left( \tau _1,\theta _1;\tau _2,\theta _2 \right) =\left( D_{\theta _1}\theta _{1}^{'} \right) ^{\frac{1}{q}}\left( D_{\theta _2}\theta _{2}^{'} \right) ^{\frac{1}{q}}\mathcal{G} \left( \tau _{1}^{'},\theta _{1}^{'};\tau _{2}^{'},\theta _{2}^{'} \right) ,
\end{equation}
which is invariant under super-transformations

\begin{equation}
\tau'=\tau +\epsilon +\theta \eta ,
\
\theta'=\theta +\eta .
\end{equation}
Note that the parameter $\epsilon$ represents an arbitrary bosonic translation, and $\eta$ is the corresponding Grassmann variable. We additionally perform a specific reparameterization of the $\mathcal{N}=1$ superalgebra 
\begin{equation}
\tau \rightarrow \tau '=f\left( \tau \right) ,
\
\theta \rightarrow \theta '=\sqrt{\partial _{\tau}f\left( \tau \right)}\theta .\label{RP}
\end{equation}
Furthermore, the superconformal transformations can be conveniently generated through reparameterized inversion.

In the IR limit, the superconformal symmetry in supersymmetric SYK allows construction of a coupling operator connecting two decoupled models via superconformal reparameterizations $h$ in the single-sided SSYK. Moreover, we can incorporate the coupling action through hyperbolic reparameterization transformations\cite{OFF} 
\begin{equation}
S_{int}=\frac{\mu}{2}\int{d\tau d\theta \left( \frac{bD_{\theta}\left( \theta ' \right) _LD_{\theta}\left( \theta ' \right) _R}{\cosh ^2\frac{h_L\left( \tau +\theta \eta \left( \tau \right) \right) -h_R\left( \tau +\theta \eta \left( \tau \right) \right) -\left( \theta ' \right) _L\left( \theta ' \right) _R}{2}} \right)}^{\frac{1}{q}}.
\end{equation}
Thanks to the superspace mathematical relations and the included terms, the low-energy interaction terms preserve superconformal symmetry. Moreover, the correlation functions reduce to the non-supersymmetric case when the superspace prolongation vanishes.

In this subsection, we extend these properties to the replica system. The theory decomposes into independent sectors where each part maintains the original superconformal invariance. (We use $n$ to label first-order terms, use $nm\cdots l$ to label arbitrary-order terms.)
\begin{align}
G_{\psi \psi}\left( \tau \right) \sim& \frac{sgn\left( \tau \right)}{c_{\psi}\left( \tau \right)}\sim \frac{sgn\left( \tau \right)}{\left| \tau \right|^{2\varDelta}},
\\
G_{bb}^{\left( n \right)}\left( \tau \right) \sim& -\frac{\delta \left( \tau \right)}{c_b\left( \tau \right)}\sim \frac{1}{\left| \tau \right|^{2\varDelta +1}},
\\
G_{bb}^{\left( nm...l \right)}\left( \tau \right)& \sim -\frac{\delta \left( \tau \right)}{c_b\left( \tau \right)}\sim \frac{1}{\left| \tau \right|^{2\varDelta +1}}.
\end{align}
Because bosons of different orders merely modify the fermionic regions without altering their fundamental properties, they should share the same superconformal dimensions. The functions $c$ represent the conformal factors in the propagators and are determined by the conformal dimension, which is related to the equation $G_{\psi \psi}\left( \tau \right) = -\partial _{\tau}G_{bb}\left( \tau \right)$. This conformal dimension should remain invariant under replica expansion.

In the weak-coupling limit, the superconformal reparameterization symmetry remains preserved 
\begin{equation}
\begin{split}
\mathcal{G} ^{\left( n \right)}\left( \tau _1,\theta _1;\tau _2,\theta _2 \right) &=\left( D_{\theta _1}\theta _{1}^{\prime} \right) ^{\frac{1}{q}}\left( D_{\theta _2}\theta _{2}^{\prime} \right) ^{\frac{1}{q}}\mathcal{G} ^{\left( n \right)}\left( \tau _{1}^{\prime},\theta _{1}^{\prime};\tau _{2}^{\prime},\theta _{2}^{\prime} \right) .
\\
\mathcal{G} ^{\left( nm...l \right)}\left( \tau _1,\theta _1;\tau _2,\theta _2 \right)& =\left( D_{\theta _1}\theta _{1}^{\prime} \right) ^{\frac{1}{q}}\left( D_{\theta _2}\theta _{2}^{\prime} \right) ^{\frac{1}{q}}\mathcal{G}^{\left( nm...l \right)}\left( \tau _{1}^{\prime},\theta _{1}^{\prime};\tau _{2}^{\prime},\theta _{2}^{\prime} \right) .
\end{split}
\end{equation}
However, there is no global superconformal function $\mathcal{G}$ exists that simultaneously covers
\begin{equation}
\begin{split}
&\mathcal{G} \left[ \mathcal{G} ^{\left( n \right)}\left( \tau _1,\theta _1;\tau _2,\theta _2 \right) ;\mathcal{G} ^{\left( nm \right)}\left( \tau _1,\theta _1;\tau _2,\theta _2 \right) ;\mathcal{G} ^{\left( nm...l \right)}\left( \tau _1,\theta _1;\tau _2,\theta _2 \right) \right] 
\\
&=\left( D_{\theta _1}\theta _{1}^{'} \right) ^{\frac{1}{q}}\left( D_{\theta _2}\theta _{2}^{'} \right) ^{\frac{1}{q}}\mathcal{G} \left[ \mathcal{G} ^{\left( n \right)}\left( \tau _{1}^{'},\theta _{1}^{'};\tau _{2}^{'},\theta _{2}^{'} \right) ;\mathcal{G} ^{\left( nm \right)}\left( \tau _{1}^{'},\theta _{1}^{'};\tau _{2}^{'},\theta _{2}^{'} \right) ;\mathcal{G} ^{\left( nm...l \right)}\left( \tau _{1}^{'},\theta _{1}^{'};\tau _{2}^{'},\theta _{2}^{'} \right) \right] ,
\end{split} \nonumber
\end{equation}
when contributions from finite-ordered Green functions 
$
\mathcal{G}^{(nm)}\left(\tau_1',\theta_1';\tau_2',\theta_2'\right),\cdots,\mathcal{G}^{(nm\cdots l)}\left(\tau_1',\theta_1';\tau_2',\theta_2'\right)
$
are non-vanishing.

This condition directly demonstrates supersymmetry breaking, and we will provide a detailed gravitational analysis of this breaking later. The reparametrization can be treated perturbatively, yielding an effective super-Schwarzian action directly from the original SSYK model.

We may further perform specialized reparameterizations of $\mathcal{N}=1$ supersymmetry for components with single indices $n$, index pairs $nm$, and multiple indices $nm\cdots l$. For instance, the first-order reparameterization is parametrized by index $n$ from \eqref{RP}
\begin{equation}
\tau \rightarrow \tau '=f^{\left( n \right)}\left( \tau \right) ,
\
\theta \rightarrow \theta '=\sqrt{\partial _{\tau}f^{\left( n \right)}\left( \tau \right)}\theta .
\end{equation}
We can also write down arbitary solutions with the indices $ nm...l $
\begin{equation}
\tau \rightarrow \tau '=f^{\left( nm...l \right)}\left( \tau \right) ,
\
\theta \rightarrow \theta '=\sqrt{\partial _{\tau}f^{\left( nm...l \right)}\left( \tau \right)}\theta .\label{rep3}
\end{equation}
The super-reparameterizations $f$ are mutually independent of order. To briefly review our discussion, the superconformal symmetry transformations can be conveniently generated through inversion reparameterization, as previously established \cite{9}. In the low-energy limit, the superconformal symmetry in the supersymmetric SYK model enables construction of a correlation operator between two decoupled models through superconformal reparameterizations $h$ in single-sided SSYK. The correlation function in this limit admits an ansatz
\begin{equation}
\begin{split}
\mathcal{G} ^{\left( n \right)}\left( \tau _1,\theta _1;\tau _2,\theta _2 \right) =\frac{b_n}{\left| \tau _1-\tau _2-\theta _1\theta _2 \right|^{2\varDelta}},
\\
\mathcal{G} ^{\left( nm...l \right)}\left( \tau _1,\theta _1;\tau _2,\theta _2 \right) =\frac{b_{nm...l}}{\left| \tau _1-\tau _2-\theta _1\theta _2 \right|^{2\varDelta}},
\end{split}
\end{equation}
where the conformal dimension is constrained to $\Delta=1/q$. The interaction action can be incorporated through hyperbolic reparameterizations
\begin{equation}
\begin{split}
S_{int}^{(n)}&=\frac{\mu ^{(n)}}{2}\int{d\tau d\theta \left( \frac{b_{n}D_{\theta}(\theta ^{'})_LD_{\theta}(\theta ^{'})_R}{\cosh ^2\!\:\frac{h_{L}^{\left( n \right)}(\tau +\theta \eta (\tau ))-h_{L}^{\left( n \right)}(\tau +\theta \eta (\tau ))-(\theta ^{'})_L(\theta ^{'})_R}{2}} \right) ^{\frac{1}{q}},}
\\
S_{int}^{(nm...l)}&=\frac{\mu ^{(nm...l)}}{2}\int{d\tau d\theta \left( \frac{b_{nm...l}D_{\theta}(\theta ^{'})_LD_{\theta}(\theta ^{'})_R}{\cosh ^2\!\:\frac{h_{L}^{\left( nm...l \right)}(\tau +\theta \eta (\tau ))-h_{L}^{\left( nm...l \right)}(\tau +\theta \eta (\tau ))-(\theta ^{'})_L(\theta ^{'})_R}{2}} \right) ^{\frac{1}{q}}.}
\end{split}
\end{equation}
We obtain the ordered Schwarzian-like effective action by summing over the single parts
\begin{equation}
\begin{split}
S_A=&-\int{d\tau d\theta}N\alpha _{S}^{(n)}S\left[ \tanh \left( \frac{h_{A}^{\left( n \right)}}{2} \right) ,\theta _{A}^{\left( n \right)\prime};\tanh \left( \frac{\tau _{A}^{\left( n \right)}}{2} \right) ,\theta _{A}^{\left( n \right)} \right] 
\\
&+N\alpha _{S}^{(nm)}S\left[ \tanh \left( \frac{h_{A}^{\left( nm \right)}}{2} \right) ,\theta _{A}^{\left( nm \right)\prime};\tanh \left( \frac{\tau _{A}^{\left( nm \right)}}{2} \right) ,\theta _{A}^{\left( nm \right)} \right] 
\\
&+...+N\alpha _{S}^{(nm...l)}S\left[ \tanh \left( \frac{h_{A}^{\left( nm...l \right)}}{2} \right) ,\theta _{A}^{\left( nm...l \right)\prime};\tanh \left( \frac{\tau _{A}^{\left( nm...l \right)}}{2} \right) ,\theta _{A}^{\left( nm...l \right)} \right] .
\end{split}
\end{equation}
Parameter $\alpha_{S}$ originates from the four-point function. The contributions arise from different orders, with a key distinction being the effective action's loss of global superconformal invariance under reparameterization\eqref{rep3}. 

\subsection{The ordered JT gravity with replicas}
\quad In this subsection, we examine the holographic duality of supersymmetric SYK models extended to replicas. First, we review low-energy $\mathcal{N}=1$ supersymmetric SYK models and their holographic duality~\cite{12,13,14}, described through JT gravity extensions~\cite{46,47}. Previous work has established the traversability of supersymmetric wormholes \cite{OFF,461,471}. We analyze $\mathcal{N}=1$ super JT gravity following \cite{12}
\begin{equation}
S=-\frac{1}{16\pi G}\left[ i\int{d^2zd^2\theta}E\varPhi \left( R_{+-}-2 \right) +2\int_{\partial M}{dud\theta}\varPhi K \right] ,\label{sjt1}
\end{equation}
where $E$ is the superdeterminant of vielbein in superspace, $\varPhi$ is the dilaton superfield. $R_{+-}$ is a natural supersymmetrization of the curvature, which is corresponding to the two fermionic generators of $OSp(1|2)$. This framework imposes a specific superconformal gauge condition 
\begin{equation}
\frac{du^2+2\theta d\theta du}{4\epsilon ^2}=dz^{\xi}E_{\xi}^{1}dz^{\pi}E_{\pi}^{\bar{1}}.
\end{equation}
The superconformal gauge condition is the most important and widely studied among those relevant to quantum gravity and supersymmetric field theories. To obtain the effective action, the vielbeins in both dimensions must satisfy
\begin{equation}
Dz=\theta D\theta ,
\end{equation}
this condition with infinitesimal variations 
\begin{equation}
z=t+i\epsilon \left( D\xi \right) ^2,
\
Dt\left( u,\theta \right) =\xi \left( u,\theta \right) D\xi \left( u,\theta \right) .
\end{equation}

Supersymmetric gravity introduces additional spinor components. Consistency between these spinors and the scalar superfield superspace requires incorporation of a supervielbein. We denote by $E$ the vielbein components associated with inverse density. Considering a direction $A$, yielding
\begin{equation}
K=\frac{T^AD_Tn_A}{T^AT_A},
\end{equation}
where $T^A$ is the boundary tangent vector satisfying $T^A n_A = 0$ 
\begin{equation}
K=4\epsilon ^2S\left[ t,\xi ;u,\theta \right] .
\end{equation}
The covariant derivative $D_T$ incorporates contributions from both the superderivative and spin connection, as expressed in $D_T n_A = D n_A + D \Omega_A$. The boundary action exhibits Schwarzian-derivative-like properties
\begin{equation}
S_{bdy}=\int{dud\theta}\varPhi _r\left( u,\theta \right) S\left[ t,\xi ;u,\theta \right] .
\end{equation}
To describe the wormhole configuration, we introduce paired interaction operators at both left and right boundaries 
\begin{equation}
S_{int}=g\sum_i{\int{dud\theta}O_{L}^{i}\left( u,\theta \right)}O_{R}^{i}\left( u,\theta \right) .
\end{equation}
Note that ${O}$ comprises $N$ operators with superconformal dimension $\Delta$. The coupling $g$ has dimension $[\text{energy}]^{2\Delta-1}$. 

The interaction terms admit holographic construction 
\begin{equation}
\left< O\left( t_{P}^{1} \right) O\left( t_{P}^{2} \right) \right> =\left| t_{P}^{1}-t_{P}^{2}-\theta _{P}^{1}\theta _{P}^{2} \right|^{-2\varDelta}.
\end{equation}
After applying thermal reparametrization
\begin{equation}
h=\tanh \frac{\left( \tau +\theta \eta \right)}{2},
\
\theta ^{'}=\left[ \partial _{\tau}\tanh \left( \frac{\left( \tau +\theta \eta \right)}{2} \right) \right] ^{\frac{1}{2}}\left( \theta +\eta +\frac{1}{2}\eta \partial _{\tau}\eta \right) .
\end{equation}
The interactions clearly maintain the superconformal form of the low-energy $\mathcal{N}=1$ SYK model. From this action's mathematical form, we derive Schwarzian theories order-by-order through $\mathcal{N}=1$ JT supergravity with replicas.

The replica component enters via the boundary-mapping parameter established previously. Interaction operators are confined to distinct reparameterization boundaries, with their couplings constrained by different dilaton field orders. The replicated gravitational actions follow an ansatz formulation

\begin{equation}
\begin{split}
S_{bdy}&=-\int{d\tau d\theta}\varPhi _{r}^{(n)}S\left[ \tanh \left( \frac{h_{A}^{\left( n \right)}}{2} \right) ,\theta ';\tanh \left( \frac{u}{2} \right) ,\theta \right] +\varPhi _{r}^{(nm)}S\left[ \tanh \left( \frac{h_{A}^{\left( nm \right)}}{2} \right) ,\theta ';\tanh \left( \frac{u}{2} \right) ,\theta \right] 
\\
&+...+\varPhi _{r}^{(nm...l)}S\left[ \tanh \left( \frac{h_{A}^{\left( nm...l \right)}}{2} \right) ,\theta ';\tanh \left( \frac{u}{2} \right) ,\theta \right]. \label{tot}
\end{split}
\end{equation}

In \eqref{tot}, interaction strengths at saddle points manifest through coupling constants within each copy. Since $\mathcal{N}=1$ JT supergravity combines the original bosonic sector with its fermionic partner via extended supersymmetry, we simply apply this property into ordered theory. The higher-order terms should also indicate supersymmetry breaking, which manifests on hyperbolic surfaces through potentially undefined boundary contours, with resulting discontinuities partitioning the surface into disjoint regions. The topological linking of these regions defines the replica wormhole geometry. Pure cross-terms are prohibited by the absence of non-trivial solutions, making boundaries indistinguishable beyond reparametrization. However, a natural topological correlation persists between dilaton fields.

\begin{figure}[!t]
\begin{minipage}{0.48\linewidth}
\centerline{\includegraphics[width=8cm]{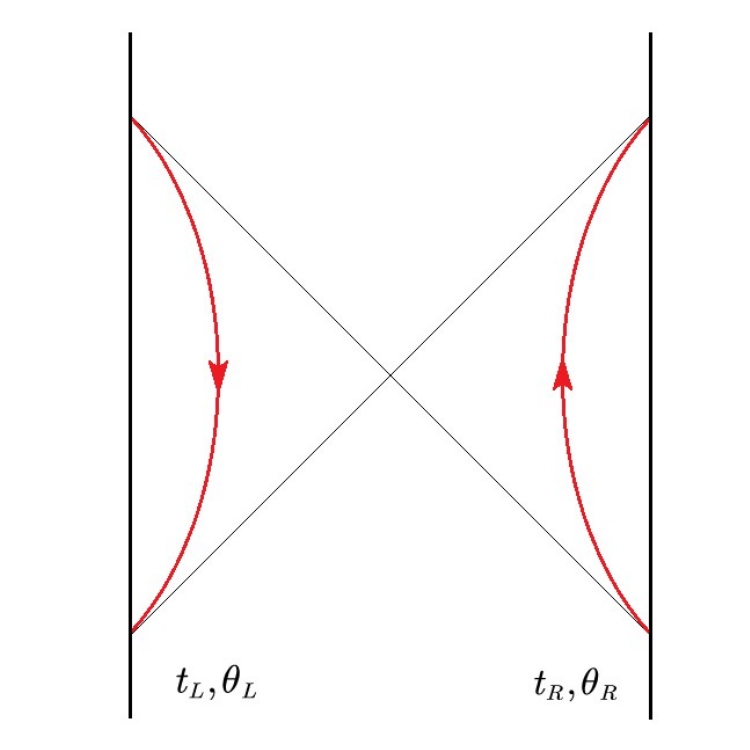}}
\centerline{(a)}
\end{minipage}
\hfill
\begin{minipage}{0.48\linewidth}
\centerline{\includegraphics[width=8cm]{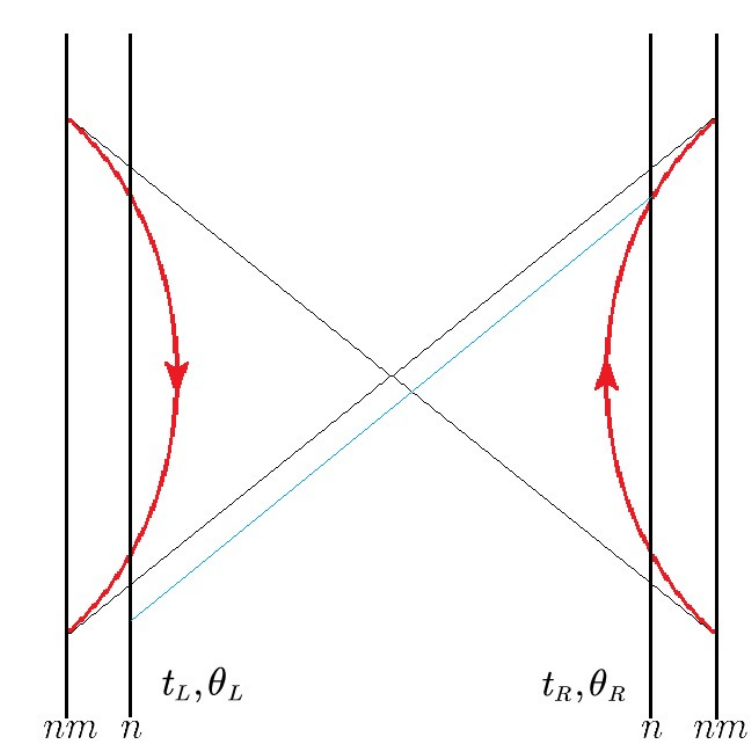}}
\centerline{(b)}
\end{minipage}
\hfill
\begin{minipage}{0.48\linewidth}
\centerline{\includegraphics[width=8cm]{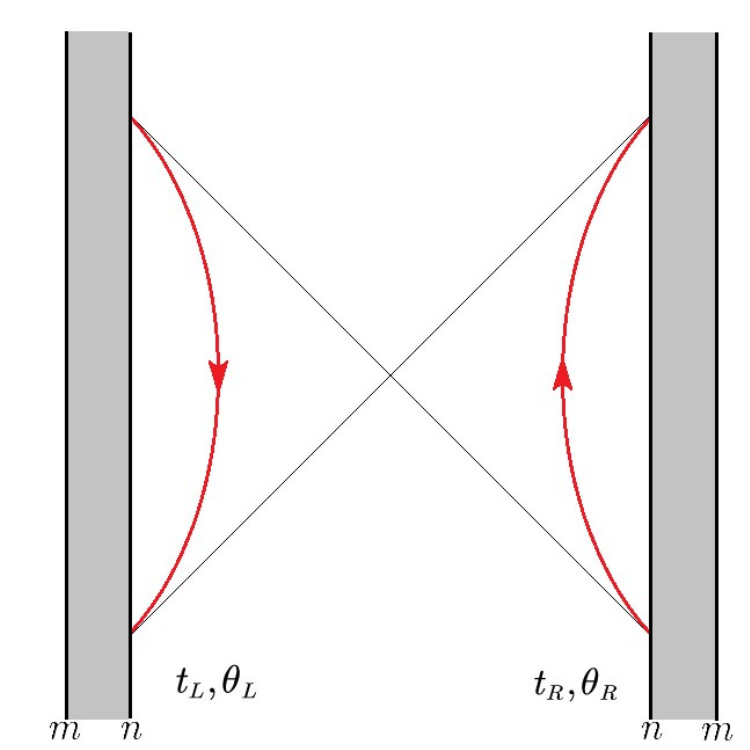}}
\centerline{(c)}
\end{minipage}
\hfill
\begin{minipage}{0.48\linewidth}
\centerline{\includegraphics[width=8cm]{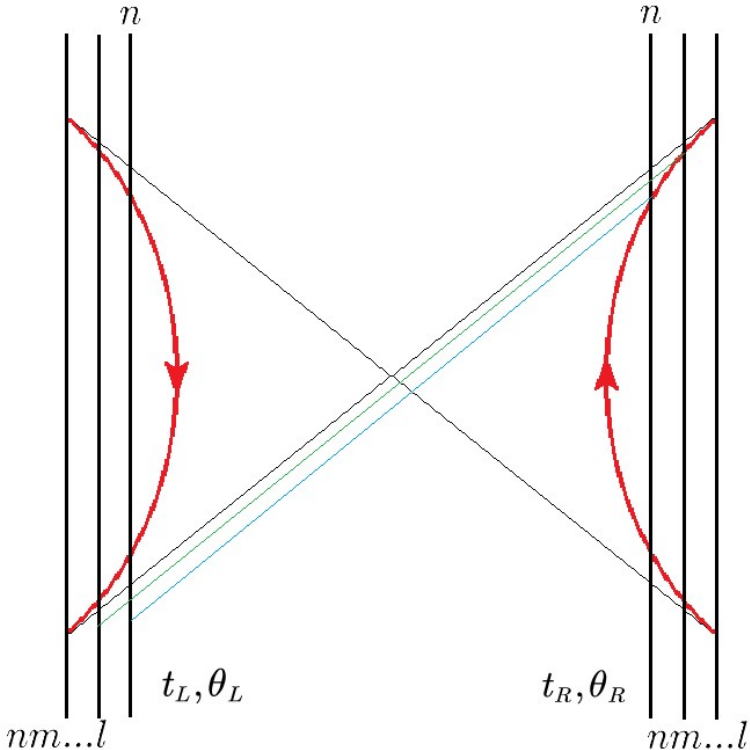}}
\centerline{(d)}
\end{minipage}
\caption{\label{fig:8} (a) In extended $\text{NAdS}_2$ spacetime with SJT gravity, boundaries are reparameterized using bosonic coordinate $t$ and Grassmann parameter $\theta$. (b) Boundary extensions employ replica indices $n$ and $nm$ with multi-order trick, though these reparameterizations remain fundamentally distinct. (c) Supersymmetry restoration occurs when replica indices coincide, unifying boundary reparameterizations with the same coordinate(the gray area). (d) We generalize the supersymmetric diagram with multi-order trick from Fig.~(b) to arbitrary order. }
\end{figure}

As shown in Figure~\ref{fig:8}(a), we plot region within the $\text{NAdS}_2$ spacetime \eqref{sjt1}. Figure~\ref{fig:8}(b) introduces additional degrees of freedom on the $\text{NAdS}_2$ boundaries(The first two terms in \eqref{sjt2}), where we represent the two dilaton types from reparametrization as $n$ and $nm$. In this configuration, no global super-reparametrization simultaneously yields the dilaton on the same boundary(but the coordinate with indices $n$ and $nm$), this feature creates a gap between $n$ and $nm$ boundary lines that signals supersymmetry breaking. Figure~\ref{fig:8}(c) demonstrates that when selecting lines with identical orders(The first term in \eqref{sjt2}), a unified super-reparametrization becomes possible, preserving supersymmetry. Figure~\ref{fig:8}(d) extends the gap properties that we have discussed in Figure~\ref{fig:8}(b) to an arbitrary order $nm...l$(the last term of \eqref{sjt2}). 

Additionally, if we have a signal from the left side surface $nm$ in Figure~\ref{fig:8}(b), it would reach both $n$ surface and $nm$ surface on the right side. However, if we have a signal from the left-hand side surface $n$, it would reach the $n$ surface on the right-hand side, its effects would vanish on the $nm$ surface due to vanishing reparameterization (shown as the blue line in Figure~\ref{fig:8}(b)). There exists the similar properties in the Figure~\ref{fig:8}(d)(The Green line)). We will further discuss this problem in this Section later and in Appendix~\ref{ap2}. We have also offered an additional discussion on the replicated geometry of supersymmetric $AdS_2$ in Appendix~\ref{ap3}. 

\begin{figure}[!t]
\begin{minipage}{0.48\linewidth}
\centerline{\includegraphics[width=8cm]{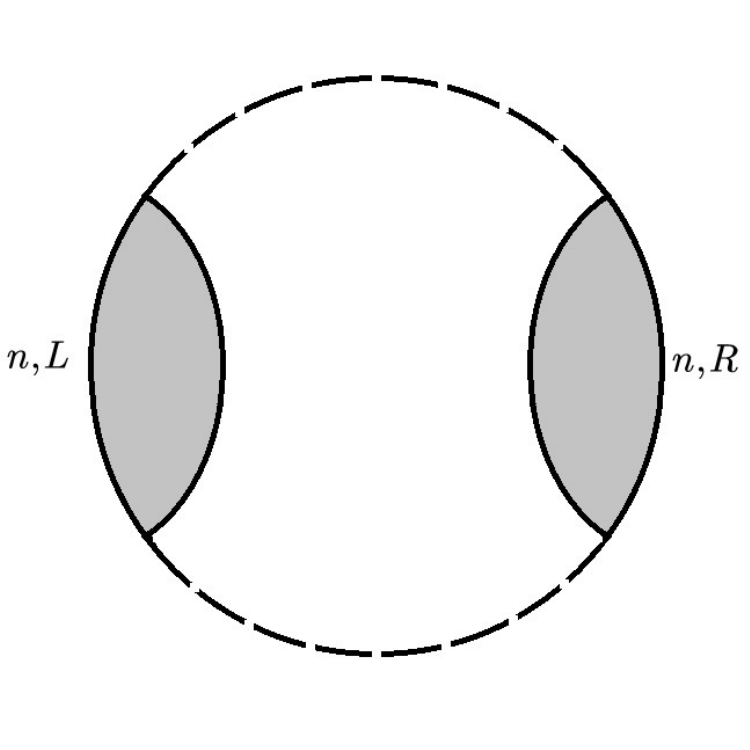}}
\centerline{(a)}
\end{minipage}
\hfill
\begin{minipage}{0.48\linewidth}
\centerline{\includegraphics[width=8cm]{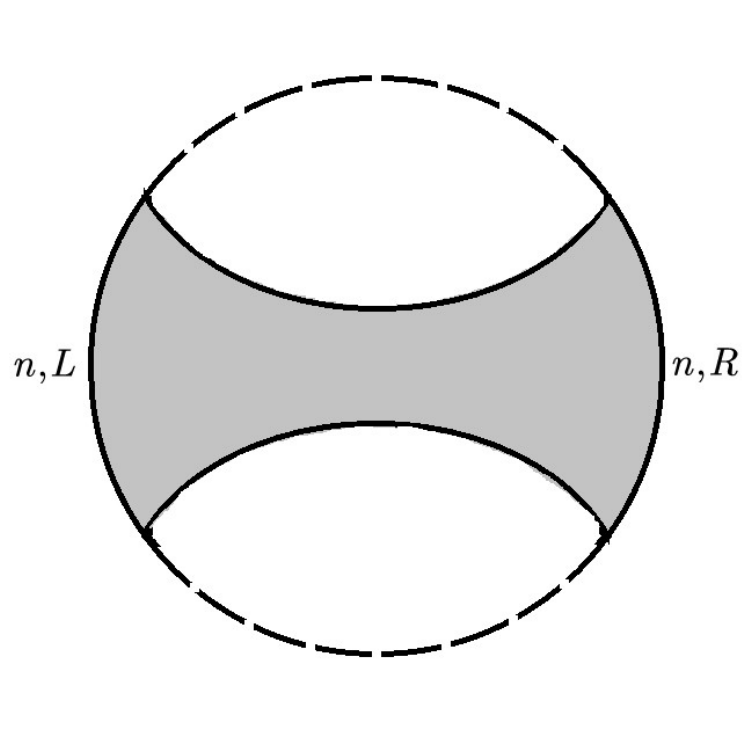}}
\centerline{(b)}
\end{minipage}
\hfill
\begin{minipage}{0.48\linewidth}
\centerline{\includegraphics[width=8cm]{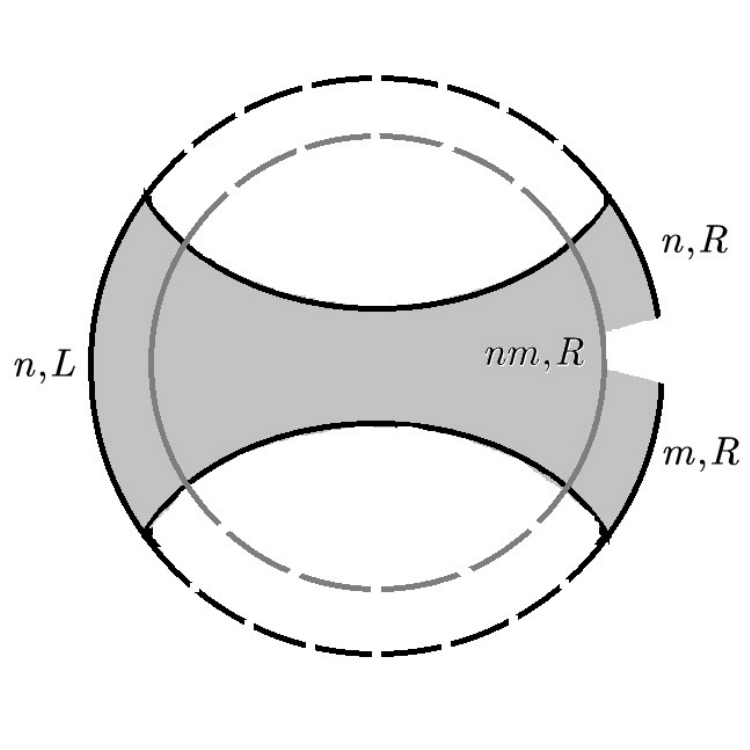}}
\centerline{(c)}
\end{minipage}
\hfill
\begin{minipage}{0.48\linewidth}
\centerline{\includegraphics[width=8cm]{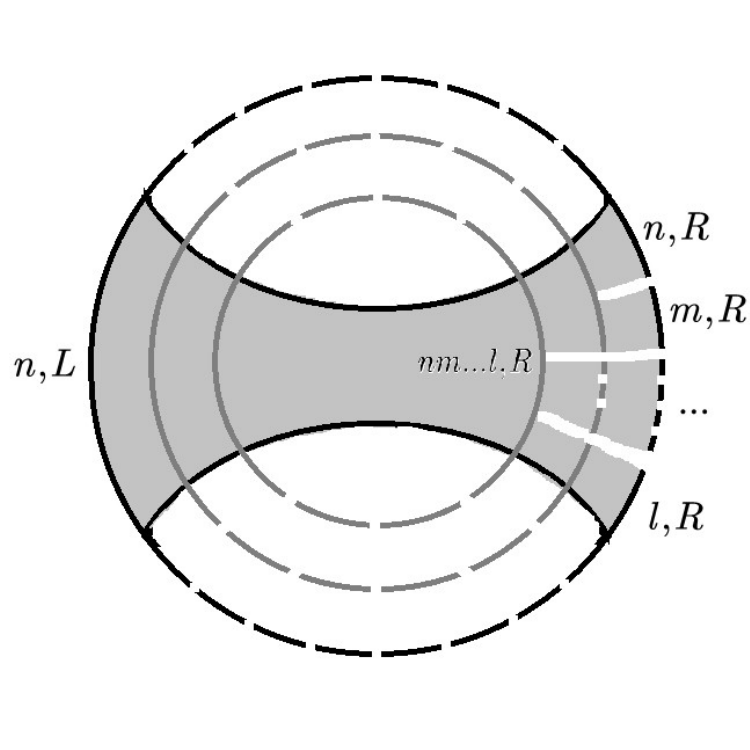}}
\centerline{(d)}
\end{minipage}
\caption{\label{fig:9} The action with replicas is represented via circle diagrams
(a) Disconnected solution for 2 replicas,
(b) Connected wormhole solution for 2 replicas,
(c) Second-order extension produces a 3-body connected wormhole solution, with a ring gap between circles $nm$ and $n$ indicating supersymmetry breaking,
(d) $l$-th order extension yields an $(l+1)$-body connected solution.}
\end{figure}

We provide detailed interpretation in Figure~\ref{fig:9}. The circle diagrams are discrete replicas, with the action's path integral corresponding to summation over gray area boundaries. Figures~\ref{fig:9}(a)-(b) show connected/disconnected diagrams representing Hawking and wormhole saddles. Figure~\ref{fig:9}(c) adds another circle describing both the $nm$-indexed sector and $n$-indexed sectors. The inner circle in Figure~\ref{fig:9}(c) denotes the $nm$ sector, which exactly have the same properties as Figures~\ref{fig:9}(b).The external circle conclude both the $n$ sector and $m$ sector, that have direct mapping onto the $nm$ sector(this is due to the supercharge structures of supersymmetric SYK model). The shadow sector interaction starts at left ($n,L$), passes through the $nm,R$, and ends at right ($n,R$ and $m,R$). However, there is no complete mapping exists between right-side $n,R$, $m,R$ and $nm,R$ due to supersymmetry-breaking regions(i.e. there exists undefined circle between $n,R$ and $m,R$). Figure~\ref{fig:9}(d) includes replica sectors $n$, $m...l$ (right) and reparametrization dilatons (inner circles $n$, $nm...nm...l$), revealing various replica-interacting solutions. Left copies ($b_{L}\psi_{R}$) appear alongside the depicted $\psi_{L}b_{R}$.

In Figure~\ref{fig:9}, the gray area boundaries consist of black lines (summed first-order effective action copies), gray lines (summed arbitrary-order effective action copies), and vertical boundaries (generated from different-order coordinate reparameterizations), which we will analyze further in following subsections.

\subsection{Super conformal SYK action on each site}
\quad The introduction of SJT gravity with indices alone is insufficient to fully capture the properties of the replicated supersymmetric SYK model. To properly describe its holographic dual, we must incorporate additional constraints. The core approach involves extracting the global structure of the replicated theory and imposing corresponding conditions on the SJT gravity side. In this section, we derive three fundamental relationships in the superconformal limit.

A crucial requirement is that the fermions at each site retain the structure of the original $\mathcal{N}=1$ supersymmetric SYK model. This ensures the preservation of the superconformal branch per site, which can be verified via $Z(n)$ symmetry.

\begin{itemize}
    \item The foundational principle requires the $\mathcal{N}=1$ supersymmetric SYK model in this theory to maintain identical properties both site-wise and order-wise, enabling construction of similarly-formed effective actions for each component. This \textit{principle of parallelism} guarantees the parallel structure of the aforementioned Schwarzian actions.
\end{itemize}
Since the effective action of the SYK model reduces to the super-Schwarzian action,
\begin{equation}
S[\tau', \theta';\tau, \theta] = \frac{D^4 \theta'}{D \theta'} - 2 \frac{D^3 \theta' D^2 \theta'}{(D \theta')^2} = S_f(\tau', \theta';\tau, \theta) + \theta S_b(\tau', \theta';\tau, \theta).
\end{equation}

The bosonic piece $S_b$ reduces to the standard Schwarzian derivative under conventional reparameterizations. These actions must be expressed in replicated coordinates. For instance, following our previous analysis, the first-order terms
\begin{equation}
S_A=-\int{d\tau d\theta}N\alpha _{S}^{(n)}S\left[ \tanh \left( \frac{h_{A}^{\left( n \right)}}{2} \right) ,\theta _{A}^{\left( n \right)\prime};\tanh \left( \frac{\tau _{A}^{\left( n \right)}}{2} \right) ,\theta _{A}^{\left( n \right)} \right] .
\end{equation}
These terms maintain identical properties and preserve invariance under coordinate transitions between labeled copies. The derived higher-order Schwarzian action simultaneously transforms superspace coordinates and introduces a unique effective correlation parameter, as shown in \eqref{tot}.
\begin{equation}
S_A=-\int{d\tau d\theta}N\alpha _{S}^{(nm)}S\left[ \tanh \left( \frac{h_{A}^{\left( nm \right)}}{2} \right) ,\theta _{A}^{\left( nm \right)'};\tanh \left( \frac{\tau _{A}^{\left( nm \right)}}{2} \right) ,\theta _{A}^{\left( nm \right)} \right] ,
\end{equation}
and
\begin{equation}
S_A=-\int{d\tau d\theta}N\alpha _{S}^{(nm...l)}S\left[ \tanh \left( \frac{h_{A}^{\left( nm...l \right)}}{2} \right) ,\theta _{A}^{\left( nm...l \right)\prime};\tanh \left( \frac{\tau _{A}^{\left( nm...l \right)}}{2} \right) ,\theta _{A}^{\left( nm...l \right)} \right] .
\end{equation}
Our analysis reveals that considering only the parallel direction, the $Z(n)$ symmetry extends to $Z(n) \times Z(C_{n}^{2}) \times \cdots \times Z(C_{n}^{n-1})$ symmetry.

\subsection{Super-Schwarzian action flow}

\quad In the previous subsection, we obtain the total action by summing over the indices of the single parts. However, these integrals are defined with different coordinates, and the relationship between these coordinates remains unknown. In this subsection, we seek to use a unified the coordinates(For example, $\tau ^{\left( n \right)}$ and $\theta ^{\left( n \right)}$) on each ordered surfaces.

Recall that finite reparameterizations obey the chain rule
\begin{equation}
S[\tau'', \theta'';\tau, \theta] = \left(D \theta' \right)^3 S[\tau'', \theta'';\tau', \theta'] + S[\tau', \theta';\tau, \theta]. \label{FLO}
\end{equation}
Therefore, the Schwarzian actions of SYK models can be unified through coordinate reparameterization into a single Schwarzian action a transformed variable and fixed coordinate system 
\begin{equation}
\begin{split}
I^{\left( n \right)}&=-\sum_n{\alpha _{S}^{\left( n \right)}}\int_{\partial \widetilde{M_n}}{d\tau d\theta}S\left[ \tau ^{\left( n \right)\prime},\theta ^{\left( n \right)\prime};\tau ,\theta \right] -...
\\
&-\sum_{nm...l}{\alpha _{S}^{\left( nm...l \right)}}\int_{\partial \widetilde{M_{nm...l}}}{d\tau d\theta}S\left[ \tau ^{\left( nm...l \right)\prime},\theta ^{\left( nm...l \right)\prime};\tau ,\theta \right] 
\\
&=-\sum_n{\sum_{nm...l}{\left( \alpha _{S}^{\left( n \right) \prime}+\alpha _{S}^{\left( nm \right) \prime}+...+\alpha _{S}^{\left( nm...l \right) \prime} \right)}}\int_{\partial \widetilde{M_n}}{d\tau d\theta}S\left[ \tau ^{\left( n \right)\prime}\theta ^{\left( n \right)\prime};\tau ,\theta \right] 
\\
&+\left( D_{\theta}\theta '' \right) ^3S\left[ \tau ^{\left( n \right)\prime},\theta ^{\left( n \right)\prime};\tau ^{\left( nm...l \right)\prime},\theta ^{\left( nm...l \right)\prime} \right] +...\label{pr2}
\end{split}
\end{equation}
where $\alpha_{S}$ represents the result of the higher-order correlation function, with prime notation absorbing permutation constants 
\begin{equation}
\alpha _{S}^{\left( n \right) '}=n\alpha _{S}^{\left( n \right)},
\
\alpha _{S}^{\left( nm \right) '}=C_{n}^{2}\alpha _{S}^{\left( nm \right)}.
\end{equation}

In this case, we have summarized the second property is that the effective super-Schwarzian actions should obey transformation laws. 
\begin{itemize}
    \item This means knowledge of a single parameter in one action component (expressed in original coordinates) determines its form in other coordinate systems. We call this the \textit{principle of perpendicularity}, guaranteeing the super-Schwarzian action transforms by maintaining its original mathematical form with new coordinates while incorporating required external terms from the transformation.
\end{itemize}

However, the 'principle of parallelism' and 'principle of perpendicularity' would not be enough to fully describe the fractal structure quantitatively. We need to further determine the relationship between the ordered surfaces.

The third property governs group structure. The fractal relationship restricts perpendicular displacements non-arbitrarily, physically constraining coordinate transitions. Additional symmetries beyond the naive $OSp(1|2) \times Z(n)$ group will be analyzed.

\begin{itemize}
    \item Since we need to determine the relationship between the coordinates of ordered surfaces, we can designate this as the \textit{principle of finite displacement}. In this framework, super-Schwarzian action coordinate transitions in \eqref{pr2} are constrained to finite surfaces to preserve both the superconformal symmetries and replica structure. The equivalent statement holds: higher-order surface actions influence lower-order surfaces, but not conversely(as hown in \eqref{INV1}).
\end{itemize}

This issue is generated from the group multiplication of $OSp(1|2)$ with replicas. Physically, if we have a perturbation (which will not break the superconformal boundaries) on the first-order coordinate with index $n$ or $m$, the mathematical form of this effect should be restricted, and the signal that ejects from the first-order boundary would be fully perpendicular to the higher-order boundary $nm$. If we have such a signal ejecting from $n$ states, it would not change any state on $nm$. However, the perturbation on higher-order state $nm$ can influence the states on both $n$ and $m$.

Another equivalent expression is that the coordinates of each ordered boundaries can not be given casually, but to restricted to satisfy the rules we have discussed in the previous paragraph(in another word, the displacement of coordinates between the surfaces can not be given casually). This is due to the anticommutation of supersymmetric SYK model. A detailed proof with group theory analysis is available in Appendix~\ref{ap2}.

\subsection{Supersymmetric JT gravity on single surface}

\quad In this subsection, we will give the constraint on SJT gravity base on the first principle that we have mentioned in the previous subsection. The supervielbeins admit multiple solutions in $\mathcal{N}=1$ supersymmetric JT gravity, yet the constraints produce effective Schwarzian actions. The three laws governing the fractal and off-diagonal coupled $\mathcal{N}=1$ supersymmetric SYK model (established previously) enable investigation of additional constraints on $\mathcal{N}=1$ supersymmetric JT gravity.

First, each site and surface's effective action must replicate the single-site $\mathcal{N}=1$ supersymmetric JT gravity behavior. Since the previous section presented qualitative analysis of supersymmetric JT gravity, we now complement this with quantitative analysis, starting from the three conformal limit principles established earlier. The primary feature requires all supersymmetric SYK model copies to maintain identical properties. Consequently, supersymmetric JT gravity exhibits transformation invariance across all replica regions and fractal surfaces. Additional indices can label gravitational action regions. The action with replica indices takes the form:
\begin{equation}
\begin{split}
S&=-\frac{1}{16\pi G}\sum_n{\left[ i\int{d^2z^{\left( n \right)}d^2\theta ^{\left( n \right)}}E^{\left( n \right)}\varPhi ^{\left( n \right)}\left( R_{+-}^{\left( n \right)}-2 \right) +2\int_{\partial M^{\left( n \right)}}{du^{\left( n \right)}d\theta ^{\left( n \right)}}\varPhi ^{\left( n \right)}K^{\left( n \right)} \right]}
\\
&-\frac{1}{16\pi G}\sum_{nm}{\left[ i\int{d^2z^{\left( nm \right)}d^2\theta ^{\left( nm \right)}}E^{\left( nm \right)}\varPhi ^{\left( nm \right)}\left( R_{+-}^{\left( nm \right)}-2 \right) +2\int_{\partial M^{\left( nm \right)}}{du^{\left( nm \right)}d\theta ^{\left( nm \right)}}\varPhi ^{\left( nm \right)}K^{\left( nm \right)} \right]}
\\
&
-...-\frac{1}{16\pi G}\sum_{nm...l}{\left[ i\int{d^2z^{\left( nm..l \right)}d^2\theta ^{\left( nm..l \right)}}E^{\left( nm..l \right)}\varPhi ^{\left( nm..l \right)}\left( R_{+-}^{\left( nm...l \right)}-2 \right) \right.}
\\
&\left. +2\int_{\partial M^{\left( nm...l \right)}}{du^{\left( nm..l \right)}d\theta ^{\left( nm..l \right)}}\varPhi ^{\left( nm..l \right)}K^{\left( nm..l \right)} \right].\nonumber \label{sjt2}
\end{split}
\end{equation}
In summary, we have written down the total action by summing over the action of single SJT gravity for each index and order. In this case, since we have changed nothing but labeled the coordinates with indices, each indexed action component must retain the original model's properties with identical coordinates.

\subsection{Super reparameterization relationship and restriction at the finite region}

\quad To match gravitational and SYK effective actions, several additional constraints must be imposed on supersymmetric JT gravity. The super-reparameterization between $\mathcal{N}\!=\!1$ supersymmetric JT gravity requires an extra constraint derived from both global and low-energy superconformal symmetries, based on the principles in the previous subsection(as shown in \eqref{tot}, \eqref{pr2} and \eqref{INV1}).

First, the total action ansatz sums over indices and ordered surfaces:

\begin{align}
S&=-\frac{1}{16\pi G}\sum_n{\sum_{nm}{\sum_{nm...l}{\left[ \left[ i\int{d^2zd^2\theta}\left( E^{\left( n \right)}\varPhi ^{\left( n \right)}+E^{\left( nm \right)}\varPhi ^{\left( nm \right)}+...E^{\left( nm...l \right)}\varPhi ^{\left( nm...l \right)} \right) \left( R_{+-}-2 \right) \right] \right.}}}\nonumber
\\
&
+2\int_{\partial M^{\left( n \right)}}{du^{\left( n \right)}d\theta ^{\left( n \right)}}\varPhi ^{\left( n \right)}K^{\left( n \right)}+2\int_{\partial M^{\left( nm \right)}}{du^{\left( nm \right)}d\theta ^{\left( nm \right)}}\varPhi ^{\left( nm \right)}K^{\left( nm \right)}\nonumber
\\
&\left. +...+2\int_{\partial M^{\left( nm...l \right)}}{du^{\left( nm..l \right)}d\theta ^{\left( nm..l \right)}}\varPhi ^{\left( nm..l \right)}K^{\left( nm..l \right)} \right] ,
\end{align}
where the super-vielbein component $E$ inherently incorporates permutation constants. According to the second principle, the surface-restricted action on $M$ becomes:
\begin{equation}
\begin{split}
S=-\frac{1}{16\pi G}\sum_n{\sum_{nm}{\sum_{nm...l}{\left[ \left[ i\int{d^2zd^2\theta}\left( E^{\left( n \right)}\varPhi ^{\left( n \right)}+E^{\left( nm \right)}\varPhi ^{\left( nm \right)}+...E^{\left( nm...l \right)}\varPhi ^{\left( nm...l \right)} \right) \left( R_{+-}-2 \right) \right] \right.}}}
\\
\left. +2\int_{\partial M^{\left( n \right)}}{du^{\left( n \right)}d\theta ^{\left( n \right)}}\left( \varPhi ^{\left( n \right)}K^{\left( n \right)}+\left( \varPhi ^{\left( n \right)}+\mathcal{F} ^{\left( nm,n \right)} \right) K^{\left( nm \right)}+...+\left( \varPhi ^{\left( n \right)}+\mathcal{F} ^{\left( nm..l,n \right)} \right) K^{\left( nm..l \right)} \right) \right], \nonumber
\end{split}
\end{equation}
where flow terms $\mathcal{F}^{(nm,n)}$ represent direct paths from $nm$ to $n$ surface. Consequently, all surface coordinate choices require action notation consistency 
\begin{equation}
\int_{\partial M^{\left( nm...l \right)}}{du^{\left( nm..l \right)}d\theta ^{\left( nm..l \right)}}\varPhi ^{\left( nm..l \right)}K^{\left( nm..l \right)}=\int_{\partial M^{\left( n \right)}}{du^{\left( n \right)}d\theta ^{\left( n \right)}}\left( \varPhi ^{\left( n \right)}+\mathcal{F} ^{\left( nm..l,n \right)} \right) K^{\left( nm..l \right)}.\nonumber
\end{equation}
This formula derives from chain reparametrizations in the Schwarzian action
\begin{equation}
\begin{split}
I^{\left( n \right)}&=-\sum_n{\phi _{r}^{(n)}}\int_{\partial \widetilde{M_n}}{d\tau d\theta}S\left[ \tau ^{\left( n \right)\prime},\theta ^{\left( n \right)\prime};\tau ,\theta \right] -...
\\
&-\sum_{nm...l}{\phi _{r}^{(nm...l)}}\int_{\partial \widetilde{M_{nm...l}}}{d\tau d\theta}S\left[ \tau ^{\left( nm...l \right)\prime},\theta ^{\left( nm...l \right)\prime};\tau ,\theta \right] 
\\
&=-\sum_n{\sum_{nm...l}{\left( \phi _{r}^{\prime(n)}+\phi _{r}^{\prime(nm)}+...+\phi _{r}^{\prime(nm...l)} \right)}}\int_{\partial \widetilde{M_n}}{d\tau d\theta}S\left[ \tau ^{\left( n \right)\prime},\theta ^{\left( n \right)\prime};\tau ,\theta \right] 
\\
&+\left( D_{\theta}\theta '' \right) ^3S\left[ \tau ^{\left( n \right)\prime},\theta ^{\left( n \right)\prime};\tau ^{\left( nm...l \right)\prime},\theta ^{\left( nm...l \right)\prime} \right] +...\label{gef}
\end{split}
\end{equation}
where the derivative $\left( D_{\theta}\theta '' \right) ^3S\left[ \tau ^{(n)\prime},\theta ^{(n)\prime};\tau ^{(nm\cdots l)\prime},\theta ^{(nm\cdots l\prime)} \right] $ corresponds to the flow term $\mathcal{F}^{(nm\cdots l,n)}K^{(nm\cdots l)}$ in our ansatz, which is corresponding to \eqref{FLO}. Here we have unified the coordinates of the total dilaton action at the cost of introducing the flow term (the last term in \eqref{gef}).

The general coefficients transform under reparameterization 
\begin{equation}
\phi _{r}'^{(n)}=n\phi _{r}^{(n)}, \phi _{r}'^{(nm)}=C_{n}^{2}\phi _{r}^{(nm)},
\end{equation}
incorporating exchange coefficients. An additional external curvature constraint ensures each copy's supersymmetric coordinates in the effective action satisfy 
\begin{equation}
Dz^{\left( n \right)}=\theta ^{\left( n \right)}D\theta ^{\left( n \right)},
\end{equation}
Under infinitesimal perturbation, the expression becomes 
\begin{equation}
z^{\left( n \right)}=t^{\left( n \right)}+i\epsilon \left( D\xi ^{\left( n \right)} \right) ^2,
\
Dt^{\left( n \right)}\left( u^{\left( n \right)},\theta ^{\left( n \right)} \right) =\xi \left( u^{\left( n \right)},\theta ^{\left( n \right)} \right) D\xi ^{\left( n \right)}\left( u^{\left( n \right)},\theta ^{\left( n \right)} \right) ,
\end{equation}
yielding the super-Schwarzian action. The effective coordinates on higher-order surfaces can also be expressed as
$
z^{\left( nm...l \right)},\theta ^{\left( nm...l \right)},t^{\left( nm...l \right)},\xi ^{\left( nm...l \right)}
$

The action must incorporate higher-order surface reparameterization contributions (see Appendix~\ref{ap2}). To align gravitational and SYK actions, coordinate choices require an additional constraint ensuring consistency with Appendix~\ref{ap2}'s group multiplication structure. This yields the higher-order term 
\begin{equation}
\begin{split}
&Dz^{\prime\left( nm...l \right)}=\theta ^{\prime\left( nm...l \right)}D\theta e^{\prime\left( nm...l \right)},
\\
&z^{\prime\left( nm...l \right)}=t^{\prime\left( nm...l \right)}+i\epsilon \left( D\xi^{\prime\left( nm...l \right)} \right) ^2,
\\
&Dt^{\prime\left( nm...l \right)}\left( u^{\prime\left( nm...l \right)},\theta ^{\prime\left( nm...l \right)} \right) =\xi '\left( u^{\prime\left( nm...l \right)},\theta ^{\prime\left( nm...l \right)} \right) D\xi ^{\prime\left( nm...l \right)}\left( u^{\prime\left( nm...l \right)},\theta ^{\prime\left( nm...l \right)} \right) ,
\end{split}
\end{equation}
and the lower ordered terms
\begin{equation}
\begin{split}
&Dz'^{\left( n \right)}=\theta '^{\left( n \right)}D\theta '^{\left( n \right)},
\\
&z'^{\left( n \right)}=t'^{\left( n \right)}+i\epsilon \left( D\xi '^{\left( n \right)} \right) ^2,
\\
&Dt'^{\left( n \right)}\left( u'^{\left( n \right)},\theta '^{\left( n \right)} \right) =\xi \left( u'^{\left( n \right)},\theta '^{\left( n \right)} \right) D\xi '^{\left( n \right)}\left( u'^{\left( n \right)},\theta '^{\left( n \right)} \right),
\end{split}
\end{equation}
where the reparameterization coordinates 
$$
z'^{\left( n \right)},\theta '^{\left( n \right)},t'^{\left( n \right)},\xi '^{\left( n \right)},
$$
which should conclude the contributions of
$$
z^{\left( n \right)},\theta ^{\left( n \right)},t^{\left( n \right)},\xi ^{\left( n \right)},
$$
and
$$
z^{\left( nm...l \right)},\theta ^{\left( nm...l \right)},t^{\left( nm...l \right)},\xi ^{\left( nm...l \right)},
$$
which summing over remaining terms with indices $n$. This shows the $n$-reparameterization naturally incorporates higher-order contributions while maintaining consistency. 

Now we can provide a physical interpretation of the signals shown in Figure~\ref{fig:10}. Suppose there is a perturbation—such as matter fields that slightly affect the metric—on the lower-order boundaries (associated with the SJTs in the super-$NAdS_2$ spacetime) on the left-hand side. This perturbation does not manifest explicitly in the higher-order coordinate frame on the right-hand side. In other words, a gravitational response originating from the lower-order boundaries must satisfy certain consistency constraints, and cannot produce physical effects in the higher-order coordinate system. For instance, the influence of matter fields is suppressed under the higher-order coordinate constraint given by Equation~\eqref{INV1}. This phenomenon is purely quantum mechanical in nature and has no analogue in classical gravitational theory.

\section{Finite element solution and replicated thermodynamic}\label{S6}

\quad This section analyzes the internal supercharge structure with the finite element method(Here we use $N$ to represent the number of fermions). The crucial difference between ``modular thermodynamics`` test replicas and our model emerges from multiple supercharges coupled via random couplings.

The density matrix definition incorporates Majorana fermions and supercharges as 
\begin{equation}
Z\left( \beta \right) =Tr\left[ \exp \left( -\beta H \right) \right], 
\end{equation}
where
\begin{equation}
\beta \leftrightarrow n,
\
H=-\log \left( \rho \right). \label{modu}
\end{equation}
Since we can combine the inverse temperature $\beta$ and the replica parameter $n$, the $H$ is defined as the modular Hamiltonian\cite{1606.08443}. The finite-$N$ Hamiltonian is constructed from Majorana fermions $\psi _{n}^{i}$.
\begin{equation}
\psi _{n}^{i}=c_{n}^{i\dagger}+c_{n}^{i},
\end{equation}
where we have used the Pauli matrix $c_{n}^{i\dagger}$ and $c_{n}^{i}$ for numerical calculations.

Under the condition \eqref{modu}, we can also define the corresponding free energy and total energy in modular thermodynamics
\begin{equation}
F=-\frac{1}{n}\log \left( Tr\left[ \rho ^n \right] \right) ,
\
E=-\frac{\partial}{\partial n}\log \left( Tr\left[ \rho ^n \right] \right) .
\end{equation}
We can also define the modular entropy as in thermodynamics.
\begin{equation}
S_{\rm{thermal}}=n \left( E-F \right) .
\end{equation}
It depend on the parameter $n$ rather than $\beta$.

\subsection{Modular thermodynamic}

\quad This subsection studies random matrix ensembles with replicated degrees of freedom, fully described within modular thermodynamics. The replica wormhole method calculates only the von Neumann entropy (e.g., radiation entropy) in the $n\to1$ limit. Alternative physical measures exist for replicas, including the improved Rényi entropy (modular entropy)~\cite{1606.08443}.

The modular entropy is given by
\begin{equation}
S_m=\left( 1-n\frac{\partial}{\partial n} \right) \log Z\left( n \right) =\log \left( Tr\left[ \rho ^n \right] \right) +n\left( Tr\left[ \rho ^nH \right] /Tr\left[ \rho ^n \right] \right) .
\end{equation}

\begin{figure}[!t]
\begin{minipage}{0.48\linewidth}
\centerline{\includegraphics[width=8cm]{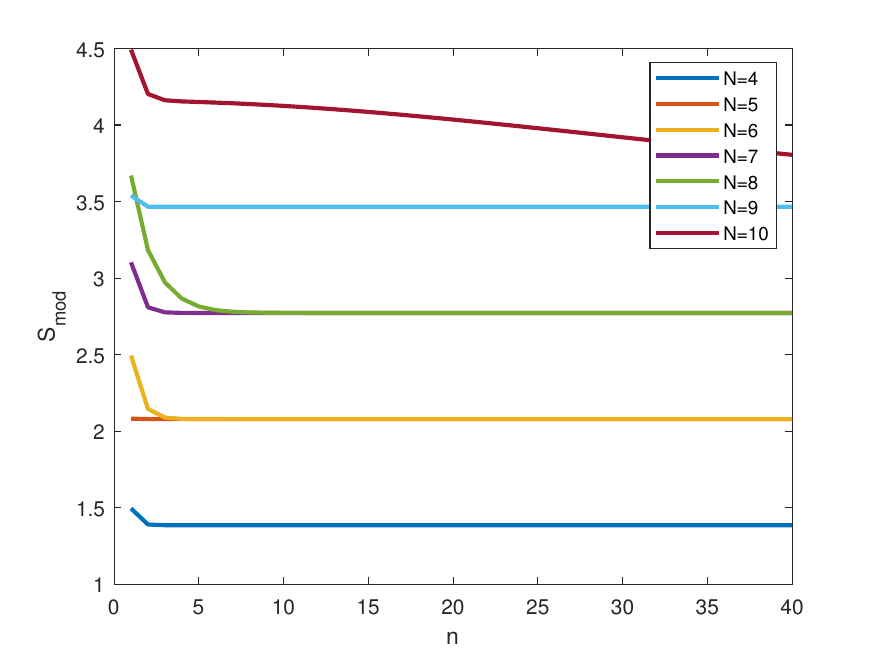}}
\centerline{(a)}
\end{minipage}
\hfill
\begin{minipage}{0.48\linewidth}
\centerline{\includegraphics[width=8cm]{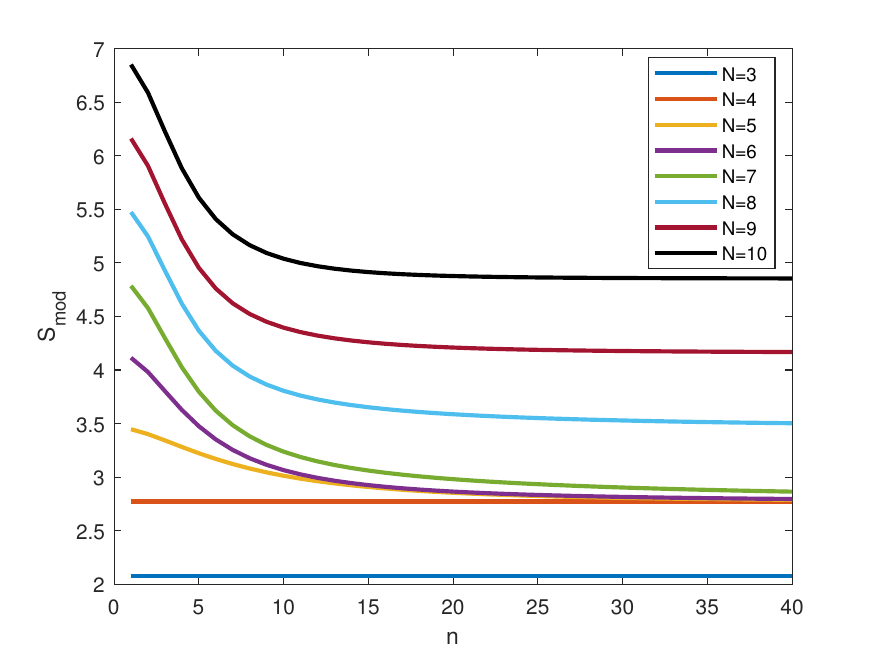}}
\centerline{(b)}
\end{minipage}
\caption{\label{fig:10} Modular entropy versus replica number:(a) Non-supersymmetric model(b) Supersymmetric model}
\end{figure}

Figure~\ref{fig:10} shows the modular entropy versus replica number for both non-supersymmetric and $\mathcal{N}=1$ supersymmetric models. The modular entropy carries more physical significance than Rényi entropy~\cite{XiDong}. As replica number increases, the modular entropy decreases and converges to a constant. Additional fermions increase the modular entropy value. The key difference between Figure~\ref{fig:10}(a) (non-supersymmetric) and Figure~\ref{fig:10}(b) ($\mathcal{N}=1$ supersymmetric) appears in convergence behavior: adjacent integers in the non-supersymmetric model converge to identical constants, while the supersymmetric case shows different convergence. This stems from the supersymmetric Hamiltonian being the square of supercharges, restricting the system to even fermion numbers. The interaction terms require separate consideration. Physically, since the parameter $n$ combine with thermal temperature, the behavior of modular entropy shows that the ground states converge to a certain number as $n$ is large enough(which is correspending to the low-temperature limit).

\subsection{$n$-dependent relative entropy}

\quad This subsection examines the distinction between single- and multiple-supercharge purity by analyzing the relative entropy between differently ordered replicated supercharges and Hamiltonians. The relative entropy across multiple replica components reveals an inverse process relation. Physically, the strong subadditivity—arising from the monotonicity of relative entropy \cite{LR}—captures the inequality between a single subsystem and the full replicated system. Moreover, the non-vanishing relative entropy implies non-zero mutual information, which in turn supports the existence of connected entanglement wedges in the holographic context. Importantly, the positivity of the relative entropy holds for arbitrary shapes of the end-of-the-world (EOW) branes. Finally, the monotonicity of relative entropy is also closely related to the Bekenstein bound \cite{BB}.

While numerical calculation proves difficult with direct coupling, we incorporate fermionic contributions from each part, enabling full diagonal matrix treatment. This approach ultimately returns to the order-summing problem. Following our established laws, two-body interactions suffice to model the complete system. We present both fully-coupled and fully-decoupled replicated models. Intermediate cases with partial coupling/decoupling exist (their statistical mechanics require future development). The relative entropy is defined as 
\begin{equation}
S\left( \rho \middle| \sigma \right) \equiv Tr\left[ \rho \left( \log \rho -\log \sigma \right) \right] .
\end{equation}
This section analyzes the relative entropy in surface systems composed of multiple replica subsystems 
\begin{equation}
S\left( \rho ^n \middle| \rho ^{(nm)} \right) =Tr\left[ \rho ^n\left( \log \rho ^n-\log \rho ^{(nm)} \right) \right] .
\end{equation}
For modular thermodynamics and subactivity analysis, these relative entropy inequalities serve as crucial quantum information measures, equivalently representing mutual information.

\begin{figure}[!t]
\begin{minipage}{0.48\linewidth}
\centerline{\includegraphics[width=8cm]{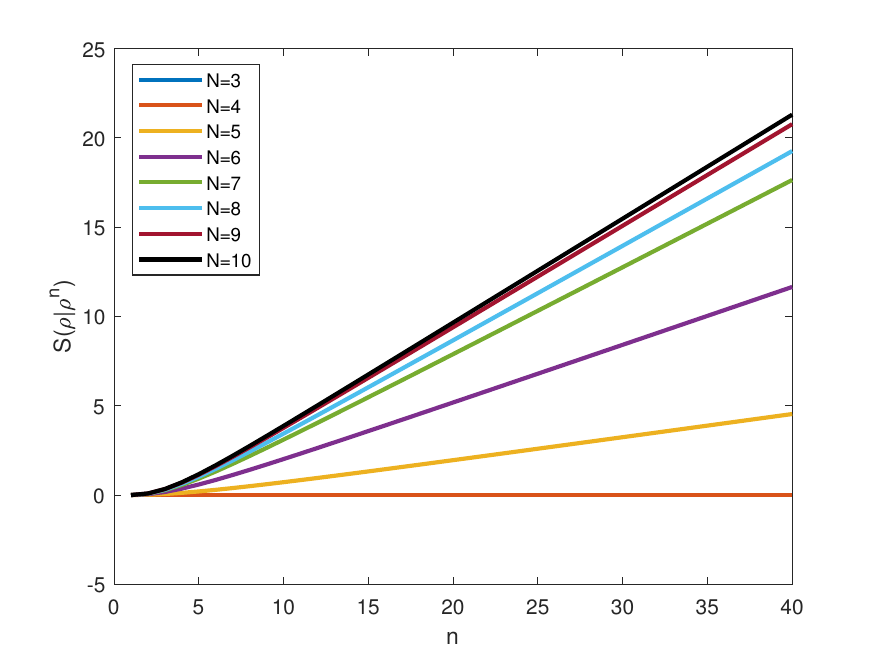}}
\centerline{(a)}
\end{minipage}
\hfill
\begin{minipage}{0.48\linewidth}
\centerline{\includegraphics[width=8cm]{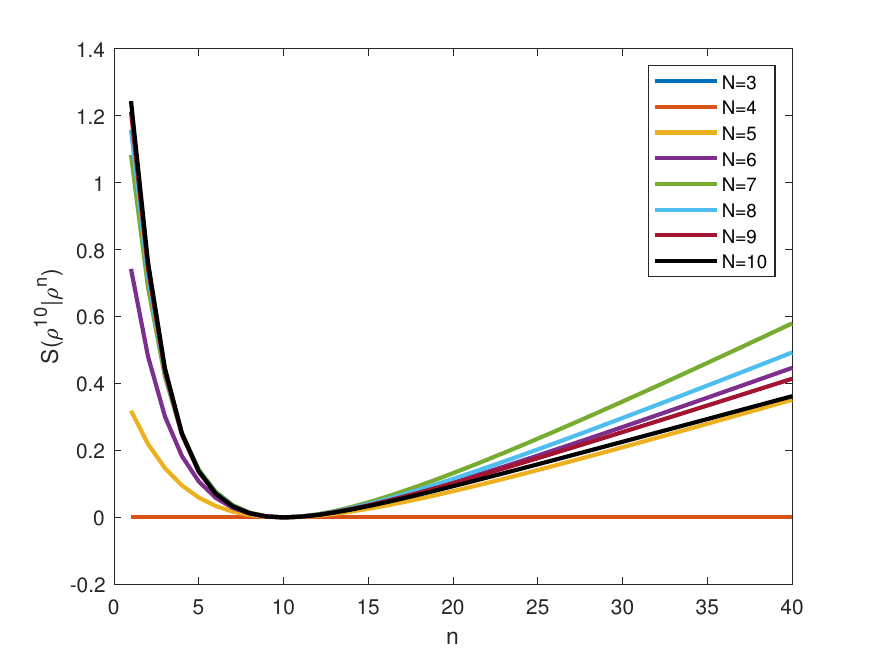}}
\centerline{(b)}
\end{minipage}
\hfill
\begin{minipage}{0.48\linewidth}
\centerline{\includegraphics[width=8cm]{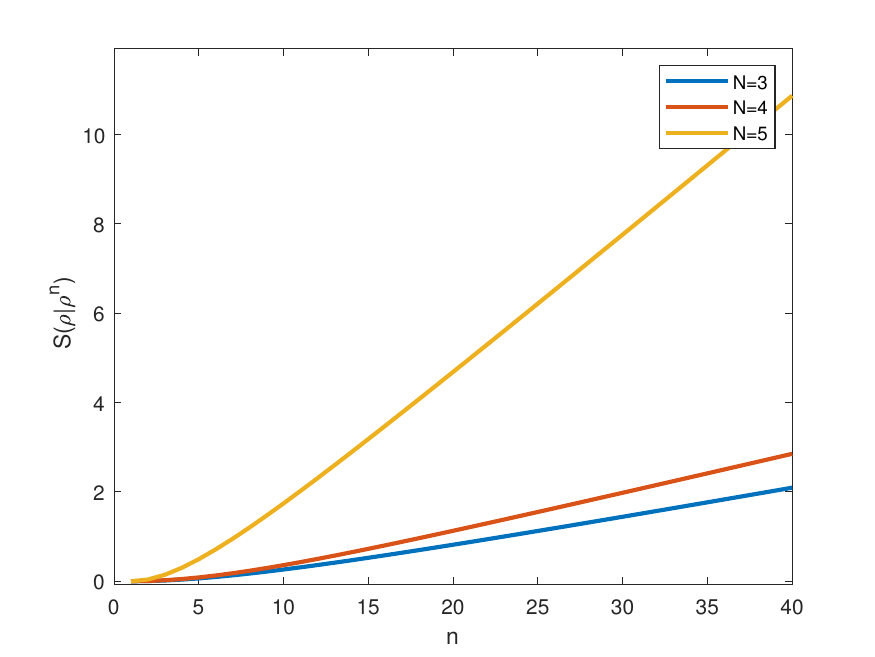}}
\centerline{(c)}
\end{minipage}
\hfill
\begin{minipage}{0.48\linewidth}
\centerline{\includegraphics[width=8cm]{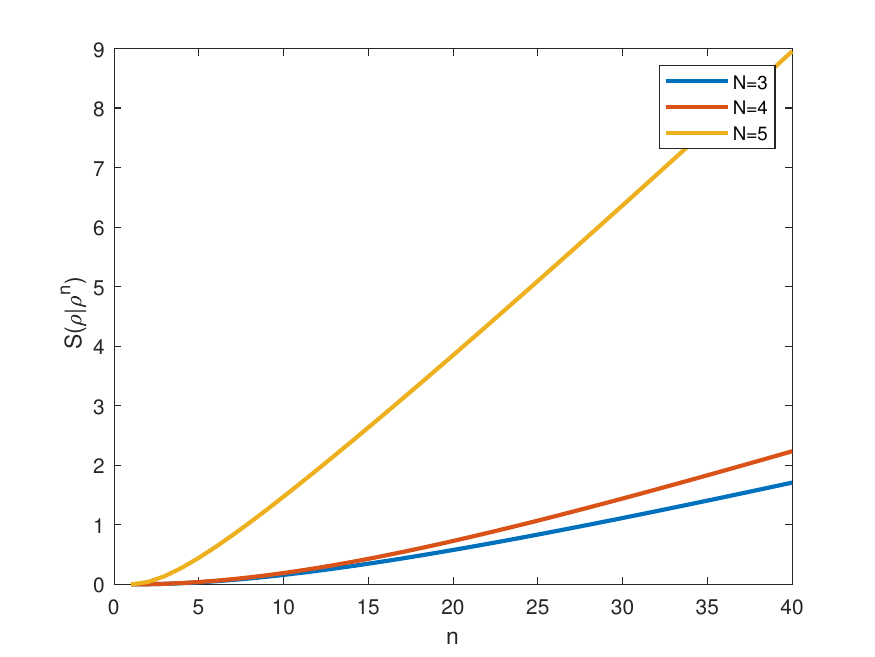}}
\centerline{(d)}
\end{minipage}

\caption{\label{fig:11} Relative entropy as a function of the replica parameter. (a) single systems and $n$-replica systems(b) 10 single systems and $n$-replica systems  (c) single systems ($\mu^{(n)}=0.1$ coupling) and $n$-replica systems(d) single systems with partial coupling (half with $\mu^{(n)}=0.1$ coupling and $\mu^{(n)}=0$ for anther) and $n$-replica systems}
\end{figure}

In Figure~\ref{fig:11}(a), we plot the relative entropy between a single SSYK system and $n$-replica systems. The relative entropy $S\left( \rho \middle| \rho ^n \right) $ increases with $n$, while it has an increasing slope. In Figure~\ref{fig:11}(b), we plot the relative entropy $S\left( \rho^{10} \middle| \rho ^n \right) $ between between 10 single systems and $n$-replica systems, numerically prove the relative entropy is always positive, consistent with theoretical arguments from previous work~\cite{1606.08443}. Figure~\ref{fig:11}(b) provides an analytical extension of the relative entropy when the number of copies is non-integer($n\ne 10,20,30...$) or smaller than a single copy($n<10$).

The fermion number must exceed 4($N>4$); otherwise, the relative entropy, entanglement capacity, and modular entropy become indistinguishable with $N$. We also consider cases with secondary-order coupling and half-model secondary-order coupling. The coupled solutions in Figure~\ref{fig:11}(c) and Figure~\ref{fig:11}(d) for $N=3,N=4,N=5$ correspond to decoupled $N=6,N=8,N=10$ systems. For $n=1$ entanglement entropy, these inequalities and irreversibility (as relative entropy) correspond to holographic properties.

We can also give a provement about the n-dependent second law of the relative entropy. Since we can define the modular free energy
\begin{equation}
F\left( \rho \right) =Tr\left( \rho H \right) -S_{vN}\left( \rho \right) /n
\end{equation}
to further represent the relative entropy
\begin{equation}
S\left( \rho \middle| \sigma \right) =n\left( F\left( \rho \right) -F\left( \sigma \right) \right). 
\end{equation}
As shown in Figure~\ref{fig:11}, the slope of the relative entropy increases with the parameter $n$. Since we can define the certain fix point on $n=1$, we have 
\begin{equation}
F\left( \rho ^n \right) -F\left( \rho \right) \leqslant 0,
\end{equation}
which means
\begin{equation}
\left( \left< H \right> _n-S_n \right) -\left( \left< H \right> _1-S_1 \right) \leqslant 0.
\end{equation}
we can further obtain the second law of relative entropy
\begin{equation}
\left( S_n-S_1 \right) -n\left( \left< H \right> _n-\left< H \right> _1 \right) \geqslant 0.
\end{equation}

\subsection{Capacity of entanglement}

\quad In modular thermodynamics, the entanglement capacity corresponds to heat capacity, showing Hamiltonian dependence on replica number. The entanglement capacity derives from the first derivative of modular entropy with respect to the replica parameter $n$.

\begin{equation}
\begin{split}
C=n^2\frac{\partial ^2}{\partial n^2}\log Z\left( n \right) =n^2\left( \left< H^2 \right> _n-\left< H \right> _{n}^{2} \right) 
\\
=n^2\left( Tr\left[ \rho ^nH^2 \right] -Tr\left[ \rho ^nH \right] ^2 \right) /Tr\left[ \rho ^n \right].
\end{split}
\end{equation}

\begin{figure}[!t]
\begin{minipage}{0.22\linewidth}
\centerline{\includegraphics[width=6cm]{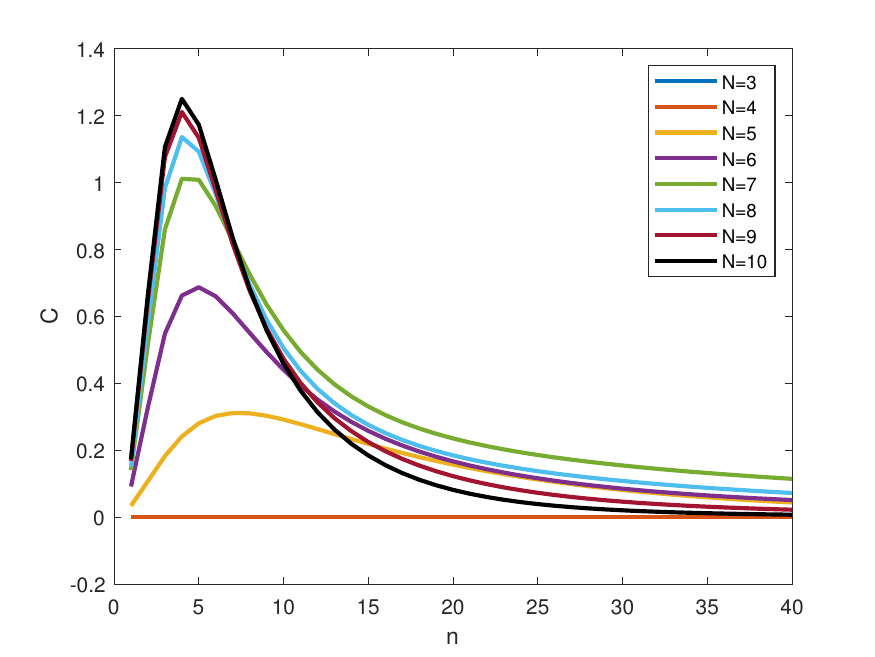}}
\centerline{(a)}
\end{minipage}
\hfill
\begin{minipage}{0.22\linewidth}
\centerline{\includegraphics[width=6cm]{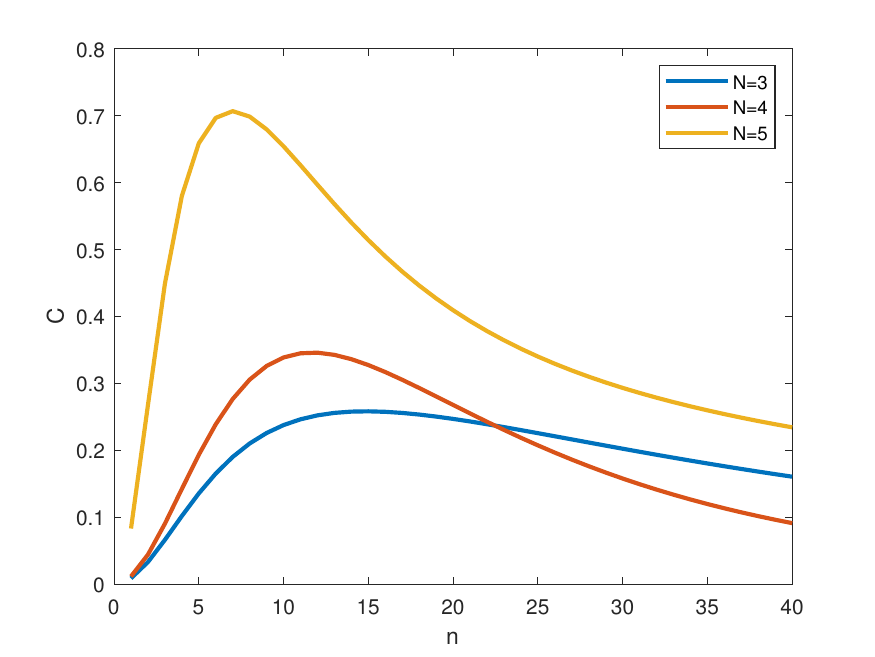}}
\centerline{(b)}
\end{minipage}
\hfill
\begin{minipage}{0.22\linewidth}
\centerline{\includegraphics[width=6cm]{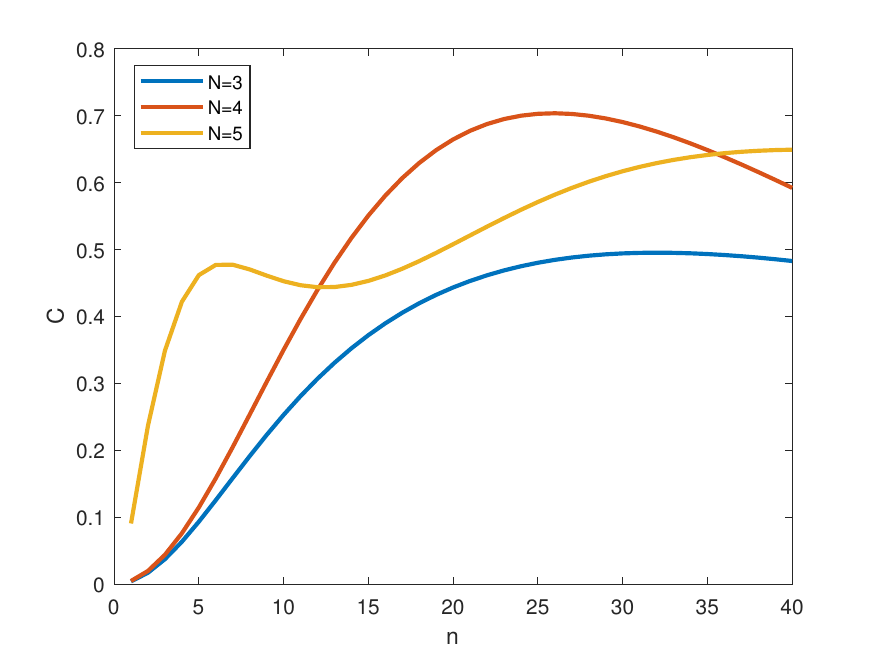}}
\centerline{(c)}
\end{minipage}

\caption{\label{fig:12} (a) Capacity of entanglement for the supersymmetric SYK model versus replica number, (b) Capacity of entanglement with coupling $\mu^{(n)}=0.1$,  (c) Capacity of entanglement with coupling $\mu^{(n)}=0.1$ and partial coupling ($\mu^{(n)}=0$ for half of systems). }
\end{figure}

As shown in Fig.~\ref{fig:12}, we present the entanglement capacity of the supersymmetric SYK model as a function of the replica parameter $n$. Three distinct configurations are considered: 
\begin{enumerate}
    \item[(i)] Uncoupled models (independent couplings)
    \item[(ii)] Fully coupled models 
    \item[(iii)] Hybrid models with half pair-coupled and half uncoupled degrees of freedom
\end{enumerate}
In Fig.~\ref{fig:12}(a), the entanglement capacity exhibits a maximum at $n=5$. 
Fig.~\ref{fig:12} (b) shows that introducing pairwise couplings shifts this maximum toward smaller $n$ values. 
For the hybrid configuration in Fig.~\ref{fig:12}(c), the splitting of the peak emerges at $n=5$, originating from the distinct contributions of the uncoupled and coupled subsystems.

Additionally, the capacity of  entanglement approaches zero in the limit $n \rightarrow 0$ and $n \rightarrow \infty$, as does the standard heat capacity for many-body systems in thermodynamics. The introduction of the wormhole in Figure~\ref{fig:12}(d) sticks the copies, and increases the maximum location in dependence on replica copies $n$. In other words, this setting decreases the degrees of freedom.

\section{Conclusion}\label{S7}

\quad In summary, we investigate SYK models that support $n$-th replica wormholes by constructing ordered supersymmetric SYK models with inter-replica couplings. We begin by examining a straightforward extension from the two-copy setup to general $n$-replica configurations. Prior studies typically coupled two SYK systems at specific Euclidean times, forming so-called ``thermometer diagrams.`` Our initial approach reformulates the path integral between coupled replicas using simple permutation group structures, aiming to generalize to arbitrary $n$. However, we find that the solvability of this naive method is limited. In particular, computing the ``half`` Green's functions arising from the interaction Hamiltonian proves to be intractable except under special constraints. To overcome this challenge, we develop a new framework that resolves the replica structure via supersymmetric SYK models.

The $\mathcal{N}=1$ supersymmetric SYK model exhibits a form of fractal symmetry across fermionic sectors, which we exploit to construct ordered fermionic configurations representing the replicas. We introduce supersymmetric fermion-boson interactions across distinct sectors, interpreting these as effective wormhole connections. This construction renders the ordering nontrivial: fermionic subsystems embedded within specific regions become dynamically distinct from their environment, leading to spontaneous supersymmetry breaking.

Subsequently, we analyze the model in the thermal regime, beginning with the effective action and solving the corresponding Schwinger-Dyson equations. The resulting Green's functions encode temperature-dependent quantities such as the free energy and internal energy. The presence of additional ordered couplings manifests dynamically as free-particle-like contributions, effectively resetting initial conditions and inducing deformations in the supersymmetric two-point functions. We examine the resulting free energy landscape and thermal phase structure, finding that the contributions from these ordered couplings closely resemble those observed in previous coupled-SYK constructions. Similar effects are also observed in the energy spectrum.

By analytically continuing to Lorentzian time, we recover both the wormhole-like and black hole-like solutions that appeared in earlier analyses. The derived effective action exhibits a nontrivial time dependence that is directly related to the dynamics of thermodynamic entanglement entropy in SYK-type systems. A crucial aspect of our construction concerns the low-energy behavior of the supersymmetric SYK (SSYK) model and its implications for super-Jackiw–Teitelboim (SJT) gravity. Our analysis demonstrates that each ordered replica sector behaves effectively as an independent $\mathcal{N}=1$ SSYK model with identical dynamical properties. We further characterize the key features of these replicated submodels.

An important component of the analysis involves the superconformal reparameterization symmetry, governed by the $OSp(1|2)$ group, and the structure of inter-replica interactions, which we interpret as relations between group elements. We identify three fundamental constraints for the emergence of a consistent superconformal effective action. These constraints are found to match precisely with the boundary conditions required in the effective action of super JT gravity.

In the final part of this paper, we perform a finite-$N$ analysis using a fully diagonalized numerical approach. We investigate the modular entropy—an improved version of Rényi entropy—for both the non-supersymmetric SYK model and the $\mathcal{N}=1$ supersymmetric version. Our results show that in the large-replica limit, the ground state degeneracies converge to finite values. 
We also numerically compute the relative entropy and the capacity of entanglement within the framework of modular thermodynamics. These computations confirm the second law of relative entropy and reveal a consistent relationship between the standard heat capacity and the effective number of degrees of freedom across replicas.

\begin{acknowledgments}
We would like to thank Jianxin Lu, Sangjin Sin, Cheng Peng and Jieqiang Wu for helpful discussions.  This work was partially supported by NSFC, China (Grant No. 12275166 and No. 12311540141).
\end{acknowledgments}

\appendix
\section{Numerical approach}\label{ap1}
\subsection{Solving the Schwinger-Dyson equation}
\quad In this regime, we can define the supersymmetric Schwinger-Dyson equation with orders. The equations of motion are given by
\begin{equation}
\partial _{\tau}G_{\psi \psi ,AB}\left( \tau ,\tau ' \right) -\sum_C{\left( \int{d\tau ''}\varSigma _{\psi \psi ,AC}\left( \tau ,\tau '' \right) G_{\psi \psi ,AB}\left( \tau ,\tau '' \right) \right)}=\delta _{AB}\delta \left( \tau -\tau ' \right) .
\end{equation}
\begin{equation}
-G_{bb,AB}^{\left( n \right)}\left( \tau ,\tau ' \right) -\sum_C{\left( \int{d\tau ''}\varSigma _{bb,AC}^{\left( n \right)}\left( \tau ,\tau '' \right) G_{bb,AB}^{\left( n \right)}\left( \tau ,\tau '' \right) \right)}=\delta _{AB}\delta \left( \tau -\tau ' \right) .
\end{equation}
\begin{equation}
-G_{bb,AB}^{\left( nm...l \right)}\left( \tau ,\tau ' \right) -\sum_C{\left( \int{d\tau ''}\varSigma _{bb,AC}^{\left( nm...l \right)}\left( \tau ,\tau '' \right) G_{bb,AB}^{\left( nm...l \right)}\left( \tau ,\tau '' \right) \right)}=\delta _{AB}\delta \left( \tau -\tau ' \right) .
\end{equation}
\begin{equation}
\sum_C{\left( -i\mu ^{(n)}\epsilon _{AC}G_{b\psi ,AB}^{\left( n \right)}\left( \tau ,\tau ' \right) -\int{d\tau ''}\varSigma _{b\psi ,AC}^{\left( n \right)}\left( \tau ,\tau '' \right) G_{b\psi ,AB}^{\left( n \right)}\left( \tau ,\tau '' \right) \right)}=\delta _{AB}\delta \left( \tau -\tau ' \right) .
\end{equation}
\begin{eqnarray}
    &\sum_C&{\left( i\mu ^{(n)}\epsilon _{AC}G_{\psi b,AB}^{\left( n \right)}\left( \tau ,\tau ' \right) -\int{d\tau ''}\varSigma _{\psi b,AC}^{\left( n \right)}\left( \tau ,\tau '' \right) G_{\psi b,AB}^{\left( n \right)}\left( \tau ,\tau '' \right) \right)}=\delta _{AB}\delta \left( \tau -\tau ' \right) .\\
&\sum_C &{\left( -i\mu ^{(nm...l)}\epsilon _{AC}G_{b\psi ,AB}^{\left( nm...l \right)}\left( \tau ,\tau ' \right) -\int{d\tau ''}\varSigma _{b\psi ,AC}^{\left( nm...l \right)}\left( \tau ,\tau '' \right) G_{b\psi ,AB}^{\left( nm...l \right)}\left( \tau ,\tau '' \right) \right)}=\delta _{AB}\delta \left( \tau -\tau ' \right) .\nonumber\\
&\sum_C&{\left( i\mu ^{(nm...l)}\epsilon _{AC}G_{\psi b,AB}^{\left( nm...l \right)}\left( \tau ,\tau ' \right) -\int{d\tau ''}\varSigma _{\psi b,AC}^{\left( nm...l \right)}\left( \tau ,\tau '' \right) G_{\psi b,AB}^{\left( nm...l \right)}\left( \tau ,\tau '' \right) \right)}=\delta _{AB}\delta \left( \tau -\tau ' \right) .\nonumber
\end{eqnarray}
We provide exact definitions of the self-energy terms and their variational forms in the matrix equations
\small{
\begin{equation}
\begin{split}
&\varSigma _{\psi \psi ,AB}\left( \tau \right) =\left( q-1 \right) \left( -1 \right) ^{\left( q-1 \right) /2}\mathcal{J}^{(n)}G_{\psi \psi ,AB}^{q-2}\left( \tau \right) G_{bb,AB}^{\left( n \right)}\left( \tau \right) -\frac{\left( q-1 \right)}{2}\left( q-2 \right) \left( -1 \right) ^{\left( q-1 \right) /2}
\\
&\mathcal{J}^{(n)}G_{\psi \psi ,AB}^{\left( n \right) q-3}\left( \tau \right) G_{\psi b,AB}^{\left( n \right) ;A}\left( \tau \right) G_{b\psi ,AB}^{\left( n \right) ;S}\left( \tau \right) -\frac{\left( q-1 \right)}{2}\left( q-2 \right) \left( -1 \right) ^{\left( q-1 \right) /2}\mathcal{J}^{(n)}G_{\psi \psi ,AB}^{q-2}\left( \tau \right) 
\\
&G_{\psi b,AB}^{\left( n \right) ;S}\left( \tau \right) G_{b\psi ,AB}^{\left( n \right) ;A}\left( \tau \right) +...+\left( q-1 \right) \left( -1 \right) ^{\left( q-1 \right) /2}\mathcal{J}^{(nm...l)}G_{\psi \psi ,AB}^{q-2}\left( \tau \right) G_{bb,AB}^{\left( nm...l \right)}\left( \tau \right) 
\\
&-\frac{\left( q-1 \right)}{2}\left( q-2 \right) \left( -1 \right) ^{\left( q-1 \right) /2}\mathcal{J}^{(nm...l)}G_{\psi \psi ,AB}^{\left( nm...l \right) q-3}\left( \tau \right) G_{\psi b,AB}^{\left( nm...l \right) ;A}\left( \tau \right) G_{b\psi ,AB}^{\left( nm...l \right) ;S}\left( \tau \right) 
\\
&-\frac{\left( q-1 \right)}{2}\left( q-2 \right) \left( -1 \right) ^{\left( q-1 \right) /2}\mathcal{J}^{(nm...l)}G_{\psi \psi ,AB}^{q-2}\left( \tau \right) G_{\psi b,AB}^{\left( nm...l \right) ;S}\left( \tau \right) G_{b\psi ,AB}^{\left( nm...l \right) ;A}\left( \tau \right) ,
\end{split}
\end{equation}
The corresponding self-energies are given by
\begin{align}
&\varSigma _{bb,AB}^{\left( n \right)}\left( \tau \right) =\mathcal{J}^{(n)}G_{\psi \psi ,AB}^{q}\left( \tau \right) ,
\\
&\varSigma _{bb,AB}^{\left( nm...l \right)}\left( \tau \right) =\mathcal{J}^{(nm...l)}G_{\psi \psi ,AB}^{q}\left( \tau \right) ,
\\
&\varSigma _{b\psi ,AB}^{\left( n \right) ;S}\left( \tau \right) =-\frac{\left( q-1 \right)}{2}\mathcal{J}^{(n)}G_{\psi b,AB}^{\left( n \right) ;A}\left( \tau \right) G_{\psi \psi ,AB}^{q-2}\left( \tau \right) ,
\\
&\varSigma _{b\psi ,AB}^{\left( nm...l \right) ;S}\left( \tau \right) =-\frac{\left( q-1 \right)}{2}\mathcal{J}^{(nm...l)}G_{\psi b,AB}^{\left( nm...l \right) ;A}\left( \tau \right) G_{\psi \psi ,AB}^{q-2}\left( \tau \right) ,
\\
&\varSigma _{b\psi ,AB}^{\left( nm...l \right) ;S}\left( \tau \right) =-\frac{\left( q-1 \right)}{2}\mathcal{J}^{(nm...l)}G_{\psi b,AB}^{\left( nm...l \right) ;A}\left( \tau \right) G_{\psi \psi ,AB}^{q-2}\left( \tau \right) ,
\\
&\varSigma _{b\psi ,AB}^{\left( n \right) ;A}\left( \tau \right) =-\left( -1 \right) ^{\left( q-1 \right) /2}\frac{\left( q-1 \right)}{2}\mathcal{J}^{(n)}G_{\psi b,AB}^{\left( n \right) ;S}\left( \tau \right) G_{\psi \psi ,AB}^{q-2}\left( \tau \right) ,
\\
&\varSigma _{b\psi ,AB}^{\left( nm...l \right) ;A}\left( \tau \right) =-\left( -1 \right) ^{\left( q-1 \right) /2}\frac{\left( q-1 \right)}{2}\mathcal{J}^{(nm...l)}G_{\psi b,AB}^{\left( nm...l \right) ;S}\left( \tau \right) G_{\psi \psi ,AB}^{q-2}\left( \tau \right) ,
\\
&\varSigma _{\psi b,AB}^{\left( n \right) ;S}\left( \tau \right) =\frac{\left( q-1 \right)}{2}\mathcal{J}^{(n)}G_{b\psi ,AB}^{\left( n \right) ;A}\left( \tau \right) G_{\psi \psi ,AB}^{q-2}\left( \tau \right) ,
\\
&\varSigma _{\psi b,AB}^{\left( nm...l \right) ;S}\left( \tau \right) =\frac{\left( q-1 \right)}{2}\mathcal{J}^{(nm...l)}G_{b\psi ,AB}^{\left( nm...l \right) ;A}\left( \tau \right) G_{\psi \psi ,AB}^{q-2}\left( \tau \right) ,
\\
&\varSigma _{\psi b,AB}^{\left( n \right) ;A}\left( \tau \right) =\left( -1 \right) ^{\left( q-1 \right) /2}\frac{\left( q-1 \right)}{2}\mathcal{J}^{(n)}G_{b\psi ,AB}^{\left( n \right) ;S}\left( \tau \right) G_{\psi \psi ,AB}^{q-2}\left( \tau \right) .
\\
&\varSigma _{\psi b,AB}^{\left( nm...l \right) ;A}\left( \tau \right) =\left( -1 \right) ^{\left( q-1 \right) /2}\frac{\left( q-1 \right)}{2}\mathcal{J}^{(nm...l)}G_{b\psi ,AB}^{\left( nm...l \right) ;S}\left( \tau \right) G_{\psi \psi ,AB}^{q-2}\left( \tau \right) .
\end{align}
}
The equations of motion in frequency space are derived via Fourier transform.
\small{
\begin{align}
G_{\psi \psi ,AB}\left( \omega \right) &=\frac{Det\left( A_{\psi \psi ,AB}\left( D_{half\,\,integer} \right) \right)}{Det\left( D_{half\,\,integer} \right)},
\
G_{bb,AB}^{\left( n \right)}\left( \omega \right) =-\frac{Det\left( A_{bb,AB}^{\left( n \right)}\left( D_{integer} \right) \right)}{Det\left( D_{integer} \right)},\nonumber
\\
G_{bb,AB}^{\left( nm...l \right)}\left( \omega \right) &=-\frac{Det\left( A_{bb,AB}^{\left( nm...l \right)}\left( D_{integer} \right) \right)}{Det\left( D_{integer} \right)},
\
G_{\psi b,AB}^{\left( n \right);A}\left( \omega \right) =\frac{Det\left( A_{\psi b,AB}^{\left( n \right);A}\left( D_{half\,\,integer} \right) \right)}{Det\left( D_{half\,\,integer} \right)},\nonumber
\\
G_{\psi b,AB}^{\left( nm...l \right);A}\left( \omega \right) &=\frac{Det\left( A_{\psi b,AB}^{\left( nm...l \right);A}\left( D_{half\,\,integer} \right) \right)}{Det\left( D_{half\,\,integer} \right)},
\
G_{b\psi ,AB}^{\left( n \right);S}\left( \omega \right) =-\frac{Det\left( A_{b\psi ,AB}^{\left( n \right);S}\left( D_{integer} \right) \right)}{Det\left( D_{integer} \right)},\nonumber
\\
G_{b\psi ,AB}^{\left( nm...l \right);S}\left( \omega \right) &=-\frac{Det\left( A_{b\psi ,AB}^{\left( nm...l \right);S}\left( D_{integer} \right) \right)}{Det\left( D_{integer} \right)},
\
G_{b\psi ,AB}^{\left( n \right);A}\left( \omega \right) =\frac{Det\left( A_{b\psi ,AB}^{\left( n \right);A}\left( D_{half\,\,integer} \right) \right)}{Det\left( D_{half\,\,integer} \right)},\nonumber
\\
G_{b\psi ,AB}^{\left( nm...l \right);A}\left( \omega \right) &=\frac{Det\left( A_{b\psi ,AB}^{\left( nm...l \right);A}\left( D_{half\,\,integer} \right) \right)}{Det\left( D_{half\,\,integer} \right)},
\
G_{\psi b,AB}^{\left( n \right);S}\left( \omega \right) =-\frac{Det\left( A_{\psi b,AB}^{\left( n \right);S}\left( D_{integer} \right) \right)}{Det\left( D_{integer} \right)}.\nonumber
\\
G_{\psi b,AB}^{\left( nm...l \right);S}\left( \omega \right) &=-\frac{Det\left( A_{\psi b,AB}^{\left( nm...l \right);S}\left( D_{integer} \right) \right)}{Det\left( D_{integer} \right)}.\nonumber
\end{align}
}
\subsection{Real time Schwinger-Dyson equation}
\quad The retarded component can be expressed using both bosonic and fermionic sector decompositions.
\begin{equation}
\begin{split}
G_{\psi \psi ,AB}^{\left( n \right) >}\left( \omega \right) =\frac{G_{\psi \psi ,AB}^{\left( n \right) R}\left( \omega \right) -\left( G_{\psi \psi ,AB}^{\left( n \right) R}\left( \omega \right) \right) ^*}{1+\exp \left( -\beta \omega \right)},
\
G_{bb,AB}^{\left( n \right) >}\left( \omega \right) =\frac{G_{bb,AB}^{\left( n \right) R}\left( \omega \right) -\left( G_{bb,AB}^{\left( n \right) R}\left( \omega \right) \right) ^*}{-1+\exp \left( -\beta \omega \right)},\nonumber
\\
G_{b\psi ,AB}^{\left( n \right) >S}\left( \omega \right) =\frac{G_{b\psi ,AB}^{\left( n \right) RS}\left( \omega \right) -\left( G_{b\psi ,AB}^{\left( n \right) RS}\left( \omega \right) \right) ^*}{-1+\exp \left( -\beta \omega \right)},
\
G_{\psi b,AB}^{\left( n \right) >A}\left( \omega \right) =\frac{G_{\psi b,AB}^{\left( n \right) RA}\left( \omega \right) -\left( G_{\psi b,AB}^{\left( n \right) RA}\left( \omega \right) \right) ^*}{1+\exp \left( -\beta \omega \right)},\nonumber
\\
G_{b\psi ,AB}^{\left( n \right) >A}\left( \omega \right) =\frac{G_{b\psi ,AB}^{\left( n \right) RA}\left( \omega \right) -\left( G_{b\psi ,AB}^{\left( n \right) RA}\left( \omega \right) \right) ^*}{1+\exp \left( -\beta \omega \right)},
\
G_{\psi b,AB}^{\left( n \right) >S}\left( \omega \right) =\frac{G_{\psi b,AB}^{\left( n \right) RS}\left( \omega \right) -\left( G_{\psi b,AB}^{\left( n \right) RS}\left( \omega \right) \right) ^*}{-1+\exp \left( -\beta \omega \right)}.\nonumber
\end{split}
\end{equation}

and
\begin{equation}
\begin{split}
G_{\psi \psi ,AB}^{\left( nm \right) >}\left( \omega \right) =\frac{G_{\psi \psi ,AB}^{\left( nm \right) R}\left( \omega \right) -\left( G_{\psi \psi ,AB}^{\left( nm \right) R}\left( \omega \right) \right) ^*}{1+\exp \left( -\beta \omega \right)},
\
G_{bb,AB}^{\left( nm \right) >}\left( \omega \right) =\frac{G_{bb,AB}^{\left( nm \right) R}\left( \omega \right) -\left( G_{bb,AB}^{\left( nm \right) R}\left( \omega \right) \right) ^*}{-1+\exp \left( -\beta \omega \right)},\nonumber
\\
G_{b\psi ,AB}^{\left( nm \right) >S}\left( \omega \right) =\frac{G_{b\psi ,AB}^{\left( nm \right) RS}\left( \omega \right) -\left( G_{b\psi ,AB}^{\left( nm \right) RS}\left( \omega \right) \right) ^*}{-1+\exp \left( -\beta \omega \right)},
\
G_{\psi b,AB}^{\left( nm \right) >A}\left( \omega \right) =\frac{G_{\psi b,AB}^{\left( nm \right) RA}\left( \omega \right) -\left( G_{\psi b,AB}^{\left( nm \right) RA}\left( \omega \right) \right) ^*}{1+\exp \left( -\beta \omega \right)},\nonumber
\\
G_{b\psi ,AB}^{\left( nm \right) >A}\left( \omega \right) =\frac{G_{b\psi ,AB}^{\left( nm \right) RA}\left( \omega \right) -\left( G_{b\psi ,AB}^{\left( nm \right) RA}\left( \omega \right) \right) ^*}{1+\exp \left( -\beta \omega \right)},
\
G_{\psi b,AB}^{\left( nm \right) >S}\left( \omega \right) =\frac{G_{\psi b,AB}^{\left( nm \right) RS}\left( \omega \right) -\left( G_{\psi b,AB}^{\left( nm \right) RS}\left( \omega \right) \right) ^*}{-1+\exp \left( -\beta \omega \right)}.\nonumber
\end{split}
\end{equation}
with the expanded Lorentzian matrix
{\small
\begin{equation}
\begin{split}
&D_{half\,\,integer}=\\
&\left( \begin{matrix}
	w_n-\varSigma _{LL,\psi \psi}&		-\varSigma _{LR,\psi \psi}&		-\varSigma _{LL,\psi b}^{n,A}&		-i\mu -\varSigma _{LR,\psi b}^{n,A}&		-\varSigma _{LL,\psi b}^{nm,A}&		-i\mu '^{(nm)}-\varSigma _{LR,\psi b}^{nm,A}\\
	-\varSigma _{RL,\psi \psi}&		w_n-\varSigma _{RR,\psi \psi}&		i\mu -\varSigma _{RL,\psi b}^{n,A}&		-\varSigma _{RR,\psi b}^{n,A}&		i\mu '^{(nm)}-\varSigma _{RL,\psi b}^{nm,A}&		-\varSigma _{RR,\psi b}^{nm,A}\\
	-\varSigma _{LL,b\psi}^{n,A}&		i\mu -\varSigma _{LR,b\psi}^{n,A}&		-1-0&		-0&		-0&		-0\\
	-i\mu -\varSigma _{RL,b\psi}^{n,A}&		-\varSigma _{RR,b\psi}^{n,A}&		-0&		-1-0&		-0&		-0\\
	-\varSigma _{LL,b\psi}^{nm,A}&		i\mu '^{(nm)}-\varSigma _{LR,b\psi}^{nm,A}&		-0&		-0&		-1-0&		-0\\
	-i\mu '^{(nm)}-\varSigma _{RL,b\psi}^{nm,A}&		-\varSigma _{RR,b\psi}^{nm,A}&		-0&		-0&		-0&		-1-0\\
\end{matrix} \right) ,\nonumber
\end{split}
\end{equation}
\begin{equation}
\begin{split}
&D_{Lorentz\,\,inverse}=\\
&\left( \begin{matrix}
	w_n-0&		-0&		-\varSigma _{LL,\psi b}^{n,S}&		i\mu ^{(n)}-\varSigma _{LR,\psi b}^{n,S}&		-\varSigma _{LL,\psi b}^{nm,S}&		i\mu '^{(nm)}-\varSigma _{LR,\psi b}^{nm,S}\\
	-0&		w_n-0&		-i\mu ^{(n)}-\varSigma _{RL,\psi b}^{n,S}&		-\varSigma _{RR,\psi b}^{n,S}&		-i\mu '^{(nm)}-\varSigma _{RL,\psi b}^{nm,S}&		-\varSigma _{RR,\psi b}^{nm,S}\\
	-\varSigma _{LL,b\psi}^{n,S}&		-i\mu ^{(n)}-\varSigma _{LR,b\psi}^{n,S}&		-1-\varSigma _{LL,bb}^{(n)}&		-\varSigma _{LR,bb}^{(n)}&		-0&		-0\\
	i\mu ^{(n)}-\varSigma _{RL,b\psi}^{n,S}&		-\varSigma _{RR,b\psi}^{n,S}&		-\varSigma _{RL,bb}^{(n)}&		-1-\varSigma _{RR,bb}^{(n)}&		-0&		-0\\
	-\varSigma _{LL,b\psi}^{nM,S}&		-i\mu '^{(nm)}-\varSigma _{LR,b\psi}^{nm,S}&		-0&		-0&		-1-\varSigma _{LL,bb}^{(nm)}&		-\varSigma _{LR,bb}^{(nm)}\\
	i\mu '^{(nm)}-\varSigma _{RL,b\psi}^{nm,S}&		-\varSigma _{RR,b\psi}^{nM,S}&		-0&		-0&		-\varSigma _{RL,bb}^{(nm)}&		-1-\varSigma _{RR,bb}^{(nm)}\\
\end{matrix} \right) .\nonumber
\end{split}
\end{equation}
}
and the Lorentzian equations of motion are
\begin{equation}
\begin{split}
G_{\psi \psi ,AB}\left( \tau \right) &=\frac{Det\left( A_{\psi \psi ,AB}\left( D_{Lorentz} \right) \right)}{Det\left( D_{Lorentz} \right)},
\
G_{bb,AB}^{(n)}\left( \tau \right) =-\frac{Det\left( A_{bb,AB}^{(n)}\left( D_{Lorentz\,\,inverse} \right) \right)}{Det\left( D_{Lorentz\,\,inverse} \right)},\nonumber
\\
G_{bb,AB}^{(nm)}\left( \tau \right) &=-\frac{Det\left( A_{bb,AB}^{(nm)}\left( D_{Lorentz\,\,inverse} \right) \right)}{Det\left( D_{Lorentz\,\,inverse} \right)},
\
G_{\psi b,AB}^{n,A}\left( \tau \right) =\frac{Det\left( A_{\psi b,AB}^{n,A}\left( D_{Lorentz} \right) \right)}{Det\left( D_{Lorentz} \right)},\nonumber
\\
G_{b\psi ,AB}^{n,S}\left( \tau \right) &=-\frac{Det\left( A_{b\psi ,AB}^{n,S}\left( D_{Lorentz\,\,inverse} \right) \right)}{Det\left( D_{Lorentz\,\,inverse} \right)},
\
G_{\psi b,AB}^{n,S}\left( \tau \right) =\frac{Det\left( A_{\psi b,AB}^{n,S}\left( D_{Lorentz\,\,inverse} \right) \right)}{Det\left( D_{Lorentz\,\,inverse} \right)},\nonumber
\\
G_{b\psi ,AB}^{n,A}\left( \tau \right) &=-\frac{Det\left( A_{b\psi ,AB}^{n,A}\left( D_{Lorentz} \right) \right)}{Det\left( D_{Lorentz} \right)},
\
G_{\psi b,AB}^{nm,A}\left( \tau \right) =\frac{Det\left( A_{\psi b,AB}^{nm,A}\left( D_{Lorentz} \right) \right)}{Det\left( D_{Lorentz} \right)},\nonumber
\\
G_{b\psi ,AB}^{nm,S}\left( \tau \right) &=-\frac{Det\left( A_{b\psi ,AB}^{nm,S}\left( D_{Lorentz\,\,inverse} \right) \right)}{Det\left( D_{Lorentz\,\,inverse} \right)},
\
G_{\psi b,AB}^{nm,S}\left( \tau \right) =\frac{Det\left( A_{\psi b,AB}^{nm,S}\left( D_{Lorentz\,\,inverse} \right) \right)}{Det\left( D_{Lorentz\,\,inverse} \right)},\nonumber
\\
G_{b\psi ,AB}^{nm,A}\left( \tau \right) &=-\frac{Det\left( A_{b\psi ,AB}^{nm,A}\left( D_{Lorentz} \right) \right)}{Det\left( D_{Lorentz} \right)}.\nonumber
\end{split}
\end{equation}

\section{Conformal group theories}\label{ap2}
\quad The original $OSp(1|2)$ supersymmetry in the low-energy limit of the $\mathcal{N}=1$ supersymmetric SYK model
\begin{equation}
\tau' = \frac{a \tau + \alpha \theta + b}{c \tau + \gamma \theta + d}, \qquad \qquad \theta' = \frac{\beta \tau + e \theta + \delta}{c \tau + \gamma \theta + d} 
\end{equation}
with
\begin{equation}
(\beta \tau + e \theta + \delta)(e + \theta \beta)+ (a \tau + \alpha \theta + b)(- \gamma + \theta c) - (c \tau + \gamma \theta + d)(- \alpha + \theta a) =0 .
\end{equation}
The fermionic and bosonic generators must satisfy additional quadratic constraints.
\begin{equation}
e\beta - a  \gamma+ \alpha c =0, \qquad  e^2 +\beta \delta+2 \alpha \gamma+ b c-a d =0, \qquad e \delta -\gamma b + \alpha d =0 
\end{equation}
i.e.
\begin{equation}
\begin{pmatrix}e & -\alpha & -\gamma \cr \beta & a & c \cr \delta & b & d \end{pmatrix}\begin{pmatrix}1 & 0 & 0 \cr 0&0&-1 \cr 0&1&0 \end{pmatrix}   \begin{pmatrix}e & \beta& \delta \cr \alpha & a & b \cr \gamma & c & d \end{pmatrix}= \begin{pmatrix}1 & 0 & 0 \cr 0&0&-1 \cr 0&1&0 \end{pmatrix}  .
\end{equation}

We now incorporate the $Z(n)$ symmetry into the $OSp(1|2)$ supergroup, where the coordinates are labeled as
\begin{equation}
\tau ^{\left( n \right)\prime}=\frac{a^{\left( n \right)}\tau ^{\left( n \right)}+\alpha ^{\left( n \right)}\theta ^{\left( n \right)}+b^{\left( n \right)}}{c^{\left( n \right)}\tau ^{\left( n \right)}+\gamma ^{\left( n \right)}\theta ^{\left( n \right)}+d^{\left( n \right)}},\qquad \qquad \theta ^{\left( n \right)\prime}=\frac{\beta ^{\left( n \right)}\tau ^{\left( n \right)}+e^{\left( n \right)}\theta ^{\left( n \right)}+\delta ^{\left( n \right)}}{c^{\left( n \right)}\tau ^{\left( n \right)}+\gamma ^{\left( n \right)}\theta ^{\left( n \right)}+d^{\left( n \right)}},
\end{equation}
with
\begin{equation}
\begin{split}
&e^{\left( n \right)}\beta ^{\left( n \right)}-a^{\left( n \right)}\gamma ^{\left( n \right)}+\alpha ^{\left( n \right)}c^{\left( n \right)}=0,
\\
&{e^{\left( n \right)}}^2+\beta ^{\left( n \right)}\delta ^{\left( n \right)}+2\alpha ^{\left( n \right)}\gamma ^{\left( n \right)}+b^{\left( n \right)}c^{\left( n \right)}-a^{\left( n \right)}d^{\left( n \right)}=0,
\\
&e^{\left( n \right)}\delta ^{\left( n \right)}-\gamma ^{\left( n \right)}b^{\left( n \right)}+\alpha ^{\left( n \right)}d^{\left( n \right)}=0.
\end{split}
\end{equation}
Each system on every surface must also be formulated with $OSp(1|2)$ symmetry, which extends beyond $Z(n)$. For example,
\begin{equation}
\tau ^{\left( nm \right)\prime}=\frac{a^{\left( nm \right)}\tau ^{\left( nm \right)}+\alpha ^{\left( nm \right)}\theta ^{\left( nm \right)}+b^{\left( nm \right)}}{c^{\left( nm \right)}\tau ^{\left( nm \right)}+\gamma ^{\left( nm \right)}\theta ^{\left( nm \right)}+d^{\left( nm \right)}},\qquad \qquad \theta ^{\left( nm \right)\prime}=\frac{\beta ^{\left( nm \right)}\tau ^{\left( nm \right)}+e^{\left( nm \right)}\theta ^{\left( nm \right)}+\delta ^{\left( nm \right)}}{c^{\left( nm \right)}\tau ^{\left( nm \right)}+\gamma ^{\left( nm \right)}\theta ^{\left( nm \right)}+d^{\left( nm \right)}}.
\nonumber
\end{equation}
with
\begin{equation}
\begin{split}
&e^{\left( nm \right)}\beta ^{\left( nm \right)}-a^{\left( nm \right)}\gamma ^{\left( nm \right)}+\alpha ^{\left( nm \right)}c^{\left( nm \right)}=0,
\\
&{e^{\left( nm \right)}}^2+\beta ^{\left( nm \right)}\delta ^{\left( nm \right)}+2\alpha ^{\left( nm \right)}\gamma ^{\left( nm \right)}+b^{\left( nm \right)}c^{\left( nm \right)}-a^{\left( nm \right)}d^{\left( nm \right)}=0,
\\
&e^{\left( nm \right)}\delta ^{\left( nm \right)}-\gamma ^{\left( nm \right)}b^{\left( nm \right)}+\alpha ^{\left( nm \right)}d^{\left( nm \right)}=0.
\end{split}
\end{equation}
Within the secondary surface, the system must possess $Z(C_{n}^{2})$ symmetry. The coordinates must correspond to either:
\begin{itemize}
    \item Higher-order correlation functions, or
    \item Dilaton degrees of freedom,
\end{itemize}
both of which are fundamentally distinct from the variables defined on the original (or primary) surface. Within this extended framework, one can systematically derive additional terms that capture subleading corrections or encode the effects of replica symmetry breaking and gravitational backreaction 
\begin{equation}
\begin{split}
\tau ^{\left( nm...l \right)\prime}=\frac{a^{\left( nm...l \right)}\tau ^{\left( nm...l \right)}+\alpha ^{\left( nm...l \right)}\theta ^{\left( nm...l \right)}+b^{\left( nm...l \right)}}{c^{\left( nm...l \right)}\tau ^{\left( nm...l \right)}+\gamma ^{\left( nm...l \right)}\theta ^{\left( nm...l \right)}+d^{\left( nm...l \right)}},
\\
\theta ^{\left( nm...l \right)\prime}=\frac{\beta ^{\left( nm...l \right)}\tau ^{\left( nm...l \right)}+e^{\left( nm...l \right)}\theta ^{\left( nm...l \right)}+\delta ^{\left( nm...l \right)}}{c^{\left( nm...l \right)}\tau ^{\left( nm...l \right)}+\gamma ^{\left( nm...l \right)}\theta ^{\left( nm...l \right)}+d^{\left( nm...l \right)}},
\end{split}
\end{equation}
with
\begin{equation}
\begin{split}
&e^{\left( nm...l \right)}\beta ^{\left( nm...l \right)}-a^{\left( nm...l \right)}\gamma ^{\left( nm...l \right)}+\alpha ^{\left( nm...l \right)}c^{\left( nm...l \right)}=0,
\\
&{e^{\left( nm...l \right)}}^2+\beta ^{\left( nm...l \right)}\delta ^{\left( nm...l \right)}+2\alpha ^{\left( nm...l \right)}\gamma ^{\left( nm...l \right)}+b^{\left( nm...l \right)}c^{\left( nm...l \right)}-a^{\left( nm...l \right)}d^{\left( nm...l \right)}=0,
\\
&e^{\left( nm...l \right)}\delta ^{\left( nm...l \right)}-\gamma ^{\left( nm...l \right)}b^{\left( nm...l \right)}+\alpha ^{\left( nm...l \right)}d^{\left( nm...l \right)}=0.
\end{split}
\end{equation}
The system should have $Z(C_{n}^{nm\cdots l})$ symmetry. 
We can list the generators and group multiplication.
\begin{equation}
\begin{split}
\left[ H,E^{\pm} \right] =\pm E^{\pm},
\left[ E^+,E^- \right] =2H,
\left[ H,Q^{\pm} \right] =\pm \frac{1}{2}Q^{\pm},
\\
\left[ E^{\pm},Q^{\mp} \right] =Q^{\pm},
\left\{ Q^+,Q^- \right\} =H,
\left\{ Q^{\pm},Q^{\pm} \right\} =\pm E^{\pm}.
\end{split}
\end{equation}
This $OSp(1|2)$ group should be invariant under each copy, that is  
\begin{equation}
\begin{split}
\left[ H^{\left( n \right)},{E^{\left( n \right)}}^{\pm} \right] =\pm {E^{\left( n \right)}}^{\pm},\left[ {E^{\left( n \right)}}^+,{E^{\left( n \right)}}^- \right] =2H^{\left( n \right)},\left[ H^{\left( n \right)},{Q^{\left( n \right)}}^{\pm} \right] =\pm \frac{1}{2}{Q^{\left( n \right)}}^{\pm},
\\
\left[ {E^{\left( n \right)}}^{\pm},{Q^{\left( n \right)}}^{\mp} \right] ={Q^{\left( n \right)}}^{\pm},\left\{ {Q^{\left( n \right)}}^+,{Q^{\left( n \right)}}^- \right\} =H^{\left( n \right)},\left\{ {Q^{\left( n \right)}}^{\pm},{Q^{\left( n \right)}}^{\pm} \right\} =\pm {E^{\left( n \right)}}^{\pm}.
\end{split}
\end{equation}
On each surface, we have
\begin{equation}
\begin{split}
\left[ H^{\left( nm \right)},{E^{\left( nm \right)}}^{\pm} \right] =\pm {E^{\left( nm \right)}}^{\pm},\left[ {E^{\left( nm \right)}}^+,{E^{\left( nm \right)}}^- \right] =2H^{\left( nm \right)},\left[ H^{\left( nm \right)},{Q^{\left( nm \right)}}^{\pm} \right] =\pm \frac{1}{2}{Q^{\left( nm \right)}}^{\pm},
\\
\left[ {E^{\left( nm \right)}}^{\pm},{Q^{\left( nm \right)}}^{\mp} \right] ={Q^{\left( nm \right)}}^{\pm},\left\{ {Q^{\left( nm \right)}}^+,{Q^{\left( nm \right)}}^- \right\} =H^{\left( nm \right)},\left\{ {Q^{\left( nm \right)}}^{\pm},{Q^{\left( nm \right)}}^{\pm} \right\} =\pm {E^{\left( nm \right)}}^{\pm},
\end{split}\nonumber
\end{equation}
and 
\begin{equation}
\begin{split}
&\left[ H^{\left( nm...l \right)},{E^{\left( nm...l \right)}}^{\pm} \right] =\pm {E^{\left( nm...l \right)}}^{\pm},\left[ {E^{\left( nm...l \right)}}^+,{E^{\left( nm...l \right)}}^- \right] =2H^{\left( nm...l \right)},
\\
&\left[ H^{\left( nm...l \right)},{Q^{\left( nm...l \right)}}^{\pm} \right] =\pm \frac{1}{2}{Q^{\left( nm...l \right)}}^{\pm},\left[ {E^{\left( nm...l \right)}}^{\pm},{Q^{\left( nm...l \right)}}^{\mp} \right] ={Q^{\left( nm...l \right)}}^{\pm},
\\
&\left\{ {Q^{\left( nm...l \right)}}^+,{Q^{\left( nm...l \right)}}^- \right\} =H^{\left( nm...l \right)},\left\{ {Q^{\left( nm...l \right)}}^{\pm},{Q^{\left( nm...l \right)}}^{\pm} \right\} =\pm {E^{\left( nm...l \right)}}^{\pm}.
\end{split}
\end{equation}
However, the interactions between the surfaces cannot be ignored.

The group multiplication between the elements can be divided into related ones (having the same replica labels and vanishing when one of the fermions with the same label vanishes) and unrelated ones (not having the same replica labels and vanishing due to anti-commutation). The group element generated from the group multiplication between different orders will yield the element of the greatest common divisor.
\begin{equation}
\begin{split}
\left[ H^{\left( nm...l \right)},{E^{\left( n \right)}}^{\pm} \right] &=\left[ H^{\left( n \right)},{E^{\left( nm...l \right)}}^{\pm} \right] =\pm {E^{\left( n \right)}}^{\pm},\\
\left[ {E^{\left( nm...l \right)}}^+,{E^{\left( n \right)}}^- \right] &=\left[ {E^{\left( n \right)}}^+,{E^{\left( nm...l \right)}}^- \right] =2H^{\left( n \right)},
\\
\left[ H^{\left( nm...l \right)},{Q^{\left( n \right)}}^{\pm} \right] &=\left[ H^{\left( n \right)},{Q^{\left( nm...l \right)}}^{\pm} \right] =\pm \frac{1}{2}{Q^{\left( n \right)}}^{\pm},\\ 
\left[ {E^{\left( nm...l \right)}}^{\pm},{Q^{\left( n \right)}}^{\mp} \right]& =\left[ {E^{\left( n \right)}}^{\pm},{Q^{\left( nm...l \right)}}^{\mp} \right] ={Q^{\left( n \right)}}^{\pm},
\\
\left\{ {Q^{\left( nm...l \right)}}^+,{Q^{\left( n \right)}}^- \right\} &=\left\{ {Q^{\left( n \right)}}^+,{Q^{\left( nm...l \right)}}^- \right\} =H^{\left( n \right)},\\
\left\{ {Q^{\left( nm...l \right)}}^{\pm},{Q^{\left( n \right)}}^{\pm} \right\} &=\left\{ {Q^{\left( n \right)}}^{\pm},{Q^{\left( nm...l \right)}}^{\pm} \right\} =\pm {E^{\left( n \right)}}^{\pm}.
\end{split}
\end{equation}

We can provide a detailed proof of this relationship: 
The group elements in supersymmetric SYK models on surfaces of higher order consist of two parts: the lower-order SYK models with fermions at the same sites, and the SYK models without any fermions at the same sites. After applying the commutator and anticommutator, the fermions at the same sites will vanish, while those at different sites will contribute nothing due to their commutation and anticommutation properties. In summary, these commutation and anticommutation relations originate from the global symmetries of the supersymmetric SYK models.

Any element of a single group can be replaced by elements of a larger group, making it a simple phase-weighted parameterization. 
The corresponding symmetries impose constraints as follows.
\begin{equation}
\begin{split}
\tau ^{\left( n \right)\prime}=\frac{a^{\left( n \right)}\tau ^{\left( n \right)}+a^{\left( nm...l \right)}\tau ^{\left( nm...l \right)}+\alpha ^{\left( n \right)}\theta ^{\left( n \right)}+\alpha ^{\left( nm...l \right)}\theta ^{\left( nm...l \right)}+b^{\left( n \right)}+b^{\left( nm...l \right)}}{c^{\left( n \right)}\tau ^{\left( n \right)}+c^{\left( nm...l \right)}\tau ^{\left( nm...l \right)}+\gamma ^{\left( n \right)}\theta ^{\left( n \right)}+\gamma ^{\left( nm...l \right)}\theta ^{\left( nm...l \right)}+d^{\left( n \right)}+d^{\left( nm...l \right)}},
\\
\theta ^{\left( n \right)\prime}=\frac{\beta ^{\left( n \right)}\tau ^{\left( n \right)}+\beta ^{\left( nm...l \right)}\tau ^{\left( nm...l \right)}+e^{\left( n \right)}\theta ^{\left( n \right)}+e^{\left( nm...l \right)}\theta ^{\left( nm...l \right)}+\delta ^{\left( n \right)}+\delta ^{\left( nm...l \right)}}{c^{\left( n \right)}\tau ^{\left( n \right)}+c^{\left( nm...l \right)}\tau ^{\left( nm...l \right)}+\gamma ^{\left( n \right)}\theta ^{\left( n \right)}+\gamma ^{\left( nm...l \right)}\theta ^{\left( nm...l \right)}+d^{\left( n \right)}+d^{\left( nm...l \right)}}
\\
\tau ^{\left( nm...l \right)\prime}=\frac{0^{\left( n \right)}\tau ^{\left( n \right)}+a^{\left( nm...l \right)}\tau ^{\left( nm...l \right)}+0^{\left( n \right)}\theta ^{\left( n \right)}+\alpha ^{\left( nm...l \right)}\theta ^{\left( nm...l \right)}+b^{\left( n \right)}+b^{\left( nm...l \right)}}{0^{\left( n \right)}\tau ^{\left( n \right)}+c^{\left( nm...l \right)}\tau ^{\left( nm...l \right)}+0^{\left( n \right)}\theta ^{\left( n \right)}+\gamma ^{\left( nm...l \right)}\theta ^{\left( nm...l \right)}+d^{\left( n \right)}+d^{\left( nm...l \right)}},
\\
\theta ^{\left( nm...l \right)\prime}=\frac{0^{\left( n \right)}\tau ^{\left( n \right)}+\beta ^{\left( nm...l \right)}\tau ^{\left( nm...l \right)}+0^{\left( n \right)}\theta ^{\left( n \right)}+e^{\left( nm...l \right)}\theta ^{\left( nm...l \right)}+\delta^{\left( n \right)}+\delta ^{\left( nm...l \right)}}{0^{\left( n \right)}\tau ^{\left( n \right)}+c^{\left( nm...l \right)}\tau ^{\left( nm...l \right)}+0^{\left( n \right)}\theta ^{\left( n \right)}+\gamma ^{\left( nm...l \right)}\theta ^{\left( nm...l \right)}+d^{\left( n \right)}+d^{\left( nm...l \right)}}, \label{INV1}
\end{split}
\end{equation}
with the constraint of orthogonality
\begin{equation}
\begin{split}
&e^{\left( nm...l \right)}\beta ^{\left( nm...l \right)}-a^{\left( nm...l \right)}\gamma ^{\left( nm...l \right)}+\alpha ^{\left( nm...l \right)}c^{\left( nm...l \right)}=0,
\\
&{e^{\left( nm...l \right)}}^2+\beta ^{\left( nm...l \right)}\delta ^{\left( nm...l \right)}+2\alpha ^{\left( nm...l \right)}\gamma ^{\left( nm...l \right)}+b^{\left( nm...l \right)}c^{\left( nm...l \right)}-a^{\left( nm...l \right)}d^{\left( nm...l \right)}=0,
\\
&e^{\left( nm...l \right)}\delta ^{\left( nm...l \right)}-\gamma ^{\left( nm...l \right)}b^{\left( nm...l \right)}+\alpha ^{\left( nm...l \right)}d^{\left( nm...l \right)}=0,
\\
&e^{\left( n \right)}\beta ^{\left( n \right)}-a^{\left( n \right)}\gamma ^{\left( n \right)}+\alpha ^{\left( n \right)}c^{\left( n \right)}=0,
\\
&{e^{\left( n \right)}}^2+\beta ^{\left( n \right)}\delta ^{\left( n \right)}+2\alpha ^{\left( n \right)}\gamma ^{\left( n \right)}+b^{\left( n \right)}c^{\left( n \right)}-a^{\left( n \right)}d^{\left( n \right)}=0,
\\
&e^{\left( n \right)}\delta ^{\left( n \right)}-\gamma ^{\left( n \right)}b^{\left( n \right)}+\alpha ^{\left( n \right)}d^{\left( n \right)}=0,
\end{split}
\end{equation}
where
\begin{equation}
a^{\left( nm...l \right)}\ne 0,c^{\left( nm...l \right)}\ne 0,\alpha ^{\left( nm...l \right)}\ne 0,\gamma ^{\left( nm...l \right)}\ne 0,\beta ^{\left( nm...l \right)}\ne 0,e^{\left( nm...l \right)}\ne 0,
\end{equation}

and the cross-terms
\begin{equation}
\begin{split}
&(\beta ^{\left( n \right)}\tau ^{\left( n \right)}+\beta ^{\left( nm...l \right)}\tau ^{\left( nm...l \right)}+e^{\left( n \right)}\theta ^{\left( n \right)}+e^{\left( nm...l \right)}\theta ^{\left( nm...l \right)}+\delta ^{\left( n \right)}+\delta ^{\left( nm...l \right)})
\\
&(e^{\left( n \right)}+e^{\left( nm...l \right)}+\theta ^{\left( n \right)}\beta ^{\left( n \right)}+\theta ^{\left( nm...l \right)}\beta ^{\left( nm...l \right)})+\left( a^{\left( n \right)}\tau ^{\left( n \right)}+a^{\left( nm...l \right)}\tau ^{\left( nm...l \right)}+ \right. 
\\
&\left. \alpha ^{\left( n \right)}\theta ^{\left( n \right)}+\alpha ^{\left( nm...l \right)}\theta ^{\left( nm...l \right)}+b^{\left( n \right)}+b^{\left( nm...l \right)} \right) (-\gamma ^{\left( n \right)}-\gamma ^{\left( nm...l \right)}+\theta ^{\left( n \right)}c^{\left( n \right)}+\theta ^{\left( nm...l \right)}c^{\left( nm...l \right)})
\\
&-(c^{\left( n \right)}\tau ^{\left( n \right)}+c^{\left( nm...l \right)}\tau ^{\left( nm...l \right)}+\gamma ^{\left( n \right)}\theta ^{\left( n \right)}+\gamma ^{\left( nm...l \right)}\theta ^{\left( nm...l \right)}+d^{\left( n \right)}+d^{\left( nm...l \right)})
\\
&(-\alpha ^{\left( n \right)}-\alpha ^{\left( nm...l \right)}+\theta ^{\left( n \right)}a^{\left( n \right)}+\theta ^{\left( nm...l \right)}a^{\left( nm...l \right)})=0.
\end{split}\nonumber
\end{equation}

The minimum translation group with the dependence on two sets of the supersymmetric coordinate
\begin{equation}
\begin{split}
&\left( e^{\left( n \right)}+e^{\left( nm...l \right)} \right) \beta ^{\left( nm...l \right)}-a^{\left( nm...l \right)}\left( \gamma ^{\left( n \right)}+\gamma ^{\left( nm...l \right)} \right) +\left( \alpha ^{\left( n \right)}+\alpha ^{\left( nm...l \right)} \right) c^{\left( nm...l \right)}=0,
\\
&\left( e^{\left( n \right)}+e^{\left( nm...l \right)} \right) \beta ^{\left( n \right)}-a^{\left( n \right)}\left( \gamma ^{\left( n \right)}+\gamma ^{\left( nm...l \right)} \right) +\left( \alpha ^{\left( n \right)}+\alpha ^{\left( nm...l \right)} \right) c^{\left( n \right)}=0,
\\
&e^{\left( nm...l \right)}\left( e^{\left( n \right)}+e^{\left( nm...l \right)} \right) +\beta ^{\left( nm...l \right)}\left( \delta ^{\left( n \right)}+\delta ^{\left( nm...l \right)} \right) +\alpha ^{\left( nm...l \right)}\left( \gamma ^{\left( n \right)}+\gamma ^{\left( nm...l \right)} \right) 
\\
&+\left( \alpha ^{\left( n \right)}+\alpha ^{\left( nm...l \right)} \right) \gamma ^{\left( nm...l \right)}+\left( b^{\left( n \right)}+b^{\left( nm...l \right)} \right) c^{\left( nm...l \right)}-a^{\left( nm...l \right)}\left( d^{\left( n \right)}+d^{\left( nm...l \right)} \right) =0,
\\
&e^{\left( n \right)}\left( e^{\left( n \right)}+e^{\left( nm...l \right)} \right) +\beta ^{\left( n \right)}\left( \delta ^{\left( n \right)}+\delta ^{\left( nm...l \right)} \right) +\alpha ^{\left( n \right)}\left( \gamma ^{\left( n \right)}+\gamma ^{\left( nm...l \right)} \right) 
\\
&+\left( \alpha ^{\left( n \right)}+\alpha ^{\left( nm...l \right)} \right) \gamma ^{\left( n \right)}+\left( b^{\left( n \right)}+b^{\left( nm...l \right)} \right) c^{\left( n \right)}-a^{\left( n \right)}\left( d^{\left( n \right)}+d^{\left( nm...l \right)} \right) =0,
\\
&\left( e^{\left( n \right)}+e^{\left( nm...l \right)} \right) \left( \delta ^{\left( n \right)}+\delta ^{\left( nm...l \right)} \right) -\left( \gamma ^{\left( n \right)}+\gamma ^{\left( nm...l \right)} \right) \left( b^{\left( n \right)}+b^{\left( nm...l \right)} \right)  
\\
&+\left( \alpha ^{\left( n \right)}+\alpha ^{\left( nm...l \right)} \right)\left( d^{\left( n \right)}+d^{\left( nm...l \right)} \right) =0.
\end{split}
\end{equation}

\section{Replicated geometry of supersymmetric $AdS_2$}\label{ap3}

\quad In this Appendix, we focus on introducing the replicated geometry into supersymmetric $AdS_2$. The replicated geometry can be thought of as the $n$-fold Riemann surfaces with $Z_n$ symmetry. This process mainly follows the replicate process in \cite{2406.16339}.

The metric of non-supersymmetric Lorentzian $AdS_2$ is
\begin{equation}
ds^2 = \frac{-dt^2 + dz^2}{z^2} = \frac{4\left| dW \right|^2}{\left( 1 - \left| W \right|^2 \right)^2}. \label{me}
\end{equation}
There exists a mapping to the replica symmetry that leads to $W(w)$ (and the complex conjugate $\bar{W}(\bar{w})$):
\begin{equation}
ds^2 = \frac{4\left| \partial_w W(w) \right|^2}{\left( 1 - \left| W(w) \right|^2 \right)^2} \left| dw \right|_n^2,
\end{equation}
where the index $n$ refers to the $n$-replica manifold.

We will focus on extending the supersymmetric $AdS_2$ with replicated geometry. There exist many degrees of freedom in supersymmetric $AdS_2$ that coexist with SJT; we can write down a simple form of supersymmetric $AdS_2$:
\begin{equation}
ds^2 = \frac{-dt^2 + dz^2}{z^2} + \frac{\theta \, dt \, d\theta + \bar{\theta} \, dt \, d\bar{\theta}}{z}. \label{me2}
\end{equation}
In this metric, there is a natural supersymmetrization relation derived from \eqref{me}:
\begin{equation}
\frac{dt^2}{z^2} \rightarrow \frac{dt^2}{z^2} - \frac{\theta \, dt \, d\theta}{z} - \frac{\bar{\theta} \, dt \, d\bar{\theta}}{z}, \label{su3}
\end{equation}
which is exactly a quadratic form in superspace. 
We can also verify the supersymmetrization relation for the spacelike component $z$:
\begin{equation}
\frac{dz^2}{z^2} \rightarrow \frac{dz^2}{z^2},
\end{equation}
where the fermionic component vanishes due to the summation over $z$ and $\bar{z}$. According to \eqref{me2}, there is no Grassmann coordinate $\theta$ coupled to $z$.

Since the supersymmetrization \eqref{me2} preserves homogeneity in $dt$ and $z$, we can naturally introduce $\mathbb{W}$ to represent the quadratic form in superspace \eqref{su3}:
\begin{equation}
ds^2 = -\frac{dt^2}{z^2} + \frac{\theta \, dt \, d\theta}{z} + \frac{\bar{\theta} \, dt \, d\bar{\theta}}{z} + \frac{dz^2}{z^2} = \frac{4\left| d\mathbb{W} \right|^2}{\left( 1 - \left| \mathbb{W} \right|^2 \right)^2}.
\end{equation}
We can further express this in terms of a super differential $D{W}$ 
\begin{equation}
ds^2 = \frac{4\left| d\mathbb{W} \right|^2}{\left( 1 - \left| \mathbb{W} \right|^2 \right)^2} = \frac{4\left| D{W} \right|^2}{\left( 1 - \left| \mathbb{W} \right|^2 \right)^2}, \label{differ}
\end{equation}
where the super differential is defined as
\begin{equation}
D = dt \, d\theta \, D_{\theta}.
\end{equation}
Here, the superscript $W$ indicates that the super differential is expressed in the coordinates $W$ and $\bar{W}$. In \eqref{differ}, we have absorbed the supersymmetry effects of $\mathbb{W}$ into the super differential $D^{W}$. Note that $W$ corresponds to the purely bosonic part, matching the non-supersymmetric case in \eqref{me}.

We then obtain a suitable coordinate representation for the replicated supersymmetry of $W(w)$:
\begin{equation}
ds^2 = \frac{4\left| D_w W(w) \right|^2}{\left( 1 - \left| \mathbb{W}(w) \right|^2 \right)^2} \left| Dw \right|_n^2,
\end{equation}
where $D^W$ and $D^w$ are super differentials associated with the coordinates $W$ and $w$ respectively. This equation naturally preserves reparametrization invariance through its complete dependence on the super differential $\left| D^w \right|_n^2$. 

Remarkably, this structure admits a $Z_n$ extension that can be iterated infinitely, generating a fractal-like replicated geometry. One can readily verify that this construction maintains the same reparametrization properties under flow and remains consistent with \eqref{gef}.

The supersymmetric replicated geometry must reduce to the single-copy case when $n \rightarrow 1$. Importantly, the geometric signatures of replicated supersymmetric $NAdS_2$ should mirror those of replicated supersymmetric $AdS_2$. Following \cite{12}, we can implement the $NAdS_2$ boundary conditions individually within each copy.

\section{Compare to the chain structure}\label{ap4}

\quad Another non-supersymmetric approach to resolving the replica problem is the SYK chain structure. In this appendix, we compare this method with the supersymmetric approach discussed in the main text.

\begin{table}
\centering
 \linespread{2}\selectfont
\setlength{\tabcolsep}{0.5mm}
   \begin{tabular}{|c|c|}
    \hline
     \textbf{Chain structure} & \textbf{Supersymmetric framework in this work} \\ \hline
     Based on non-supersymmetric SYK model & Based on $\mathcal{N}=1$ supersymmetric SYK model  \\ \hline
     Includes random Hamiltonian  & Includes random supercharge and Hamiltonian \\ \hline
     Multiple copies of non-supersymmetric SYK models  & Multiple copies of supersymmetric SYK models\\ \hline
     Sum over identical copied models  & Extended fermionic regions with ordering \\ \hline
     Non-supersymmetric interactions  & Supersymmetric interactions \\ \hline
     Sum over inter-copy interactions & Single structured supersymmetric interaction \\ \hline
     Off-shell multi-replica saddle & On-shell multi-replica saddle \\ \hline
   \end{tabular}
   \caption{Analogy between statistical mechanics and the replica-trick construction. The correspondence suggests that the proper thermodynamic counterpart to thermal entropy is the \emph{modular entropy} rather than the R\'enyi entropy.}
    \label{table1}
\end{table}

This chain structure permits only two-site interactions. When considering $n > 2$ copies (e.g., $n=3$), the path integrals can propagate from replica~1 to replica~2 but cannot access replica~3 without first returning to replica~1. We demonstrate that non-supersymmetric SYK chain models cannot support connected diagrams or saddle points for $n > 2$ copies.

While circular-chain configurations allow information transfer between arbitrary replica sectors, the resulting states are generically mixed and can be expressed as sums over direct-product states. This structure obscures fine-grained information about the interior geometry of multi-boundary replica wormholes, making their detailed reconstruction particularly challenging. Moreover, the corresponding phase structure lacks sharp distinctions, further complicating the identification of physically distinct regimes. From the path-integral perspective, a circular chain of replicas effectively corresponds to tracing over intermediate sectors along the chain, thereby entangling the remaining boundaries with an uncontrolled set of degrees of freedom. This partial trace converts otherwise pure multi-boundary wormhole states into mixed states, erasing phase-coherent information that would be necessary to fully probe the wormhole interior. Consequently, the gravitational saddle encodes only coarse-grained connectivity, while fine-grained structure, such as the detailed mapping of boundary operators to interior modes, is lost. In contrast, the ordered supersymmetric construction summarized in Table~\ref{table1} preserves coherent inter-replica correlations by restricting to structured on-shell interactions, thereby retaining access to a larger fraction of the interior information.

\end{document}